\documentclass[aps,prd,nofootinbib,onecolumn,10pt,superscriptaddress,notitlepage]{revtex4-2}

\usepackage[utf8]{inputenc}
\usepackage[a4paper,total={7.0in,10in}]{geometry}
\usepackage{mathrsfs}
\usepackage{amsmath}
\usepackage{amssymb}
\usepackage{epsfig}
\usepackage{epstopdf}
\usepackage{graphicx}
\usepackage{booktabs}
\usepackage{float}
\usepackage[dvipsnames]{xcolor}
\usepackage{multirow}
\usepackage{bm}
\usepackage{appendix}
\usepackage{slashed}
\usepackage{cancel}
\usepackage{morefloats}
\usepackage{siunitx}
\usepackage{subcaption}
\usepackage[colorlinks=true, urlcolor=blue, citecolor=blue, linkcolor=red]{hyperref}

\begin{document}
	
	\title{A Phenomenological Study of Semileptonic $B^+$ and $B_s^0$ Decays into Axial-Vector Mesons $\big(D_1(2420),\, D_1^\prime(2430),\, D_{s1}(2460),\, \text{and } D_{s1}^\prime(2536)\big)$ within the Standard Model}
	
	\author{Rana Khan}
	\affiliation{National Centre for Physics, Quaid-i-Azam University Campus, 45320, Islamabad, Pakistan}
	
	\author{Qazi Maaz Us Salam}
	\email{qazimaaz92@gmail.com}
	\affiliation{National Centre for Physics, Quaid-i-Azam University Campus, 45320, Islamabad, Pakistan}
	\affiliation{Department of Physics, Lahore University of Management Sciences (LUMS), Opposite Sector U, D.H.A, Lahore 54792, Pakistan}
	
	\author{Zohaib Aarfi}
	\affiliation{Department of Physics and Astronomy, School of Natural Sciences, National University of Sciences and Technology, H-12, Islamabad 44000, Pakistan}
	
	\author{Ishtiaq Ahmed}
	\affiliation{National Centre for Physics, Quaid-i-Azam University Campus, 45320, Islamabad, Pakistan}
	
	\begin{abstract}
		\begin{center}
			\textbf{Abstract}
		\end{center}
		\noindent
		We study semileptonic $B$ meson decays $B^+ \to  D_1^{(\prime)}\ell^+\nu_\ell$ and
		$B_s^0 \to D_{s1}^{-(\prime)}\ell^+\nu_\ell$, where $\ell=\mu,\tau$. The final state axial vector mesons are treated as mixtures of the heavy quark basis states with light degree of freedom angular momenta $j_\ell=1/2$ and $j_\ell=3/2$, parametrized by the mixing angle $\theta_{D_1}$. Using form factors obtained in the covariant light front quark model, we analyze the dependence of various observables on $\theta_{D_1}$ such as polarized and unpolarized branching ratios, the lepton forward-backward asymmetry, the longitudinal polarization fraction, and the lepton flavor universality ratios. In addition, we also discuss correlations among different observables. We study these observables in the experimentally motivated mixing angle regions as well as over a wider range of $\theta_{D_1}$. Our results show that branching ratios and other observables are sensitive to the axial-vector mixing structure. These predictions provide useful Standard Model benchmarks for future measurements of semileptonic $B_{(s)}$ decays into orbitally excited mesons and may help to clarify the long standing $1/2$ vs. $3/2$ puzzle through semileptonic $B$ decays.
	\end{abstract}
	\maketitle
	\renewcommand{\baselinestretch}{1.3}
	\normalsize
	\newpage
\section{INTRODUCTION}
The study of meson decays is one of the tools to test the theoretical framework of the Standard Model (SM) and to probe the internal structure of hadrons. In this context, semileptonic $B_{(s)} \to D_{(s)J}$ transitions have attracted considerable attention in recent decades, as they provide a clean environment to study heavy-quark dynamics and the properties of orbitally excited charm mesons \cite{Becirevic:2012te, Azizi:2014nta, Alomayrah:2022lne, Barakat:2022lmr, Hwang:2006cua, Li:2016tbj, Gan:2014jxa, Aliev:2008jb, Zhao:2006at, Wei:2005ag, Wang:2006fg},  
Here $D_{(s)J}$ denotes both nonstrange and strange charmed mesons: $D_J$ represents states with light-quark content $c\bar q$ $(q=u,d)$, whereas $D_{sJ}$ represents the corresponding strange states with $c\bar s$ content. The lowest $s$-wave states include the pseudoscalar
and vector mesons, while the $p$-wave multiplet contains the scalar, axial-vector, and tensor states. In the present work, we focus on the axial-vector $p$-wave states $D_{1}(2420)$, $D_{1}^\prime(2430)$, $D_{s1}(2460)$, and $D^\prime_{s1}(2536)$.

They are classified using the spectroscopic $^{(2S+1)}L_J$ notation. 
The scalar mesons $D^*_0(2300)$ and $D^*_{s0}(2317)$, which have quantum 
numbers $J^P=0^+$, correspond to the $^{3}P_{0}$ state, while the 
axial-vector mesons with $J^P=1^+$ arise from the two basis states 
$^{1}P_{1}$ and $^{3}P_{1}$. In the heavy-quark limit, these states can 
also be classified using the $P_J^{j_\ell}$ notation,\footnote{Here 
$j_\ell$ denotes the total angular momentum of the light degrees of freedom. 
For $p$-wave charmed mesons, the $j_\ell=1/2$ doublet contains the 
$J^P=(0^+,1^+)$ states, while the $j_\ell=3/2$ doublet contains the 
$J^P=(1^+,2^+)$ states.} for the axial-vector states, finite charm-quark-mass effects induce mixing  between the $j_\ell=1/2$ and $j_\ell=3/2$ basis states. Therefore, the 
physical states are not pure $j_\ell$ eigenstates. They are commonly identified according to their dominant components: $D_1(2430)$ and $D_{s1}(2460)$ are mainly $j_\ell=1/2$ states, whereas $D_1(2420)$ and $D_{s1}(2536)$ are mainly $j_\ell=3/2$ states.

In the current study, we investigate the decays corresponding to 
$D_{1}(2420)\equiv D_{1}$, $D_{1}^\prime(2430)\equiv D_{1}^\prime$, 
$D_{s1}(2460)\equiv D_{s1}$, and $D_{s1}(2536)\equiv D_{s1}^\prime$. The physical axial-vector states are 
obtained from the mixing of the $j_\ell=1/2$ and $j_\ell=3/2$ basis states, 
denoted by $|D_{1}^{1/2}\rangle$ and $|D_{1}^{3/2}\rangle$, respectively.
This mixing can be parameterized in terms of a single mixing angle
$\theta_{D_{1}}$~\cite{Wang:2018psi}
\begin{align}
|D_{(s)1}\rangle &=
|D_{1}^{1/2}\rangle \sin\theta_{D_{1}}
+ |D_{1}^{3/2}\rangle \cos\theta_{D_{1}}, \label{mixing eq 1} \\
|D_{(s)1}^{\prime}\rangle &=
-\, |D_{1}^{3/2}\rangle \sin\theta_{D_{1}}
+ |D_{1}^{1/2}\rangle \cos\theta_{D_{1}}.
\label{mixing eq 2}
\end{align}
The physical meaning of the mixing angle $\theta_{D_1}$ can be understood as
follows. In the heavy quark limit $m_c \to \infty$, the spin of the charm quark
decouples, $j_\ell$ becomes a good quantum number, and the two $J^P = 1^+$ states
are pure $|D_1^{1/2}\rangle$ and $|D_1^{3/2}\rangle$ states. In this limit the
mixing angle vanishes. For a finite charm quark mass, the two basis states mix,
and for the $c\bar{s}$ states the $SU(3)$ breaking also contributes. The mixing
angle therefore measures the amount of heavy quark spin symmetry breaking in the
$p$-wave charm sector. It is a property of the final state mesons and not a free
parameter of the decay amplitude, so that the same angle should describe the
production and the strong decays of these states. Since the $j_\ell = 1/2$ and
$j_\ell = 3/2$ contributions enter with opposite relative signs in
Eqs.~(\ref{mixing eq 1}) and (\ref{mixing eq 2}), the branching ratios of the two
physical states are anti-correlated, which is the behavior we obtain in
Sec.~\ref{corelationD1}.

In the literature, as reported in~\cite{Yang:2025abu}, the analysis of the mixing angle $\theta_{D_1}$ was performed in the range $-45^\circ \leq \theta_{D_1} \leq -10^\circ$. It was further noted that the experimentally favored window for the decays $B \to (D_1, D_1') \ell \nu_\ell$ lies within $-30.3^\circ \leq \theta_{D_1} \leq -24.9^\circ$. Motivated by this constraint, we first restrict our study to this experimentally preferred region in order to evaluate and analyze the relevant observables for the decays $B_s^0 \to (D_1^-, D_1^{-\prime}) \ell^+ \nu_\ell$. In addition, we extend our analysis to the second experimentally allowed mixing-angle window, $43.3^\circ \leq \theta_{D_1} \leq 49.9^\circ$. We also perform a wider scan over the mixing angle, which provides an alternative allowed region corresponding to the same physical classification of the two axial-vector mesons.

It is worth noting that several puzzles exist in the spectroscopy of excited
$p$-wave charmed mesons, including the $SU(3)$ mass-hierarchy problem and the
low-mass puzzle associated mainly with the $D_0^*(2300)$, $D_{s0}^*(2317)$, and $D_{s1}(2460)$ states, as discussed in Refs.~\cite{Godfrey:1985xj,Godfrey:1986wj,Belle:2003nsh,BaBar:2009pnd}.
Another related issue is the long-standing so-called ``$1/2$ vs. $3/2$''
puzzle in semileptonic $B$ decays~\cite{Bigi:2007qp,Morenas:1997nk,Scora:1995ty,Colangelo:1998ga,Zhang:2025pde}. The so-called ``$1/2$ vs. $3/2$'' puzzle refers to the tension between the heavy-quark-limit expectation and the observed pattern of semileptonic $B$ decays into orbitally excited charmed mesons. In the heavy-quark limit, transitions to the broad $j_\ell=1/2$ states are expected to be suppressed
relative to those into the narrow $j_\ell=3/2$ states. However, experimental measurements indicate that the rates into the $j_\ell=1/2$ sector are not as small as expected, leading to a discrepancy between theoretical predictions and observed branching fractions~\cite{Belle:2003nsh,BaBar:2009pnd,Belle:2022yzd}. This tension between theoretical predictions and experimental measurements has motivated various interpretations of the internal structure of these states, including compact tetra-quark configurations, hadronic molecular states, and possible mixing between conventional $\bar{c}s$ mesons and four-quark components, among other scenarios discussed in the refs~\cite{Barnes:2003dj,Hofmann:2003je,Cheng:2003kg,Nowak:2003ra,Browder:2003fk,Bardeen:2003kt,Chen:2004dy,Kim:2005gt,Kolomeitsev:2003ac,Maiani:2004vq,Guo:2006fu,Wang:2006uba,Vijande:2006hj,Xie:2010zza,Cleven:2014oka,Xiao:2016hoa}.

From a theoretical perspective, semileptonic $B_{(s)}$ decays to the excited charmed mesons $D_0(2300)$, $D_{s0}(2317)$, $D_1(2420)$, $D_1'(2430)$, $D_{s1}(2460)$, and $D_{s1}'(2536)$ have been extensively studied using various approaches, including the covariant light-front quark model (CLFQM)~\cite{Yang:2025abu,Wang:2007sxa,Cheng:2003sm,Yang:2024qij,Cheng:2004yj,Hwang:2006cua,Lu:2007sg,Wuenqi:2025tup}, the Isgur–Scora–Grinstein–Wise (ISGW) model~\cite{Cheng:2003id}, QCD sum rules~\cite{Dai:1998ca,Huang:2001qh,Zuo:2023ksq}, light-cone sum rules (LCSRs)~\cite{Gubernari:2022hrq}, and heavy quark effective theory (HQET)~\cite{Leibovich:1997tu,Leibovich:1997em,Bernlochner:2016bci,Bernlochner2017}. In the present work, we employ hadronic form factors calculated within the CLFQM framework~\cite{Yang:2025abu} to study semileptonic $b \to c$ transitions. Using these inputs, we analyze a range of physical observables for both polarized and unpolarized final-state mesons and investigate the impact of meson polarization on the full decay distributions.

In addition, a major motivation for studying flavor-changing charged-current (FCCC) processes such as $b \to c \ell \bar{\nu}$ in the $B$-meson sector is the presence of persistent tensions between experimental measurements and SM predictions. These discrepancies, typically at the level of $2\sigma$–$4\sigma$, are primarily observed in observables such as $R(D^{(*)})$, $R(J/\psi)$, and $\tau$-polarization asymmetries~\cite{Blanke:2018yud,Fedele:2022iib,Yasmeen:2024cki,Dutta:2018jxz,Li:2016vvp,Sakaki:2013bfa,FermilabLattice:2014ysv,Ligeti:2016npd,Belle:2019ewo,Belle:2019gij,LHCb:2017vlu,Kamali:2018fhr,Colangelo:2016ymy}, which provide important probes of possible new physics (NP) effects. To address these anomalies, several studies beyond the SM in the $b \to c$ sector have been performed in recent years~\cite{Chang:2018sud,Mahata:2024kxu,Mahata:2023geq,Wei:2023rzs,Karmakar:2023rdt,Sheng:2022peg,Sheng:2021iss,Zhang:2019hth,Huang:2025kof,Zafar:2025mhe}. In this context, we also study different lepton flavor universality (LFU) ratios as a complementary analysis.

In this work, we present a phenomenological analysis of the semileptonic decays $B^+ \to  D_1^{(\prime)}\ell^+\nu_\ell$ and
$B_s^0 \to D_{s1}^{-(\prime)}\ell^+\nu_\ell$, where $\ell=\mu,\tau$. We study several physical observables, including the branching ratios and lepton forward-backward asymmetry $\mathcal{A}_{\rm FB}$ with the longitudinal and transverse polarization of the final-state meson, the longitudinal polarization fraction $F_L$ and lepton flavor universality ratio $\mathcal{R}_{D_{1}^{(\prime)}}$ and $\mathcal{R}_{D_{s1}^{(\prime)}}$. Keeping in view the experimental data of the branching ratio for $B^+ \to D_1,  D_1^\prime$, it is expected that in future the results of other decay distributions and different angular observables will be available with much better statistics. Therefore, in addition to the branching ratio calculation, we have studied the impact of other polarized and unpolarized observables over the parameter $\theta_{D1}$ in wide range of angle.

The form factors used in this work are taken from Ref.~\cite{Yang:2025abu}. In
Ref.~\cite{Yang:2025abu}, the $q^2$ distributions, the forward-backward asymmetry
and the longitudinal polarization fraction were calculated at a single fixed value
of the mixing angle, $\theta_s=7^\circ$, and only the branching ratios were
studied as a function of the mixing angle, in the range
$-45^\circ\leq\theta_s\leq-14^\circ$. Moreover, the $q^2$-dependent results of
Ref.~\cite{Yang:2025abu} are shown as central curves only, without their
uncertainties. In the present work, we extend this analysis as follows:
\begin{itemize}
\item All observables are calculated within the mixing-angle windows allowed by
the experimental data, $\theta_{D_1}\in[-30.3^\circ,-24.9^\circ]$ and
$\theta_{D_1}\in[43.3^\circ,49.9^\circ]$, instead of at a single fixed angle.
\item The mixing angle is varied over a wider range,
$-45^\circ\leq\theta_{D_1}\leq80^\circ$, and in addition to the branching ratios,
the dependence of $\mathcal{A}_{\rm FB}$ and $F_{\rm L}$ on $\theta_{D_1}$ is
also studied.
\item The form-factor uncertainties are included in all figures as error bands.
\item The forward-backward asymmetry is calculated separately for the
longitudinally and transversely polarized final-state meson,
$\mathcal{A}_{\rm FBL}$ and $\mathcal{A}_{\rm FBT}$.
\item All observables, i.e.\ the unpolarized and polarized differential branching
ratios, $\mathcal{A}_{\rm FB}$, $\mathcal{A}_{\rm FBL}$, $\mathcal{A}_{\rm FBT}$,
$F_{\rm L}$ and the ratios $\mathcal{R}_{D_1^{(\prime)}}$ and
$\mathcal{R}_{D_{s1}^{(\prime)}}$, are given as functions of $q^2$ and are also
calculated bin-wise over the full $q^2$ range, with a bin width of
$\Delta q^2=1~\mathrm{GeV}^2$ (TABLEs~\ref{table3}--\ref{tableRD}). In
Ref.~\cite{Yang:2025abu}, no bin-wise results are given: $F_{\rm L}$ is given only
in two $q^2$ regions, and the other observables only as integrated values.
\item The correlations among the branching ratios, $\mathcal{A}_{\rm FB}$ and
$F_{\rm L}$ are studied.
\end{itemize}

The rest of the paper is organized as follows. In Sec.~\ref{framework}, we present the theoretical framework and define the relevant observables used in our analysis. In Sec.~\ref{PHENOMENOLOGICALANALYSIS}, we carry out a comprehensive phenomenological study. In particular, in Sec.~\ref{thetasection}, we examine the dependence of various observables on the mixing angle. The $q^2$-dependent behavior of these observables is analyzed in Sec.~\ref{q2section}. Furthermore, in Sec.~\ref {corelationD1}, we provide the correlations among different observables. Finally, in Sec.~\ref{concl}, we conclude our work and give the supporting details in the appendices.

\section{Theoretical Framework}\label{framework}
At the quark level,  the semileptonic transition $b \to c\, \bar{\ell}\, \nu_\ell$
is governed by the weak effective Hamiltonian given as follows,   
\begin{equation}
\mathcal{H}^{b\to c}_{\rm eff}
=
\frac{G_F}{\sqrt{2}}\, V_{cb} (\bar{c}\gamma_\mu P_L b)
(\bar{\ell}\gamma^\mu P_L \nu_\ell)
\label{eq:Heff_bc}
\end{equation}

where $P_L = (1-\gamma_5)/2$, $G_F$ is the Fermi constant, and $V_{\text{cb}}$ is the relevant CKM matrix element.
 It is established that the physical states $D_{(s)1}^{(\prime)} = \{D_1, D_1^\prime, D_{s1}, D_{s1}^\prime$\} are mixtures of $D_{1}^{3/2}$ and $D_{1}^{1/2}$ such as defined in Eq.~(\ref{mixing eq 1}) and Eq.~(\ref{mixing eq 2}). Consequently, the hadronic transition between the initial $B^{+}$ and $B_{(s)}$ meson and the final axial-vector meson $D_{1}^{(\prime)}$ and $D_{s1}^{(\prime)}$ can be expressed in terms of these mixed states.  Furthermore, we can write these hadronic matrix elements (HMEs) in the weak eigenstate basis as,
\begin{align}
\langle ^iD_{1}(k) | \gamma^\mu | B^{0,+}_{(s)}(p) \rangle
&=
-\, i \Bigg[
(m_{B^{0,+}_{(s)}} - m_{^iD_1})\, \varepsilon^{\mu\ast}
V_1^{^iD_1}(q^2)
- \frac{\varepsilon^{\ast} \!\cdot\! P}{m_{B^{0,+}_{(s)}}- m_{^iD_1}}\, 
P_{\mu}\,  V_2^{^iD_1}(q^2)
\nonumber \\[2mm]
\hspace{1.5cm}
&-\,  2 m_{^iD_1}\, 
\frac{\varepsilon^{\ast} \!\cdot\! P}{q^2}\, 
q^\mu
\left(
V_3^{^iD_1}(q^2)-V_0^{^iD_1}(q^2)
\right)
\Bigg], 
\label{Veq}
\\
\langle ^iD_{1}(k) | \gamma^\mu \gamma^5 | B^{0,+}_{(s)}(p) \rangle
&=
-\, \frac{1}{m_{B^{0,+}_{(s)}}- m_{^iD_1}}\, 
\epsilon^{\mu\nu\alpha\beta}\, 
\varepsilon^{\nu\ast} P^\alpha\,  q^\beta\, 
A_0^{^iD_1}(q^2),
\label{AVeq}
\end{align}

Here $p$($k$) denotes the four-momentum of the initial (final) meson and $q^2$ is the square of the momentum transfer. Here $P = p + k$,  $q = p-k$,  and the convention $\epsilon^{0123} = 1$ is adopted. The superscript $i$ labels the weak eigenstates, with $i = 1/2$ and $3/2$. Accordingly, the states are denoted as $D_1^{1/2}$ and $D_1^{3/2}$. The scalar functions $A_0(q^2)$, $V_0(q^2)$, $V_1(q^2)$, and $V_2(q^2)$ are the hadronic form factors (FFs) defined in Eq. (\ref{FFeq}),  which constitute the primary source of hadronic uncertainties and are computed within the CLFQM, as reported in~\cite{Yang:2025abu}. 

In order to calculate the decay amplitude from the effective Hamiltonian given in Eq.~(\ref{eq:Heff_bc}), the physical states $D_1$,  $D_{s1}$, $D_1^{\prime}$ and $D_{s1}^{\prime}$ will appear. These states are obtained from the weak eigenstates $D_1^{1/2}$ and $D_1^{3/2}$ through the mixing relations given in Eqs.~(\ref{mixing eq 1}) and (\ref{mixing eq 2}).
\subsection{PHYSICAL OBSERVABLE}
\label{physicalobs}
In order to compute the decay observables, we start from the transition amplitude derived from the effective Hamiltonian $\mathcal{H}^{b\to c}_{\rm eff}$ given in Eq.~(\ref{eq:Heff_bc}). Taking into account the two-body phase space, the hadronic matrix elements (HMEs) that are evaluated using the scalar products of the relevant four-momenta and the kinematic relations summarized in Appendix~\ref{APPA}--\ref{tansA}. The analysis is performed separately for the polarized and unpolarized cases. Using the HMEs defined in Eqs.~(\ref{Veq}) and (\ref{AVeq}), along with the necessary scalar products listed in Appendix~\ref{APPA}--\ref{tansA}, we obtain the final two-fold differential decay rate for the processes $B^+ \to  D_1^{(\prime)}\ell^+\nu_\ell$ and
$B_s^0 \to D_{s1}^{-(\prime)}\ell^+\nu_\ell$, where $\ell=\mu,\tau$ with a longitudinally or transversely polarized final-state meson $D^{(\prime)}_{1}$ and $D_{s1}^{(\prime)}$ as,
\begin{align}\frac{d\Gamma_{\text{L}}}{dq^2} = &{\frac{G_F^2 |V_{cb}|^2 }{192 \pi^3\, 2m^3_{B^{0,+}_{(s)}}\; q^2}} \sqrt{\lambda (m_{B_{(s)}^{0,+}}^2,m_{D^{(\prime)}_{(s)1}}^2,q^2)} \left( 1- \frac{m_\ell^2}{q^2} \right)^2 \times \nonumber \\& \Bigg\{ 3 m_\ell^2 \, \lambda(m_{B_{(s)}^{0,+}}^2, m_{D_{(s)1}^{(\prime)}}^2, q^2)\, V_0^2(q^2) + (m_\ell^2 + 2q^2) \nonumber \\& \times \left| \frac{1}{2m_{D_{(s)1}^{(\prime)}}} \left[ (m_{B_{(s)}^{0,+}}^2 - m_{D_{(s)1}^{(\prime)}}^2 - q^2)(m_{B_{(s)}^{0,+}} - m_{D_{(s)1}^{(\prime)}})\, V_1(q^2) - \frac{\lambda(m_{B_{(s)}^{0,+}}^2, m_{D_{(s)1}^{(\prime)}}^2, q^2)}{m_{B_{(s)}^{0,+}} - m_{D_{(s)1}^{(\prime)}}}\, V_2(q^2) \right] \right|^2 \Bigg\}\label{LdecayrateBc} \end{align}
For the transverse final state meson polarized the two-fold angular decay distribution is, 
\begin{align}\frac{d\Gamma_{\pm}}{dq^2} = &{\frac{G_F^2 |V_{cb}|^2 }{192 \pi^3\, 2m^3_{B^{0,+}_{(s)}}\; q^2}} \sqrt{\lambda (m_{B_{(s)}^{0,+}}^2,m_{D^{(\prime)}_{(s)1}}^2,q^2)} \left( 1- \frac{m_\ell^2}{q^2} \right)^2 \times \nonumber \\& \Bigg\{(m_\ell^2 + 2q^2)\, \lambda(m_{B_{(s)}^{0,+}}^2, m_{D_{(s)1}^{(\prime)}}^2, q^2) \left| \frac{A_0(q^2)}{m_{B_{(s)}^{0,+}} - m_{D_{(s)1}^{(\prime)}}} \mp \frac{(m_{B_{(s)}^{0,+}} - m_{D_{(s)1}^{(\prime)}})\, V_1(q^2)}{\sqrt{\lambda(m_{B_{(s)}^{0,+}}^2, m_{D_{(s)1}^{(\prime)}}^2, q^2)}} \right|^2\Bigg\}\label{TdecayrateBc} \end{align}

Here $\lambda(m_{B_{(s)}^{0,+}}^2,m_{D^{(\prime)}_{(s)1}}^2,q^2) = (m_{B_{(s)}^{0,+}}^2 + m_{D^{(\prime)}_{(s)1}}^2-q^2)^2- 4m_{B_{(s)}^{0,+}}^2m_{D^{(\prime)}_{(s)1}}^2$ is the Kallen function, and $m_\ell$ denotes the mass of the charged lepton with $\ell =  \mu $, $ \tau$.
The form factors entering Eqs.~(\ref{LdecayrateBc}) and (\ref{TdecayrateBc})
are those of the \emph{physical} states $D^{(\prime)}_{(s)1}$, which are built from
the weak-eigenstate form factors through Eqs.~(\ref{mixing eq 1}) and
(\ref{mixing eq 2}) at every value of $q^2$; this point is discussed in detail in
Sec.~\ref{PHENOMENOLOGICALANALYSIS}, see Eq.~(\ref{eq:remix}).

\begin{itemize}
\item  The polarized and unpolarized differential branching ratio are related to the decay rate as follows, 
\begin{equation}
\frac{d^2\mathcal{B}}{dq^2}=\tau_{B^{0,+}_{(s)}}\frac{d^2\Gamma}{dq^2},\quad \quad  \frac{d^2\Gamma}{dq^2}=\frac{d^2\Gamma_{\text{L}}}{dq^2}+\frac{d^2\Gamma_{+}}{dq^2}+\frac{d^2\Gamma_{-}}{dq^2},
    \label{brdecayrelation}
\end{equation}

\begin{equation}
 \frac{d^2\mathcal{B}_{\text{L}}}{dq^2}=\tau_{B^{0,+}_{(s)}}\frac{d^2\Gamma_{\text{L}}}{dq^2},\quad \quad  \frac{d^2\mathcal{B}_{\text{T}}}{dq^2}=\tau_{B^{0,+}_{(s)}}\Big(\frac{d^2\Gamma_{+}}{dq^2}+\frac{d^2\Gamma_{-}}{dq^2}\Big),
    \label{brdecayrelation1}
\end{equation}
where $\tau_{B^{0,+}_{(s)}}$ is the lifetime of the initial meson $B_s^0 $ and $B^+$.

\item For the decay the lepton forward-backward asymmetry is defined with respect to the angle $\theta$,  which denotes the angle between the three momentum of the charged lepton and that of the final state meson in the dilepton center of mass frame. It is given by,
\begin{equation}
\mathcal{A}_{\rm FB}(q^2)
=
\frac{
\displaystyle
\int_{0}^{1}
\frac{d^2\Gamma}{dq^2\,  d\cos\theta}\,  d\cos\theta
-
\int_{-1}^{0}
\frac{d^2\Gamma}{dq^2\,  d\cos\theta}\,  d\cos\theta
}
{
\displaystyle
\int_{-1}^{1}
\frac{d^2\Gamma}{dq^2\,  d\cos\theta}\,  d\cos\theta
}.
\label{AFBeq}
\end{equation}
\item The longitudinal polarization fraction $F_{\text{L}}(q^2)$ as a function of the momentum transfer is defined as
\begin{equation}
F_{\text{L}}(q^2)
=
\frac{\displaystyle
\int_{-1}^{1}
\frac{d^2\Gamma_{\text{L}}}{dq^2\,  d\cos\theta}\,  d\cos\theta}
{\displaystyle
\int_{-1}^{1}
\frac{d^2\Gamma}{dq^2\,  d\cos\theta}\,  d\cos\theta}, 
\label{Fleq}
\end{equation}
where $\Gamma_L$ denotes the partial decay width corresponding to the
longitudinal polarization of the final state meson. 

\item We also construct the ratio of the branching fractions of the tau and muon modes. In the present context for the $B^+ \to  D_1^{(\prime)}\ell^+\nu_\ell$ and
$B_s^0 \to D_{s1}^{-(\prime)}\ell^+\nu_\ell$, we define the ratio $\mathcal{R}_{D_{1}^{(\prime)}}$ and $\mathcal{R}_{D_{s1}^{(\prime)}}$ as, 
\begin{align}
\mathcal{R}_{D_{1}^{(\prime)}}(q^2)& =
\frac{
\displaystyle \int_{q_{\min}^2}^{q_{\max}^2}
\frac{d\mathcal{B}(B_{}^{+} \to D_{1}^{(\prime)}\,\tau \nu)}{dq^2}\, dq^2
}{
\displaystyle \int_{q_{\min}^2}^{q_{\max}^2}
\frac{d\mathcal{B}(B_{}^{+} \to D_{1}^{(\prime)}\,\mu \nu)}{dq^2}\, dq^2
} \quad \quad \mathcal{R}_{D_{s1}^{(\prime)}}(q^2)& =
\frac{
\displaystyle \int_{q_{\min}^2}^{q_{\max}^2}
\frac{d\mathcal{B}(B_{s}^{0} \to D_{s1}^{-(\prime)}\,\tau \nu)}{dq^2}\, dq^2
}{
\displaystyle \int_{q_{\min}^2}^{q_{\max}^2}
\frac{d\mathcal{B}(B_{s}^{0} \to D_{s1}^{-(\prime)}\,\mu \nu)}{dq^2}\, dq^2
}
\label{rdratio}
\end{align}
where the integration is over the appropriate  $q^2$ bins for comparison with the experimental data.
\item We also constructed  polarized branching ratio and lepton forward-backward asymmetry
for the decay. The combined transverse and total differential decay widths are defined
as
\begin{equation}
\frac{d\Gamma}{dq^2}
=
\frac{d\Gamma_{\text{L}}}{dq^2}
+
\frac{d\Gamma_{\text{T}}}{dq^2},
\qquad
\frac{d\Gamma_{\text{T}}}{dq^2}
=
\frac{d\Gamma_+}{dq^2}
+
\frac{d\Gamma_-}{dq^2}.
\end{equation}
 The $\Gamma_+$ and
$\Gamma_-$ represent the contributions from the two transverse helicity
states \cite{Yang:2025abu}.
\item  For the longitudinally (\text{L}) and transversely (\text{T}) polarized cases, the lepton forward-backward asymmetry is defined using the same expression as in Eq.~(\ref{AFBeq}), with the unpolarized decay rate replaced by the corresponding polarized decay rates $\Gamma_{\text{L}}$ and $\Gamma_{\text{T}}$, respectively.
\end{itemize}
\section{PHENOMENOLOGICAL ANALYSIS}
\label{PHENOMENOLOGICALANALYSIS}
In this section, we analyze the physical observables introduced in Section~\ref{physicalobs} within the framework of the SM. To do so, the input parameters taken from~\cite{ParticleDataGroup:2024cfk} relevant to our analysis are given in TABLE~\ref{tabnumerics}.

\begin{table}[h]
\centering
\caption{The values of the input parameters~\cite{ParticleDataGroup:2024cfk}, except $\tau_{B_s^0}$, which is taken from Ref.~\cite{Yang:2025abu}.}
\label{tabnumerics}
\begin{tabular}{cccccc}
\hline\hline
Constant & Value & Constant & Value & Constant & Value \\
\hline

$m_b$ & 4.183 $\text{GeV}$ 
& $m_{D_1}$ & 2.420 $\text{GeV}$ 
&$\tau_{B^+}$ & $1.638 \times 10^{-12}$ s\\

$m_\mu$ & 0.106 $\text{GeV}$ 
& $m_{B^+}$ & 5.279 $\text{GeV}$
& $V_{cb}$ & $40.3 \times 10^{-3}$ \\

$m_u$ &0.003 $\text{GeV}$
& $m_{D_1^{\prime}}$ & 2.430 $\text{GeV}$ 
& $\tau_{B_s^0}$ & $1.520 \times 10^{-12}$ s \\

 $m_s$ & 0.093 $\text{GeV}$  
&$m_{D_{s1}}$ &  2.460 $\text{GeV}$ 
& $m_{D^{\prime}_{s1}}$&2.536  $\text{GeV}$ \\
$m_\tau$ & 1.77 $\text{GeV}$
&$m_{B_s^0}$ & 5.367 $\text{GeV}$
& $G_{F}$&$1.166 \times 10^{-5} $ $\text{GeV}^{-2}$ \\

$m_c$ & 1.273 $\text{GeV}$
& & & & \\
\hline\hline
\end{tabular}
\end{table}
The hadronization of quarks and gluons is described in terms of transition FFs, which constitute the dominant source of theoretical uncertainties. 
In our numerical analysis, the FFs that we use for $B^+ \to  D_1^{(\prime)}\ell^+\nu_\ell$ and
$B_s^0 \to D_{s1}^{-(\prime)}\ell^+\nu_\ell$, where $\ell=\mu,\tau$ are calculated within
the CLFQM, as reported in~\cite{Yang:2025abu}.  These form factors are
extrapolated using a double pole parametrization of the form, 
\begin{align}
F(q^2) =
\frac{F(0)}
{1- a \left(\dfrac{q^2}{m_{B^{0,+}_{(s)}}^2}\right)
+ b \left(\dfrac{q^2}{m_{B^{0,+}_{(s)}}^2}\right)^2},
\label{FFeq}
\end{align}
where $m_{B^{0,+}_{(s)} }$ denotes the masses of the initial $B^{0,+}_{(s)}$ mesons and $F(q^2)$
represents the relevant transition FFs, namely $A_0^{D_{(s)1}^{(\prime)}}$, $V_0^{D_{(s)1}^{(\prime)}}$, $V_1^{D_{(s)1}^{(\prime)}}$ and $V_2^{D_{(s)1}^{(\prime)}}$ given in TABLE~\ref{ffvalues}.
\begin{table}[H]
\centering
\caption{The form factors of the transitions
$B^+ \to  D_1^{(\prime)}\ell^+\nu_\ell$ and
$B_s^0 \to D_{s1}^{-(\prime)}\ell^+\nu_\ell$,
in the covariant light-front quark model~\cite{Yang:2025abu}.}
\label{TableFFDs1}
\begin{tabular}{cccc||cccc}
\hline\hline
FFs & $F(0)$  & $a$ & $b$& FFs & $F(0)$  & $a$ & $b$\\
\hline
$A_{0}^{D_{1}}$ & $+0.21^{+0.028}_{-0.014}$  & $-0.23^{+0.135}_{-0.127}$  & $+0.09^{+0.028}_{-0.028}$&$A_{0}^{D_{s1}}$&$+0.20^{+0.022}_{-0.022}$&$-0.27^{+0.802}_{-0.114}$&$+0.11^{+0.028}_{-0.036}$\\
$V_{0}^{D_{1}}$ &$+0.42^{+0.050}_{-0.045}$    &$+0.16^{+0.036}_{-0.054}$   & $-0.02^{+0.010}_{-0.010}$&$V_{0}^{D_{s1}}$&$+0.40^{+0.045}_{-0.045}$&$-0.17^{+0.045}_{-0.072}$&$-0.02^{+0.010}_{-0.010}$ \\
$V_{1}^{D_{1}}$ & $+0.57^{+0.042}_{-0.041}$  &$-0.07^{+0.010}_{-0.010}$   & $+0.02^{+0.000}_{-0.000}$ &$V_{1}^{D_{s1}}$&$+0.58^{+0.032}_{-0.045}$&$-0.05^{+0.014}_{-0.014}$&$+0.02^{+0.010}_{-0.000}$\\
$V_{2}^{D_{1}}$ & $-0.06^{+0.022}_{-0.022}$   & $+0.57^{+0.180}_{-0.212}$  &  $+2.13^{+1.317}_{-1.184}$&$V_{2}^{D_{s1}}$&$-0.05^{+0.022}_{-0.010}$&$+0.56^{+0.228}_{-0.257}$&$+2.50^{+1.689}_{-1.315}$\\
$A_{0}^{D^{\prime}_{1}}$ & $+0.07^{+0.028}_{-0.028}$    & $-1.90^{+0.150}_{-0.172}$ & $+5.64^{+1.605}_{-1.303}$ &$A_{0}^{D^{\prime}_{s1}}$&$+0.08^{+0.022}_{-0.022}$&$+2.05^{+0.373}_{-0.364}$&$+5.57^{+0.559}_{-0.456}$\\
$V_{0}^{D^{\prime}_{1}}$ & $-0.08^{+0.045}_{-0.036}$  & $-0.11^{+0.061}_{-0.117}$ & $+3.67^{+0.512}_{-0.512}$ &$V_{0}^{D^{\prime}_{s1}}$&$-0.08^{+0.041}_{-0.041}$&$+1.24^{+0.235}_{-0.257}$&$+0.74^{+0.211}_{-0.171}$\\
$V_{1}^{D^{\prime}_{1}}$ & $+0.14^{+0.050}_{-0.050}$     & $-0.36^{+0.057}_{-0.085}$ &$+0.09^{+0.014}_{-0.014}$  &$V_{1}^{D^{\prime}_{s1}}$&$+0.17^{+0.045}_{-0.042}$&$-0.52^{+0.085}_{-0.078}$&$+0.36^{+0.032}_{-0.080}$\\
$V_{2}^{D^{\prime}_{1}}$ & $+0.10^{+0.036}_{-0.028}$    & $+0.19^{+0.085}_{-0.127}$ &$+0.00^{+0.014}_{-0.010}$  &$V_{2}^{D^{\prime}_{s1}}$ &$+0.11^{+0.014}_{-0.028}$&$+0.25^{+0.085}_{-0.099}$&$-0.07^{+0.032}_{-0.050}$\\
\hline\hline
\end{tabular}
\label{ffvalues}
\end{table}
The values in TABLE~\ref{ffvalues} are taken from Ref.~\cite{Yang:2025abu}.
The uncertainties in the FFs parameters originate from both statistical and systematic sources and are quoted in quadrature. Since the covariance matrix among the fitted parameters is not available, correlations among $F(0)$,  $a$,  and $b$ are neglected,
and the total uncertainty of the FFs is estimated by standard
Gaussian error propagation. Explicitly, the uncertainty of each form factor
at a given value of $q^2$ is obtained from the double pole parametrization of
Eq.~(\ref{FFeq}) as
\begin{align}
\Delta F(q^2) = \sqrt{\left(\frac{\partial F}{\partial F(0)}\,\Delta F(0)\right)^{2}
+\left(\frac{\partial F}{\partial a}\,\Delta a\right)^{2}
+\left(\frac{\partial F}{\partial b}\,\Delta b\right)^{2}}\, ,
\label{FFerr}
\end{align}
where $\Delta F(0)$, $\Delta a$ and $\Delta b$ are the uncertainties given in
TABLE~\ref{ffvalues}. These uncertainties are then propagated to the decay rates
given in Eqs.~(\ref{LdecayrateBc}) and (\ref{TdecayrateBc}), and consequently to
the branching ratios and the angular observables. It is important to emphasize
that the form factor parameters are the only source of uncertainty in our results.
Since $\mathcal{A}_{\text{FB}}$, $F_{\text{L}}$ and the ratios
$\mathcal{R}_{D_{1}^{(\prime)}}$ and $\mathcal{R}_{D_{s1}^{(\prime)}}$ are the
ratios of the decay rates, a large part of the form factor uncertainty cancels
between the numerator and the denominator. Therefore, the uncertainties in these
observables are smaller than those in the branching ratios. In the next step, with these input parameters, the task is to perform the phenomenological analysis of the $B^+ \to  D_1^{(\prime)}\ell^+\nu_\ell$ and
$B_s^0 \to D_{s1}^{-(\prime)}\ell^+\nu_\ell$ decay, with $\ell=\mu,\tau$.

The mixing is incorporated in the form factors. Using Eqs.~(\ref{mixing eq 1})
and (\ref{mixing eq 2}), the form factors of the physical states are obtained
from those of the $j_\ell=1/2$ and $j_\ell=3/2$ states at each value of $q^2$ as
\begin{align}
F^{D_{(s)1}}(q^2) &= F^{1/2}(q^2)\sin\theta_{D_1}
+F^{3/2}(q^2)\cos\theta_{D_1}, \nonumber\\
F^{D^{\prime}_{(s)1}}(q^2) &= -F^{3/2}(q^2)\sin\theta_{D_1}
+F^{1/2}(q^2)\cos\theta_{D_1},
\label{eq:remix}
\end{align}
where $F\in\{A_0,V_0,V_1,V_2\}$. These mixed form factors then enter the
decay rates of Eqs.~(\ref{LdecayrateBc}) and (\ref{TdecayrateBc}), so that the
mixing is not applied separately in the amplitude construction.

\subsection{Comparison with other theoretical frameworks and with experiment}
\label{comparison}

The form factors are the main source of uncertainty in our results. In
TABLE~\ref{tabcompareff}, we compare the form factors at $q^2=0$ used in this
work with those of other quark-model calculations, the CLFQM of
Ref.~\cite{Cheng:2003sm} and the ISGW2 model \cite{Cheng:2003id}. For the
$j_\ell=3/2$ state, the three calculations agree within about $30\%$. For the
$j_\ell=1/2$ state, the differences are larger, in particular for $V_0$. Note
that Refs.~\cite{Cheng:2003sm,Cheng:2003id} use the opposite sign convention for
$|D_1^{1/2}\rangle$, so their $j_\ell=1/2$ values should be multiplied by $-1$
for comparison.

\begin{table}[H]
\centering
\caption{Comparison of the $B\to D_1^{3/2},\,D_1^{1/2}$ transition form
factors at $q^2=0$ in different approaches. The CLFQM values of the present
analysis are taken from Ref.~\cite{Yang:2025abu}.
The entries of Refs.~\cite{Cheng:2003sm,Cheng:2003id} for the $j_\ell=1/2$ state
are quoted in the opposite sign convention for
$|D_1^{1/2}\rangle$ and should be multiplied by $-1$ for a direct comparison.}
\label{tabcompareff}
\begin{tabular}{llcccc}
\hline\hline
Transition & Approach & $A_0(0)$ & $V_0(0)$ & $V_1(0)$ & $V_2(0)$\\
\hline
$B\to D_1^{3/2}$ & CLFQM (this work, from~\cite{Yang:2025abu}) & $+0.20$ & $+0.43$ & $+0.55$ & $-0.07$\\
                        & CLFQM~\cite{Cheng:2003sm}   & $+0.23$ & $+0.47$ & $+0.55$ & $-0.09$\\
                        & ISGW2~\cite{Cheng:2003id}   & $+0.16$ & $+0.43$ & $+0.40$ & $-0.12$\\
\hline
$B\to D_1^{1/2}$ & CLFQM (this work, from~\cite{Yang:2025abu}) & $+0.09$ & $-0.02$ & $+0.21$ & $+0.09$\\
                        & CLFQM~\cite{Cheng:2003sm}   & $-0.12$ & $+0.08$ & $-0.19$ & $-0.12$\\
                        & ISGW2~\cite{Cheng:2003id}   & $-0.16$ & $+0.18$ & $-0.19$ & $-0.18$\\
\hline\hline
\end{tabular}
\end{table}

The $B\to D_1^{(\prime)}$ form factors have also been calculated with LCSRs
\cite{Gubernari:2022hrq}, QCD sum rules \cite{Dai:1998ca,Huang:2001qh}, HQEFT
\cite{Zuo:2023ksq} and the HQET-based LLSW approach
\cite{Leibovich:1997tu,Leibovich:1997em,Bernlochner:2016bci}. Since these works
use different form-factor definitions, we compare the integrated branching ratios
instead. These are given in TABLE~\ref{tabcomparebr}, together with the
experimental values \cite{ParticleDataGroup:2024cfk}. For
$B^+\to\bar D_1^0\mu^+\nu_\mu$, our result, like the other predictions, is
somewhat larger than the measured value. For $B^+\to\bar D_1^{\prime0}\mu^+\nu_\mu$,
our result and all other predictions are well below the measured value. This
difference between theory and experiment for the $j_\ell=1/2$ state is the
``$1/2$ vs.\ $3/2$'' puzzle discussed in Sec.~I.
For the $B_s^0 \to D_{s1}^{-}\mu^+\nu_\mu$ and
$B_s^0 \to D_{s1}^{\prime-}\mu^+\nu_\mu$ channels, no experimental data are
currently available, and our results are consistent with the other theoretical
predictions within the uncertainties. It is worth noting that the form factors
used in our analysis are the same as those of Ref.~\cite{Yang:2025abu}, but our
branching ratios are different from the values reported there. The reason is that
we have calculated these observables in the experimentally allowed range of the
mixing angle, $\theta_{D_1}\in[-30.3^\circ,-24.9^\circ]$, whereas in
Ref.~\cite{Yang:2025abu} a different value of the mixing angle is used. Since the
physical states are the mixtures of the $j_\ell=1/2$ and $j_\ell=3/2$ states,
their form factors, and consequently their branching ratios, depend on the mixing
angle.

\begin{table}[H]
\centering
\caption{Integrated branching ratios (in units of $10^{-3}$) for
$B^+\to \bar D_1^{(\prime)0}\mu^+\nu_\mu$ and
$B_s^0\to D_{s1}^{-(\prime)}\mu^+\nu_\mu$ in different approaches, compared with
the experimental data. Our results are obtained in the experimentally allowed
range $\theta_{D_1}\in[-30.3^\circ,-24.9^\circ]$, and the quoted uncertainties
follow from the form factor uncertainties of TABLE~\ref{ffvalues} through
Eq.~(\ref{FFerr}).}
\label{tabcomparebr}
\begin{tabular}{lcccc}
\hline\hline
Approach & $B^+\!\to\!\bar D_1^{0}\mu^+\nu$ & $B^+\!\to\!\bar D_1^{\prime0}\mu^+\nu$
 & $B_s^0\!\to\! D_{s1}^{-}\mu^+\nu$ & $B_s^0\!\to\! D_{s1}^{\prime-}\mu^+\nu$\\
\hline
This work & $5.55^{+0.11}_{-0.12}$ & $0.28^{+0.44}_{-0.35}$ & $5.70^{+0.04}_{-0.33}$ & $0.28^{+0.03}_{-0.07}$\\
CLFQM~\cite{Yang:2025abu}     & $6.16^{+1.42}_{-1.13}$ & $0.15^{+0.18}_{-0.11}$ & $6.13^{+1.03}_{-1.23}$ & $0.18^{+0.12}_{-0.13}$\\
HQEFT~\cite{Zuo:2023ksq}      & $7.19^{+3.59}_{-2.88}$ & $0.67^{+0.31}_{-0.24}$ & $6.25^{+3.06}_{-2.46}$ & $0.64^{+0.30}_{-0.24}$\\
LLSW (HQET)~\cite{Bernlochner:2016bci} & $6.40\pm0.44$ & $0.46\pm0.37$ & $-$ & $-$\\
\hline
Experiment~\cite{ParticleDataGroup:2024cfk} & $4.26\pm0.26$ & $2.55\pm0.90$ & $-$ & $-$\\
\hline\hline
\end{tabular}
\end{table}

\subsection{Mixing Angle Dependence of Observables in $B^+ \to  D_1^{(\prime)}\ell^+\nu_\ell$ and
$B_s^0 \to D_{s1}^{-(\prime)}\ell^+\nu_\ell$ decays}
\label{thetasection}
\begin{figure}[H]
\caption{ The mixing angle $\theta_{D_{1}}$ dependencies of the polarized branching ratios, lepton forward-backward asymmetry and longitudinal fraction for the semileptonic decays $B^+ \to D_{1} \mu^+ \nu$ and $B^+ \to D_{1}^{\prime} \mu^+ \nu$. }
\vspace{-0.7cm}
\centering
\includegraphics[width=2.3in]{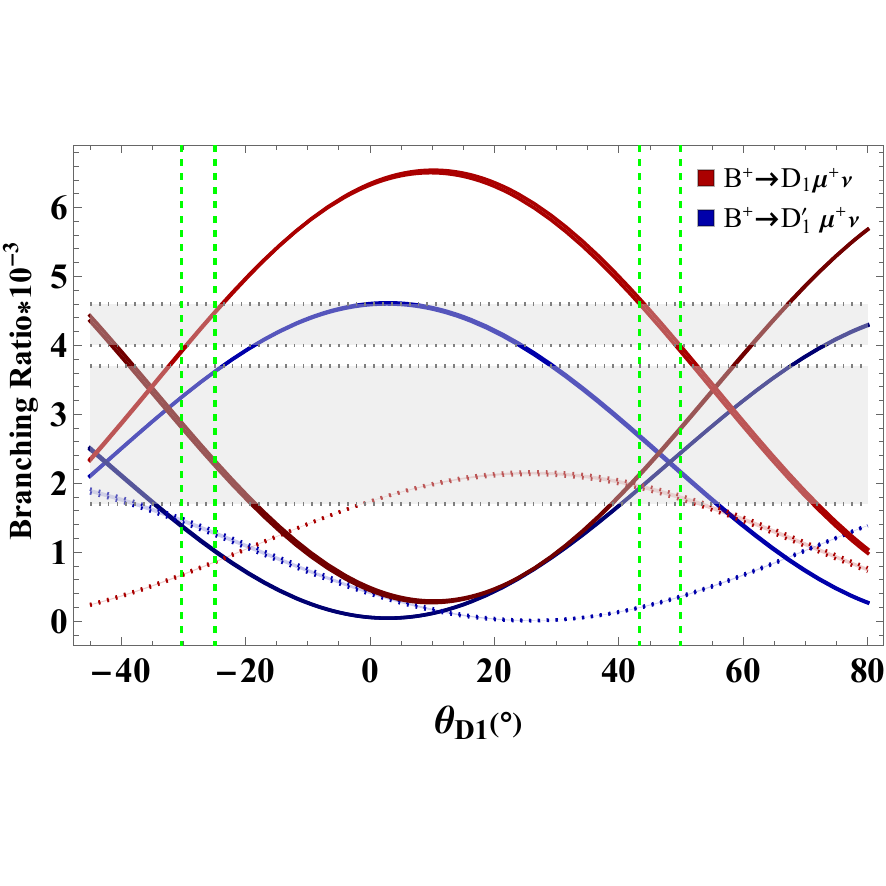} 
\includegraphics[width=2.3in]{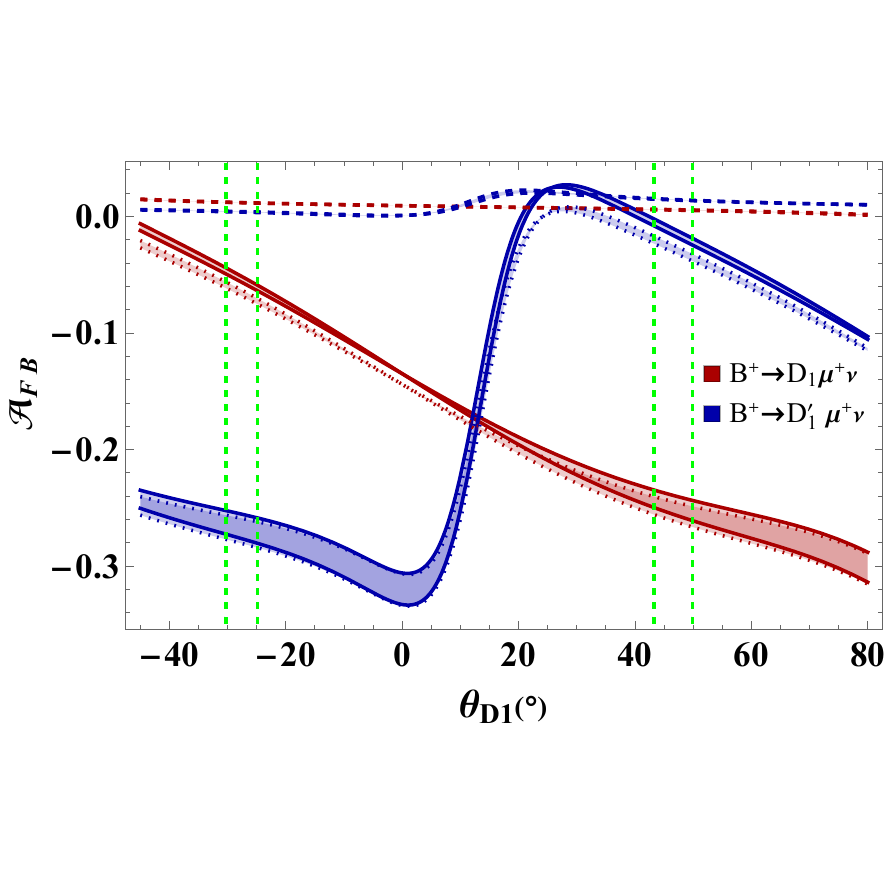}
\includegraphics[width=2.3in]{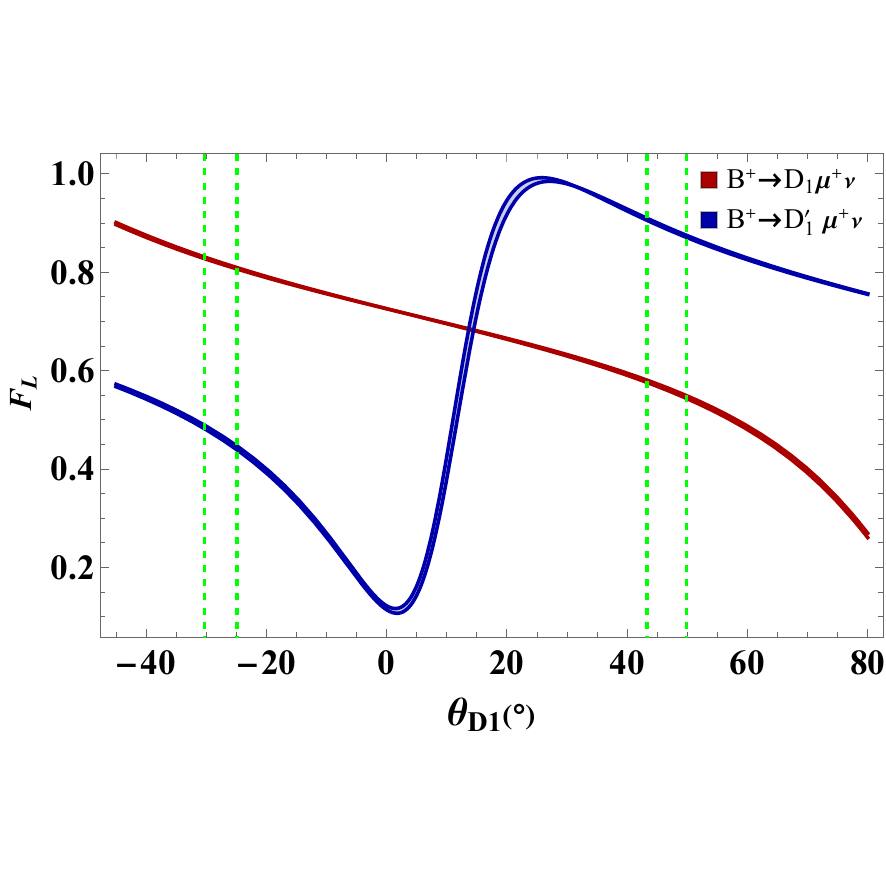}
\label{yplotD1}
\end{figure}

In FIG.~\ref{yplotD1} we present the differential branching ratios, $\mathcal{A}_{\text{FB}}$ and $F_{\text{L}}$ as functions of $\theta_{D1}$ for the decay $B^{+} \rightarrow D_{1}^{(\prime)}\mu \nu_\mu$. As indicated by the legend in the plot, the darker red band corresponds to $B^{+} \rightarrow D_{1}\mu \nu_\mu$, while the darker blue band corresponds to $B^{+} \rightarrow D_{1}^{(\prime)}\mu \nu$. The solid line shows the unpolarized behavior, while the dashed (dotted) line shows the longitudinally (transversely) polarized final state meson, respectively. The bands in these plots originate from the errors in the input parameters, mainly the FFs. For the branching ratios, we present the experimentally accessible ranges, the upper (lower) shaded band corresponds to the decay channels $B^+ \to D_{1} \mu^+ \nu$ ($B^+ \to D_{1}^{\prime} \mu^+ \nu$) respectively. The mixing angle is considered within two ranges shown in green vertical dashed lines, $\theta_{D1} \in [-30.3^\circ, -24.9^\circ]$ and $\theta_{D1} \in [43.3^\circ, 49.9^\circ]$. The obtained results lie within these ranges and are consistent within current experimental uncertainties with the currently available experimental data \cite{ParticleDataGroup:2024cfk}. At present, experimental measurements are primarily limited to the branching ratios, which alone may not be sufficient to fully resolve the existing tension between theoretical predictions and experimental data. In this work, we have also provided predictions for additional observables $\mathcal{A}_{\mathrm{\text{FB}}}$ and  $F_{\text{L}}$ within the same angle ranges. Future measurements of these observables in the $B^+ \to  D_1^{(\prime)}\ell^+\nu_\ell$ and
$B_s^0 \to D_{s1}^{-(\prime)}\ell^+\nu_\ell$ channels could provide crucial complementary information and help clarify this tension that cannot be addressed by branching ratio measurements alone. From the plots, we can read as,
\vspace{-1cm}
\begin{itemize}
\item The branching ratio of $B^+ \to D_1 \mu^+ \nu$ increases when $\theta_{D_1}$ lies in the negative range  $[-30.3^\circ, -24.9^\circ]$ and decreases in the positive range $ [43.3^\circ, 49.9^\circ]$. In contrast, the decay $B^+ \to D_1^\prime \mu^+ \nu$ shows the opposite behavior. The \text{L} and \text{T} polarized observables follow the same behavior as the unpolarized case, but with smaller magnitudes. The \text{L} contribution is larger than the \text{T} one.
\item For $\mathcal{A}_{\text{FB}}$, the $\text{T}$ polarization component is found to be dominant for both $D_1$ and $D_1^{\prime}$ channels, while the $\text{L}$ contribution remains close to zero over the full $\theta_{D1}$ range. For the $D_1$ channel, within the negative mixing angle range $\theta_{D1} \in [-30.3^\circ, -24.9^\circ]$, the forward-backward asymmetry varies as $[-0.257, -0.263]$, while in the positive range $\theta_{D1} \in [43.3^\circ, 49.9^\circ]$, it takes values $[0.002, 0.019]$. For the $D_1^{\prime}$ channel in the negative (positive) $\theta_{D1}$ range as mentioned before, the asymmetry lies within $[-0.056, -0.059]$ ($[0.250, 0.241]$), respectively.

\item Similarly, for $F_L$, we obtain $[0.828, 0.810]$ for the $D_1$ state in the mentioned negative $\theta_{D1}$ region and $[0.581, 0.543]$ in the positive $\theta_{D1}$ region. For the $D_1^{\prime}$ channel, the corresponding ranges are given by $[0.489, 0.447]$ in the negative angle region and $[0.904, 0.873]$ in the positive angle region. Since no experimental data are currently available for these observables, future experiments can probe and measure their values.
\end{itemize}
\begin{figure}[H]
\caption{ The mixing angle $\theta_{D_{1}}$ dependencies of the polarized branching ratios, lepton forward-backward asymmetry and longitudinal fraction for the semileptonic decays $B^0_s \to D_{s1}^{-} \mu^+ \nu_\ell$ and $B^0_s \to D_{s1}^{-\prime} \mu^+ \nu_\ell$. }
\vspace{-0.7cm}
\centering
\includegraphics[width=2.3in]{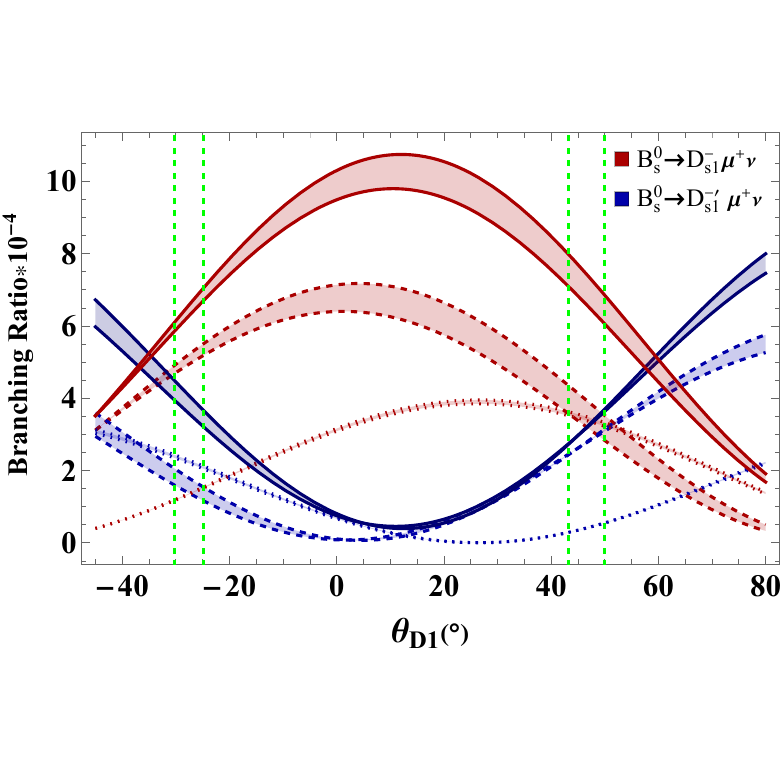} 
\includegraphics[width=2.3in]{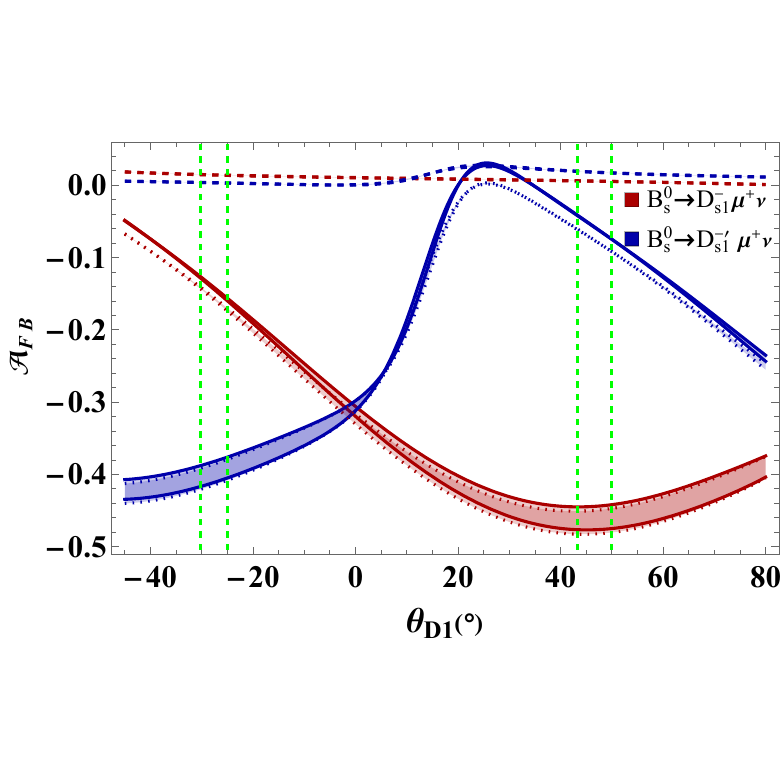}
\includegraphics[width=2.3in]{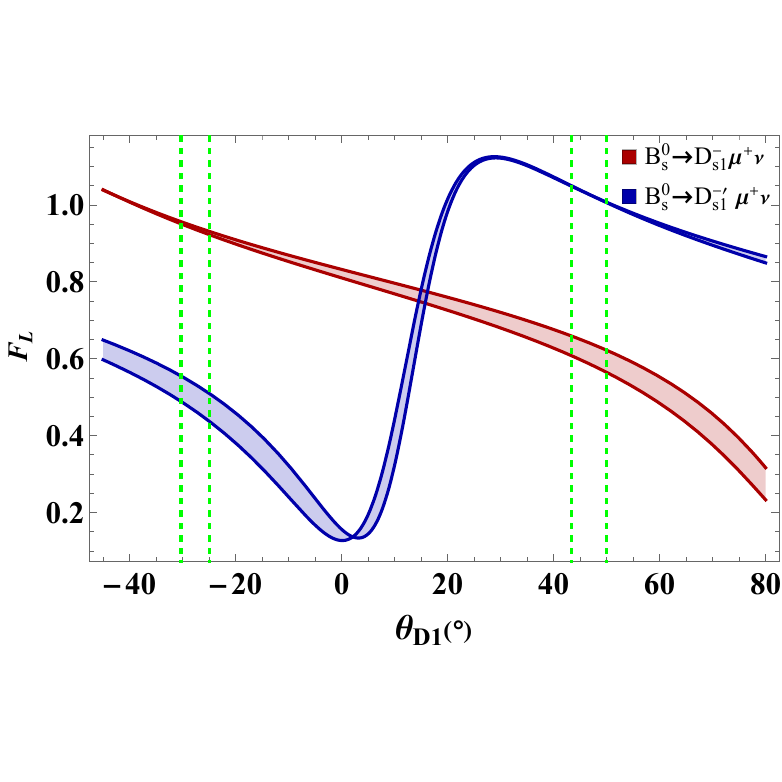}
\label{yplotDs1}
\end{figure}

\vspace{-1.0cm}
\begin{itemize}
\item  In FIG.~\ref{yplotDs1}, the branching ratio of $B_s^0 \to D^-_{s1} \mu^+ \nu$ increases in $\theta_{D1} \in [-30.3^\circ, -24.9^\circ]$ range, where it varies within $[0.610, 0.702]\times10^{-3}$, and decreases in the range $\theta_{D1} \in[43.3^\circ, 49.9^\circ]$, taking values within $[0.799, 0.606]\times10^{-3}$. In contrast, the decay $B_s^0 \to D_{s1}^{-\prime} \mu^+ \nu$ shows the opposite behavior, with the branching ratio lying in the range $[0.456, 0.324]\times10^{-3}$ for the negative $\theta_{D1}$ range mentioned above and $[0.247, 0.313]\times10^{-3}$ for positive $\theta_{D1}$. The \text{L} and \text{T} polarized observables follow the same trend as the unpolarized case, but with smaller magnitudes. The \text{L} contribution remains dominant over the \text{T} contribution across the full range of the mixing angle.

\item For $\mathcal{A}_{\text{FB}}$, the $\text{T}$ polarization component remains dominant for both $D_{s1}$ and $D_{s1}^{\prime}$ channels, while the $\text{L}$ contribution stays close to zero over the full $\theta_{D1}$ range. For the $D_{s1}$ state, within the negative mixing angle range $\theta_{D1} \in [-30.3^\circ, -24.9^\circ]$, the forward-backward asymmetry varies as $[-0.411,-0.393]$, while in the positive range $\theta_{D1} \in [43.3^\circ, 49.9^\circ]$, it takes values $[-0.042, -0.091]$. For the $D_{s1}^{\prime}$ channel, $\mathcal{A}_{\text{FB}}$ spans the ranges $[-0.128, -0.160]$ in the aforementioned negative angle region and $[-0.467, -0.458]$ in the positive angle region. The two channels show clearly different magnitudes across the given $\theta_{D1}$ range.

\item Similarly, $F_L$ shows a $\theta_{D1}$ dependence for both channels. For the $D_{s1}$ state, we obtain $[0.951, 0.921]$ in the $\theta_{D1} \in [-30.3^\circ, -24.9^\circ]$ region and $[0.635, 0.594]$ in the $\theta_{D1} \in [43.3^\circ, 49.9^\circ]$ region. For the $D_{s1}^{\prime}$ channel, the corresponding ranges are shown in the plots.\end{itemize}
In the absence of experimental data for these observables for $B^+ \to  D_1^{(\prime)}\ell^+\nu_\ell$ and
$B_s^0 \to D_{s1}^{-(\prime)}\ell^+\nu_\ell$ decays, our results may serve as theoretical predictions; furthermore, future experiments can test these predicted values. The different patterns seen for both $D_{s1}$ and $D_{s1}^{\prime}$ in the negative and positive $\theta_{D1}$ region show that these physical observables, such as the branching ratio, $\mathcal{A}_{\text{FB}}$ and $F_L$ are highly sensitive to the mixing angle. This sensitivity can help us better understand the current tension between theoretical predictions and experimental results.

\subsection{$q^2$-dependent analysis of the Observables in the $B^+ \to  D_1^{(\prime)}\ell^+\nu_\ell$ and
$B_s^0 \to D_{s1}^{-(\prime)}\ell^+\nu_\ell$ decays.}
\label{q2section}
In this section, we compute the branching ratios, $\mathcal{A}_{\text{FB}}$, and $F_{\text{L}}$ as functions of the momentum transfer $q^2$ for the decay channels $B^+ \to  D_1^{(\prime)}\ell^+\nu_\ell$ and
$B_s^0 \to D_{s1}^{-(\prime)}\ell^+\nu_\ell$, where $\ell=\mu,\tau$. We also present the polarized branching ratio and lepton forward-backward asymmetry for the cases where the final state meson is L or T polarized. The red curve corresponds to the unpolarized result, while the blue and purple curves represent the $\text{L}$ and $\text{T}$ polarization contributions, respectively, as shown in FIGs~(\ref{btod1lotmuon}-\ref{btods1ptau}). The bands in these plots arise from the uncertainties in the input parameters, mainly the form factors, as well as from the maximum and minimum values of $\theta_{D1}$, where $\theta_{D1} \in [-30.3^\circ, -24.9^\circ]$ and $\theta_{D1} \in [43.3^\circ, 49.9^\circ]$. It is important to emphasize that the spread associated with $\theta_{D1}$ does not represent a theoretical uncertainty, but instead, it arises from varying the mixing angle over its allowed range.
The uncertainties quoted in this work therefore arise only from the form factors of TABLE~\ref{ffvalues}.

We present the bin-averaged values of the observables using a bin width of $\Delta q^2 = 1~\mathrm{GeV}^2$. The corresponding binned differential branching ratio is defined by,
\begin{align}
    \left\langle \frac{d\mathcal{B}}{dq^2} \right\rangle_{[q^2_{\min},\, q^2_{\max}]} 
= \frac{\displaystyle \int_{q^2_{\min}}^{q^2_{\max}} \left( \frac{d\mathcal{B}}{dq^2} \right) dq^2}
{q^2_{\max} - q^2_{\min}} \, .
\end{align}
The binned lepton forward-backward and longitudinal fraction are defined as,
\begin{align}
\left\langle F_L \right\rangle_{[q^2_{\min},\,q^2_{\max}]}
&=
\frac{
\displaystyle \int_{q^2_{\min}}^{q^2_{\max}}
F_L(q^2)\,\frac{d\Gamma}{dq^2}\,dq^2
}{
\displaystyle \int_{q^2_{\min}}^{q^2_{\max}}
\frac{d\Gamma}{dq^2}\,dq^2
},\qquad
\left\langle \mathcal{A}_{\rm FB} \right\rangle_{[q^2_{\min},\,q^2_{\max}]}
=
\frac{
\displaystyle \int_{q^2_{\min}}^{q^2_{\max}}
\mathcal{A}_{\rm FB}(q^2)\,\frac{d\Gamma}{dq^2}\,dq^2
}{
\displaystyle \int_{q^2_{\min}}^{q^2_{\max}}
\frac{d\Gamma}{dq^2}\,dq^2
}.
\end{align}

The binned analysis of the ratio $R_{D_{1}^{(\prime)}}$ and $R_{D_{s1}^{(\prime)}}$  is performed using the following definition,
\begin{align}
\left\langle \mathcal{R}_{D_{1}^{(\prime)}} \right\rangle_{[q^2_{\min},\, q^2_{\max}]} 
= 
\frac{
\displaystyle \int_{q^2_{\min}}^{q^2_{\max}}
\frac{d\mathcal{B}(B_{}^{+} \to D_{1}^{(\prime)}\,\tau^+ \nu_\tau)}{dq^2}\, dq^2
}{
\displaystyle \int_{q^2_{\min}}^{q^2_{\max}}
\frac{d\mathcal{B}(B_{}^{+} \to D_{1}^{(\prime)}\,\mu^+ \nu_\mu)}{dq^2}\, dq^2
}, \quad \quad \left\langle \mathcal{R}_{D_{s1}^{(\prime)}} \right\rangle_{[q^2_{\min} q^2_{\max}]} 
= 
\frac{
\displaystyle \int_{q^2_{\min}}^{q^2_{\max}}
\frac{d\mathcal{B}(B_{s}^{0} \to D_{s1}^{-(\prime)}\,\tau^+ \nu_\tau)}{dq^2}\, dq^2
}{
\displaystyle \int_{q^2_{\min}}^{q^2_{\max}}
\frac{d\mathcal{B}(B_{s}^{0} \to D_{s1}^{-(\prime)}\,\mu^+ \nu_\mu)}{dq^2}\, dq^2
}
\end{align}
\subsubsection{Observable for the decay of $B^+ \to D_{1} \ell^+ \nu_\ell$.}
\begin{figure}[H]
\caption{ The polarized and unpolarized branching ratio, lepton forward-backward asymmetry, and the  longitudinal fraction as a function of $q^2$ for $B^+ \to D_{1} \mu^+ \nu$ transition.} 
\vspace{-0.7cm}
  \begin{subfigure}{0.33\textwidth}
    \includegraphics[width=\linewidth]{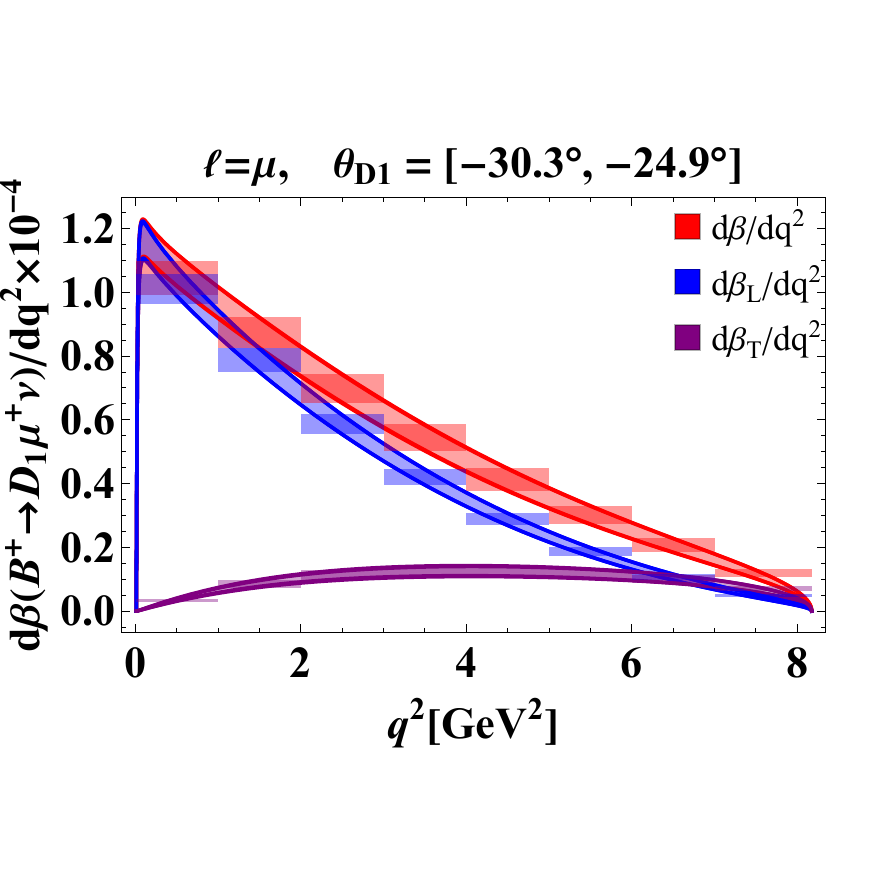}
    \vspace{-1.5cm}
    \caption{}
    \label{fig3a}
  \end{subfigure}%
  \hfill
  \begin{subfigure}{0.33\textwidth}
    \includegraphics[width=\linewidth]{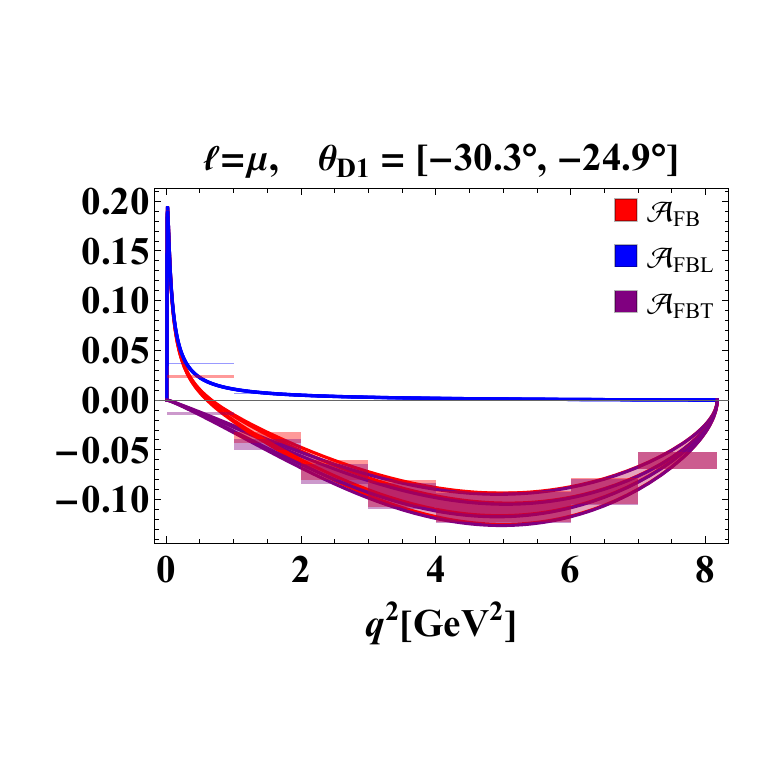}
    \vspace{-1.5cm}
    \caption{}
    \label{fig3b}
  \end{subfigure}%
  \hfill
  \begin{subfigure}{0.33\textwidth}
    \includegraphics[width=\linewidth]{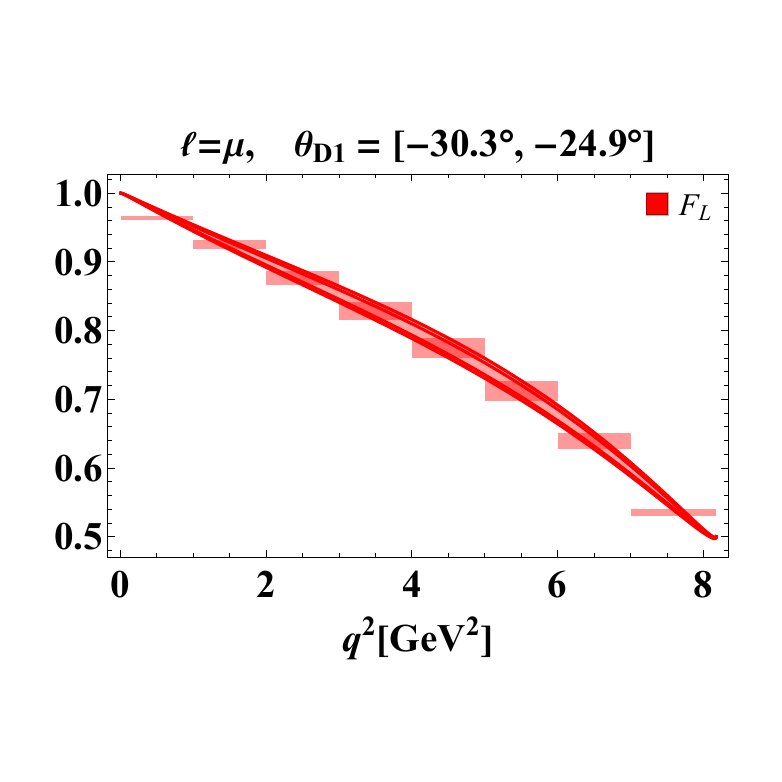}
    \vspace{-1.5cm}
    \caption{}
    \label{fig3c}
  \end{subfigure}
 \vspace{-0.9cm} \\ 
    \begin{subfigure}{0.33\textwidth}
    \includegraphics[width=\linewidth]{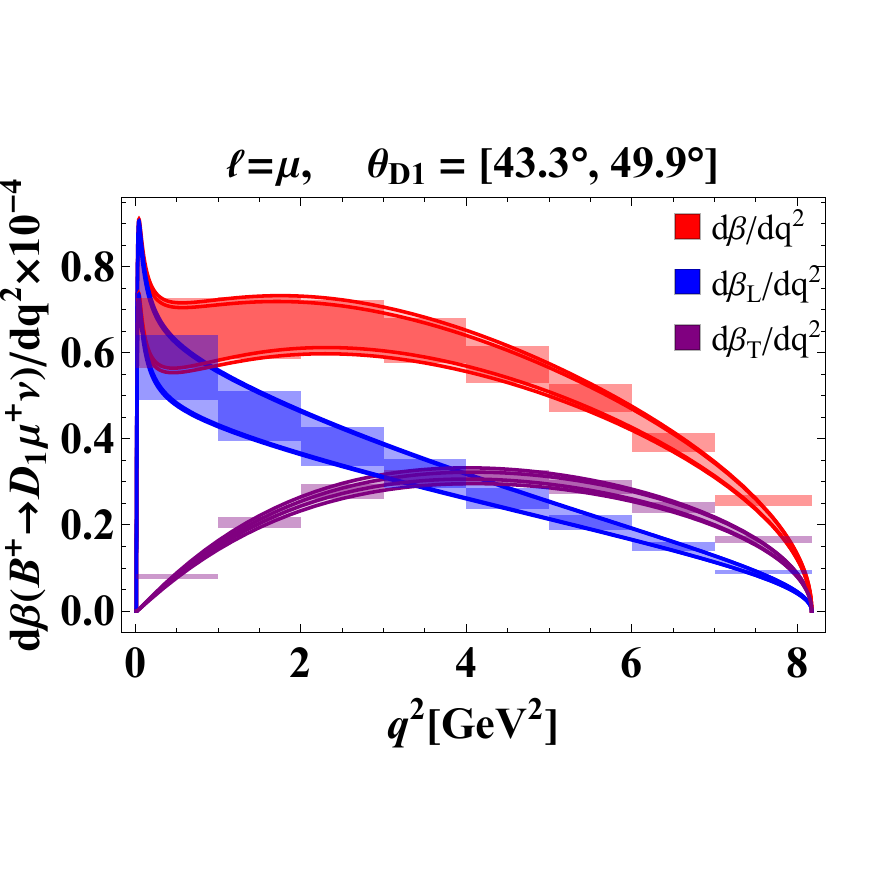}
    \vspace{-1.5cm}
    \caption{}
    \label{fig3d}
  \end{subfigure}%
  \hfill
  \begin{subfigure}{0.33\textwidth}
    \includegraphics[width=\linewidth]{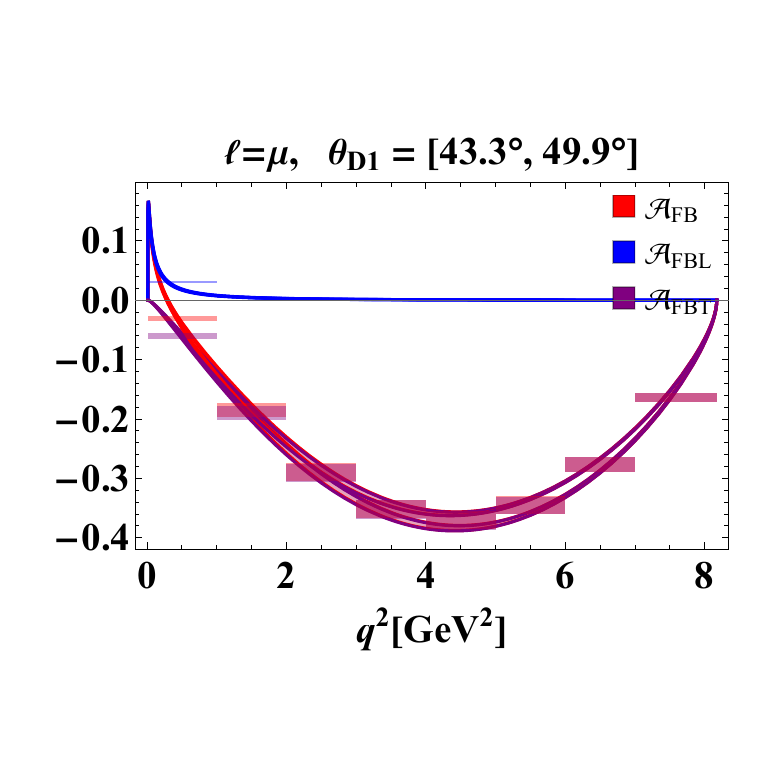}
    \vspace{-1.5cm}
    \caption{}
    \label{fig3e}
  \end{subfigure}%
  \hfill
  \begin{subfigure}{0.33\textwidth}
    \includegraphics[width=\linewidth]{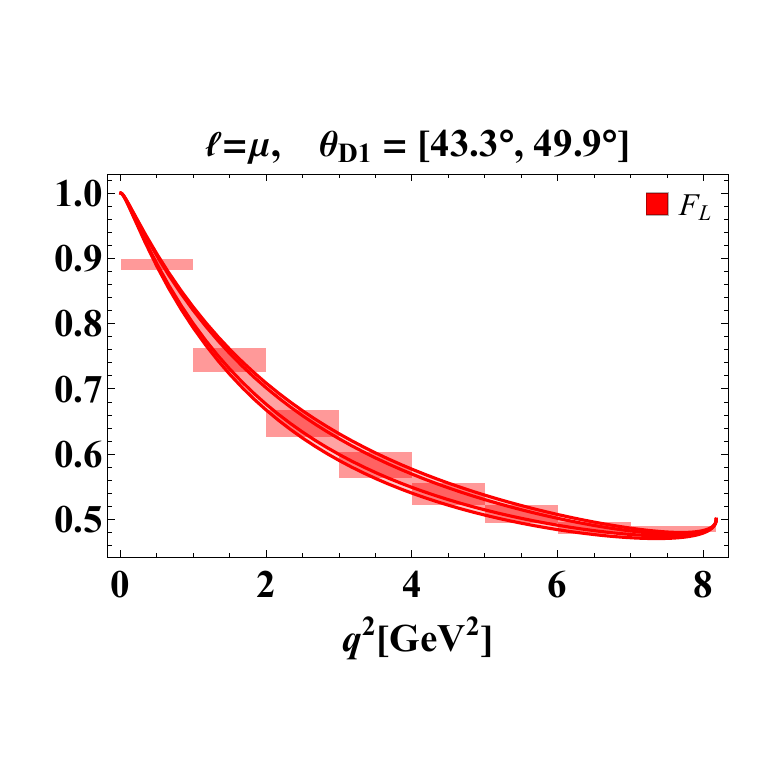}
    \vspace{-1.5cm}
    \caption{}
    \label{fig3f}
  \end{subfigure}
  \label{btod1lotmuon}
\end{figure}

 FIGs.~\ref{btod1lotmuon} shows the polarized and unpolarized branching ratio, lepton forward-backward asymmetry, and longitudinal fraction as functions of $q^2$ for the $B^+ \to D_{1} \mu^+ \nu$ transition. The first row corresponds to the fixed mixing-angle range $\theta_{D1} \in [-30.3^\circ, -24.9^\circ]$, while the second row corresponds to $\theta_{D1} \in [43.3^\circ, 49.9^\circ]$.
\begin{itemize}

\item FIG.~\ref{fig3a} shows the differential branching ratio, including both the unpolarized result and the contributions from the polarized final state meson. As we observe, over the whole kinematic region, the L polarization contribution is dominant over the T one.
The L and T contributions do not follow the same trend: over most of the kinematic range the T component remains strongly suppressed compared to the L one. For example, around $q^2\!\sim\!4~\mathrm{GeV}^2$, one finds approximately $d\mathcal{B}_L/dq^2 \!\sim\! 0.35\times10^{-4}$, whereas $d\mathcal{B}_T/dq^2 \!\sim\! 0.12\times10^{-4}$; however, near high $q^2$ or in the last bin, the T contribution becomes slightly dominant, with $d\mathcal{B}_{\rm T}/dq^2 \!\sim\! (0.61\text{--}0.78)\times10^{-5}$ and $d\mathcal{B}_{\rm L}/dq^2 \!\sim\! (0.43\text{--}0.54)\times10^{-5}$, in agreement with the last row of TABLE~\ref{table3}. In FIG.~\ref{fig3d}, the behavior of the two polarization components differs noticeably. The L contribution exhibits a smooth and relatively large distribution in the range $q^2 \lesssim 2.4~\mathrm{GeV}^2$, whereas the T component stays smaller. Beyond this point the T contribution becomes dominant, reaching roughly twice the magnitude of the L one in the range $q^2\sim6$--$8~\mathrm{GeV}^2$, as shown in TABLE~\ref{table3}.

\item FIG.~\ref{fig3b} shows the $\mathcal{A}_{\text{FB}}$ for the decay $B^+ \to D_1 \mu \nu$ as a function of $q^2$. The L contribution, $\mathcal{A}_{\rm FBL}$, is positive and sharply enhanced in the very low-$q^2$ region, reaching about $0.19$ near $q^2\simeq 0$, but then decreases rapidly and remains close to zero over most of the $q^2$ region. In contrast, the T contribution, $\mathcal{A}_{\rm FBT}$, is negative throughout the entire $q^2$ region, decreases to a minimum of about $-0.12$ around $q^2\sim 5~{\rm GeV}^2$, and then rises again toward zero near high $q^2$. The unpolarized asymmetry, $\mathcal{A}_{\rm FB}$, follows the transverse behavior closely, becoming negative after the very low-$q^2$ region, it crosses zero at approximately $q^2 \sim 0.6~\mathrm{GeV}^2$, then becomes negative and reaches about $-0.10$ in the intermediate region before approaching zero near high-$q^2$ region. For FIG.~\ref{fig3e}, the behavior follows the same trend as observed in FIG.~\ref{fig3b}, but with a comparatively larger magnitude. The distribution reaches its maximum in the bin $q^2 = 3\text{--}4~\mathrm{GeV}^2$, where it lies in the range $(-0.355,-0.368)$. This is roughly a factor of three to four larger than the corresponding values obtained in the negative angle region, which are about $(-0.083,-0.109)$ in the same $q^2$ bin.
\item FIG.~\ref{fig3c} and \ref{fig3f} shows the $F_{\text{L}}$ for both angle ranges where the bin-wise values of $F_{\text{L}}$ are listed in the TABLE~\ref{table3} show agreement with the plots. 
\end{itemize}

\begin{figure}[H]
\caption{ The polarized and unpolarized branching ratio, lepton forward-backward asymmetry, and the  longitudinal fraction as a function of $q^2$ for $B^+ \to D_{1} \tau^+ \nu$ transition.} 
\vspace{-0.7cm}
\begin{subfigure}{0.33\textwidth}
    \includegraphics[width=\linewidth]{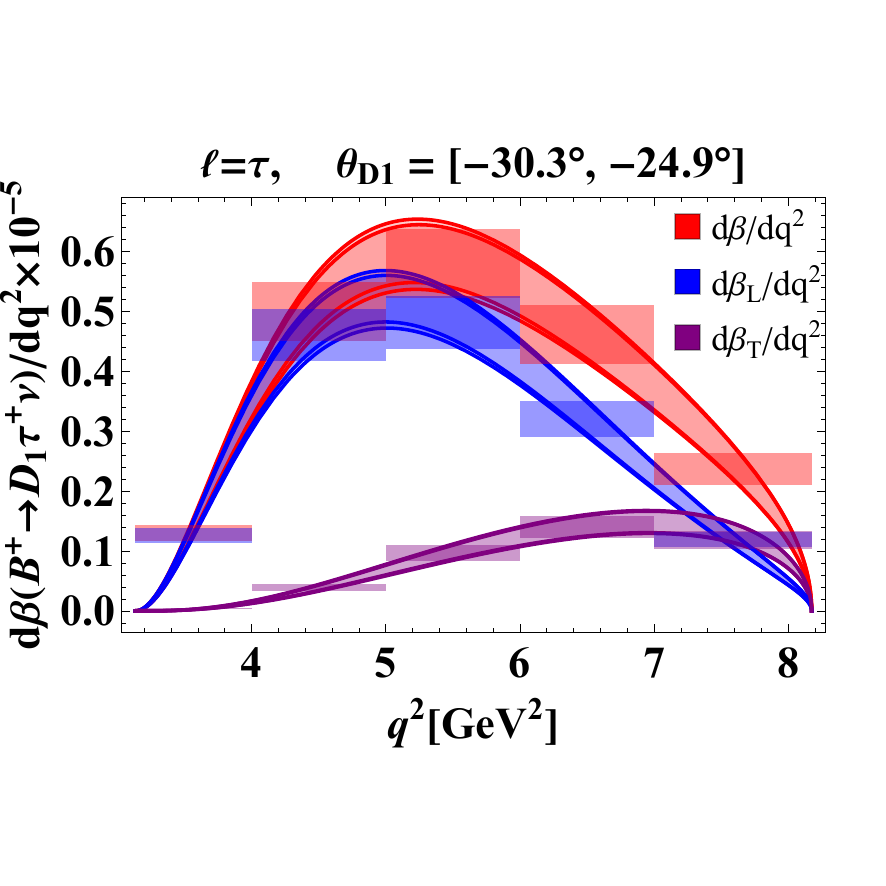}
    \vspace{-1.5cm}
    \caption{}
    \label{fig:4a}
  \end{subfigure}%
  \hfill
  \begin{subfigure}{0.33\textwidth}
    \includegraphics[width=\linewidth]{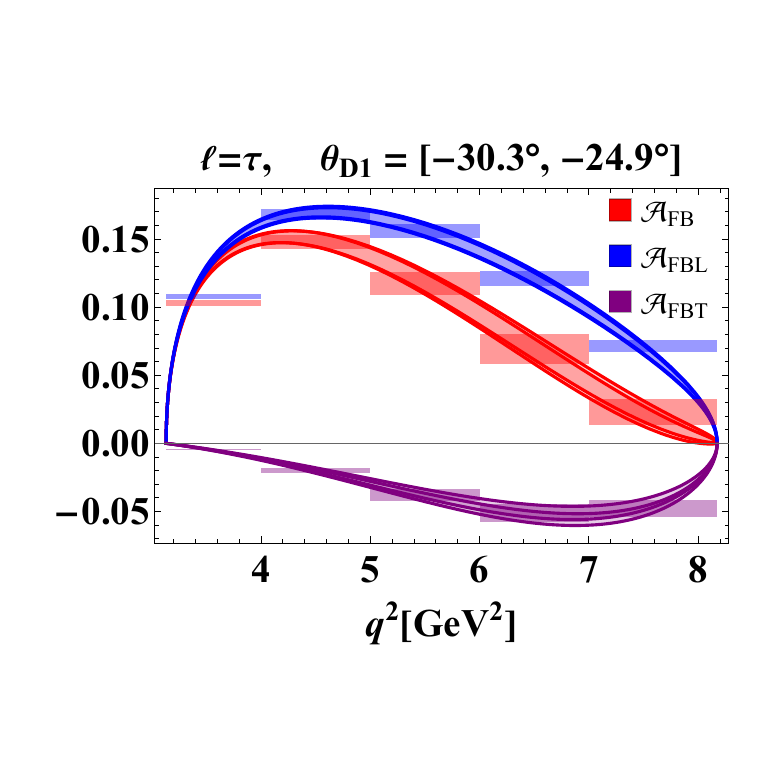}
    \vspace{-1.5cm}
    \caption{}
    \label{fig:4b}
  \end{subfigure}%
  \hfill
  \begin{subfigure}{0.33\textwidth}
    \includegraphics[width=\linewidth]{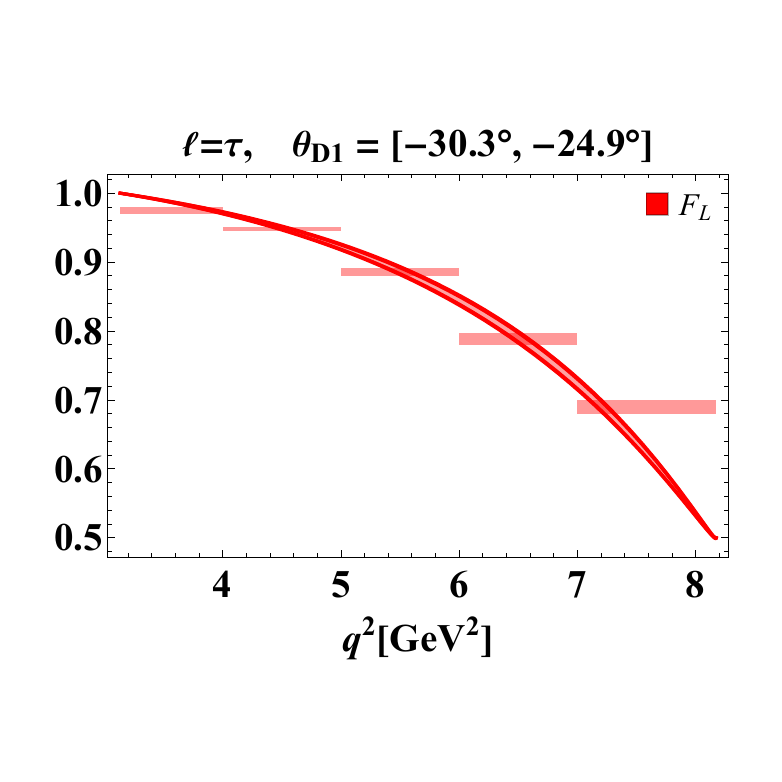}
    \vspace{-1.5cm}
    \caption{}
    \label{fig:4c}
  \end{subfigure}
 \vspace{-0.9cm} \\ 
    \begin{subfigure}{0.33\textwidth}
    \includegraphics[width=\linewidth]{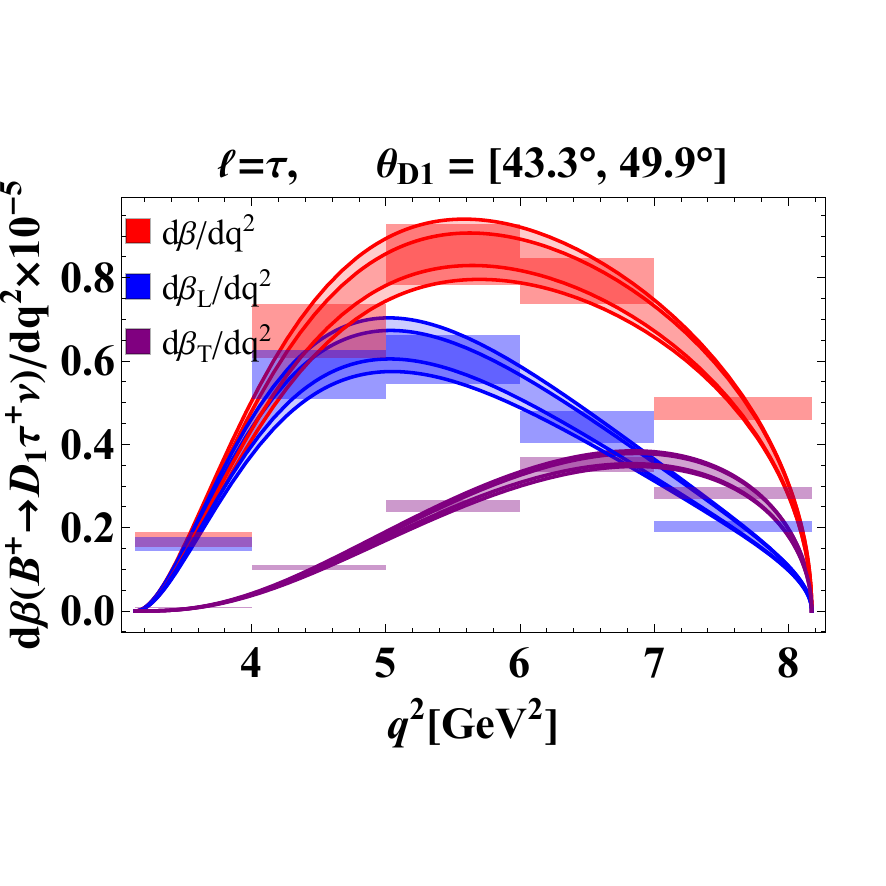}
    \vspace{-1.5cm}
    \caption{}
    \label{fig:4d}
  \end{subfigure}%
  \hfill
  \begin{subfigure}{0.33\textwidth}
    \includegraphics[width=\linewidth]{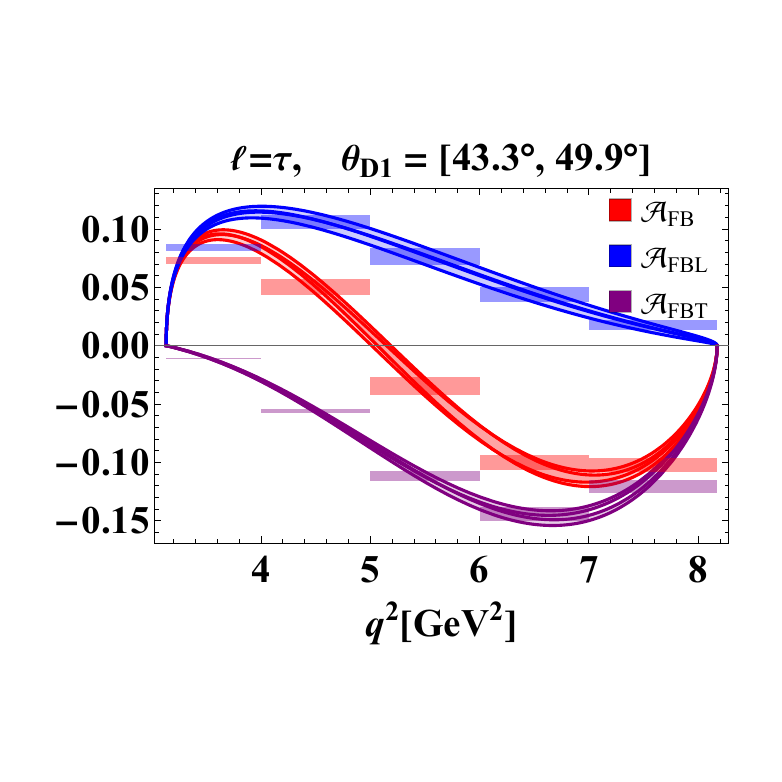}
    \vspace{-1.5cm}
    \caption{}
    \label{fig:4e}
  \end{subfigure}%
  \hfill
  \begin{subfigure}{0.33\textwidth}
    \includegraphics[width=\linewidth]{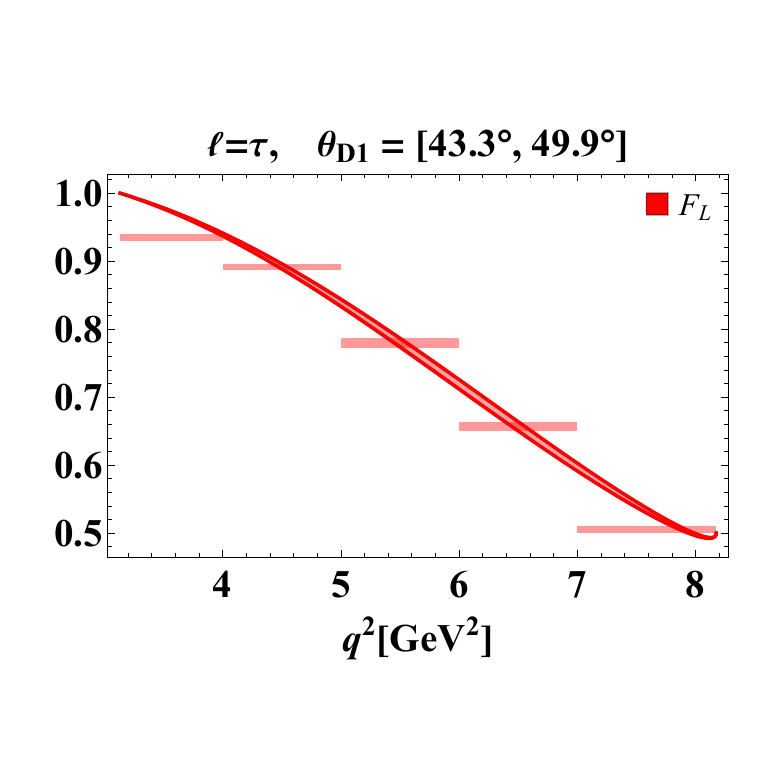}
    \vspace{-1.5cm}
    \caption{}
    \label{fig:4f}
  \end{subfigure}
\label{btod1plottuon}
\end{figure}
FIGs.~\ref{btod1plottuon} displays the $q^2$ dependence of the different observables for the decay $B^+ \to D_{1}\tau^+ \nu$, including the branching ratio, the lepton forward-backward asymmetry, and the longitudinal polarization fraction. The results are presented for two representative intervals of the mixing angle: $\theta_{D1} \in [-30.3^\circ, -24.9^\circ]$ (top row) and $\theta_{D1} \in [43.3^\circ, 49.9^\circ]$ (bottom row). 
\begin{itemize}

    \item In FIG.~\ref{fig:4a}, the differential branching ratio is shown, separating the unpolarized result from the L and T polarization. A clear difference is observed across most of the $q^2$ region, where the L component provides the leading contribution. In the highest bin, $q^2 \in [7,8]~\mathrm{GeV}^2$, the T contribution becomes comparable to the L one, with both contributions having similar integrated values in the range $[0.10, 0.13]\times10^{-5}$. For FIG.~\ref{fig:4d}, in the positive angle range corresponding to $q^2 \in [7-8]~\mathrm{GeV}^2$, the T contribution becomes larger than the L one, with integrated values $[0.26, 0.29]\times10^{-5}$ for T and $[0.18, 0.21]\times10^{-5}$ for L, indicating that the T component dominates near high $q^2$.

\item In FIG.~\ref{fig:4b}, the L and unpolarized components of 
$\mathcal{A}_{\rm FB}$ remain positive over the entire $q^2$ region, 
whereas the T contribution stays negative throughout. The T component is 
strongly suppressed at low and intermediate $q^2$, for example 
$\mathcal{A}_{{\rm FBT}}\simeq -0.01$ around 
$q^2\simeq 4~\mathrm{GeV}^2$, compared with 
$\mathcal{A}_{{\rm FBL}}\simeq 0.16$. However, near the high-$q^2$ 
region, its magnitude increases to about $0.05$, becoming comparable to the 
unpolarized contribution as the latter approaches zero. In contrast to FIG.~\ref{fig:4b}, FIG.~\ref{fig:4e} shows that the unpolarized $\mathcal{A}_{\rm FB}$ exhibits a zero crossing at approximately $q^2 \simeq 5.3~\mathrm{GeV}^2$. However, the T contribution remains negative, while the L contribution remains positive throughout the entire $q^2$ region.
\item FIG.~\ref{fig:4c} and \ref{fig:4f} shows the $F_{\text{L}}$ for both angle ranges where the bin-wise values of $F_{\text{L}}$ are listed in the TABLE~\ref{table4} show agreement with the plots. 
\end{itemize}

\subsubsection{Observable for the decay of $B^+ \to D_{1}^\prime \ell^+ \nu_\ell$.}

\begin{figure}[H]
\caption{ The polarized and unpolarized branching ratio, lepton forward-backward asymmetry, and the  longitudinal fraction as a function of $q^2$ for $B^+ \to D_{1}^\prime \mu^+ \nu$ transition.} 
\vspace{-0.7cm}
\begin{subfigure}{0.33\textwidth}
    \includegraphics[width=\linewidth]{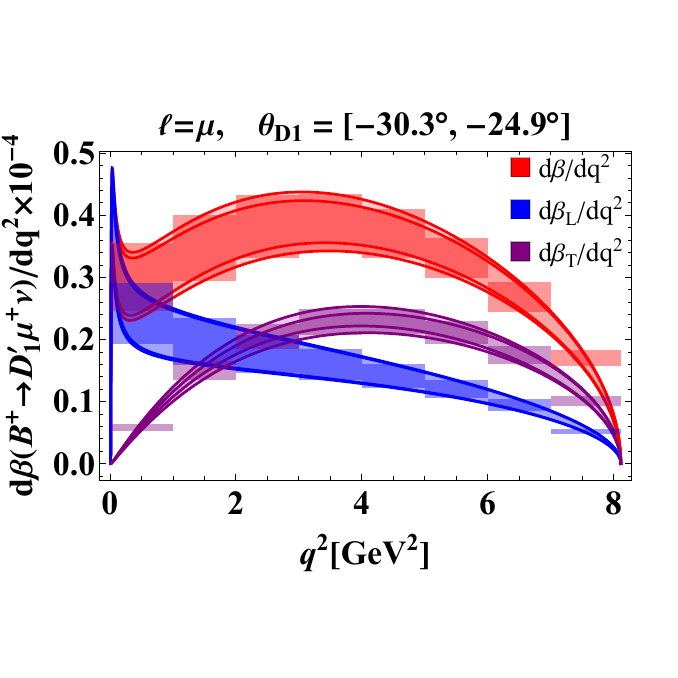}
    \vspace{-1.5cm}
    \caption{}
    \label{fig:5a}
  \end{subfigure}%
  \hfill
  \begin{subfigure}{0.33\textwidth}
    \includegraphics[width=\linewidth]{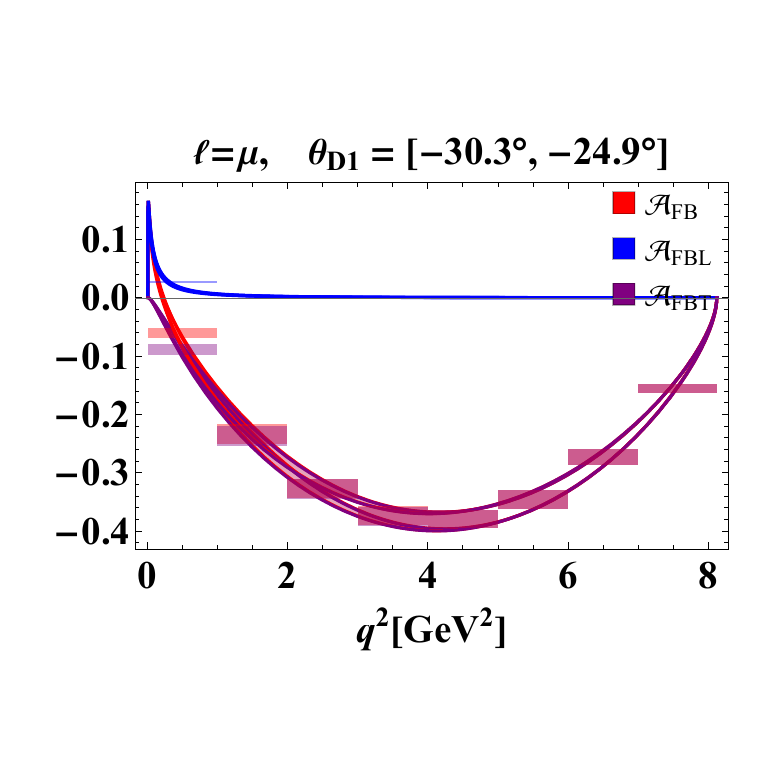}
    \vspace{-1.5cm}
    \caption{}
    \label{fig:5b}
  \end{subfigure}%
  \hfill
  \begin{subfigure}{0.33\textwidth}
    \includegraphics[width=\linewidth]{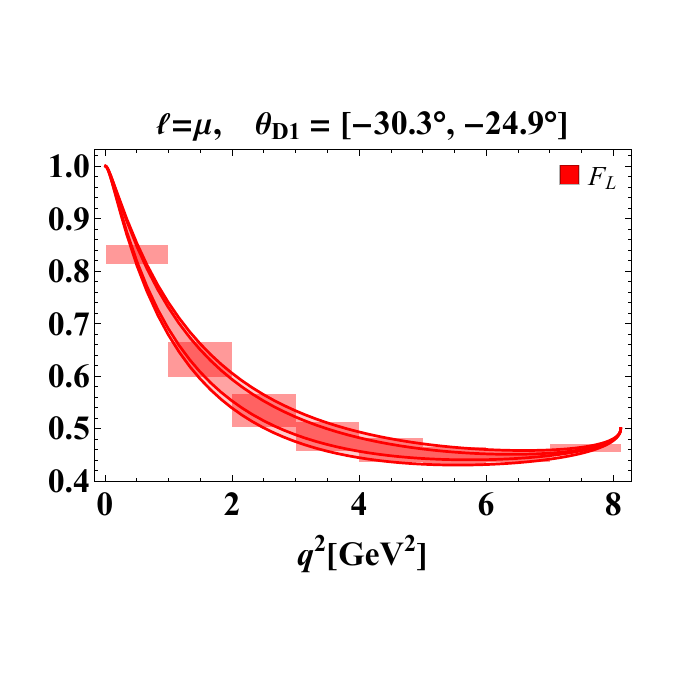}
    \vspace{-1.5cm}
    \caption{}
    \label{fig:5c}
  \end{subfigure}
 \vspace{-0.9cm} \\ 
    \begin{subfigure}{0.33\textwidth}
    \includegraphics[width=\linewidth]{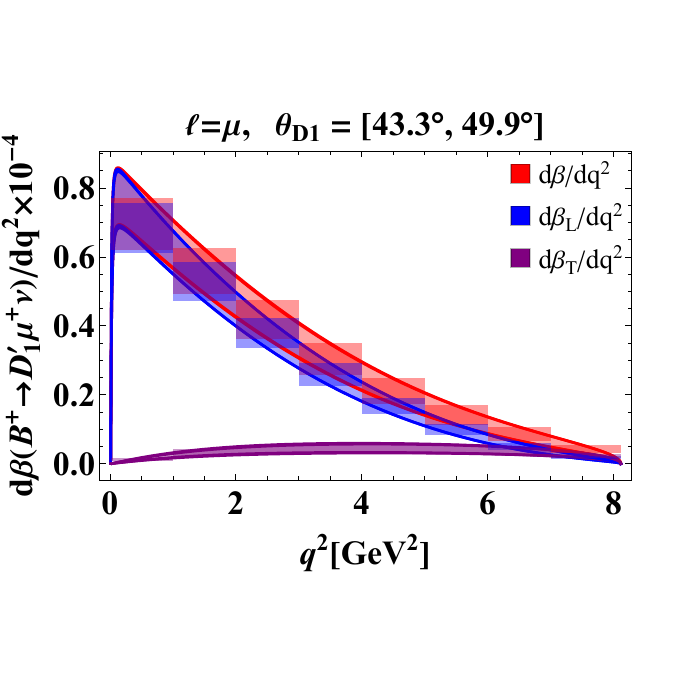}
    \vspace{-1.5cm}
    \caption{}
    \label{fig:5d}
  \end{subfigure}%
  \hfill
  \begin{subfigure}{0.33\textwidth}
    \includegraphics[width=\linewidth]{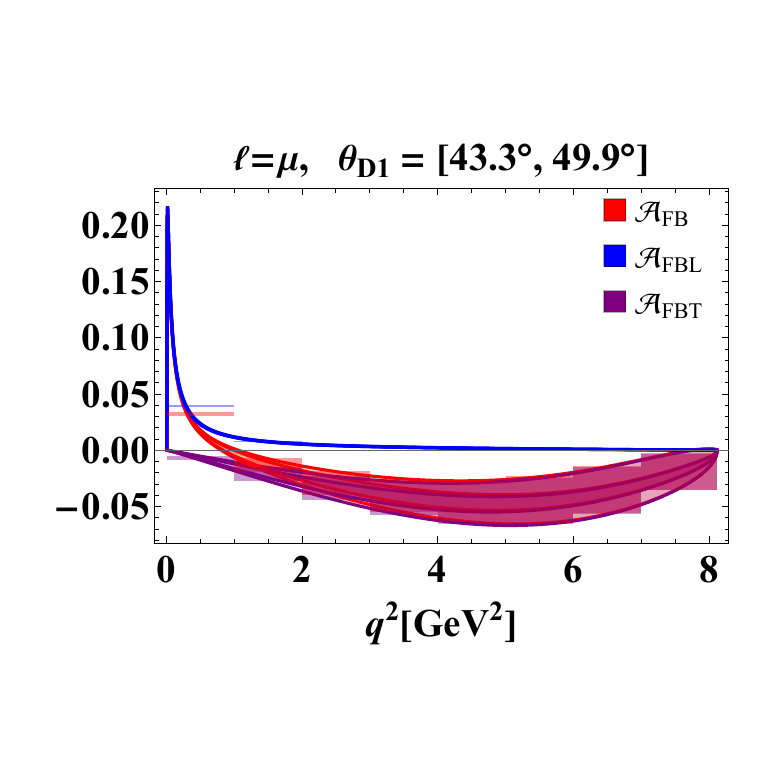}
    \vspace{-1.5cm}
    \caption{}
    \label{fig:5e}
  \end{subfigure}%
  \hfill
  \begin{subfigure}{0.33\textwidth}
    \includegraphics[width=\linewidth]{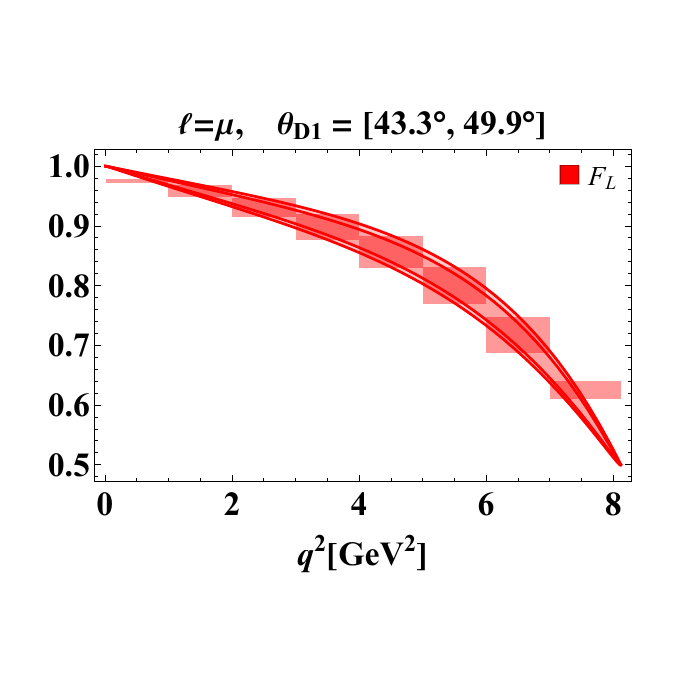}
    \vspace{-1.5cm}
    \caption{}
    \label{fig:5f}
  \end{subfigure}
\label{btod1prmuon}
\end{figure}

FIGs.~\ref{btod1prmuon} depict the $q^2$ dependence of the different observables for the decay $B^+ \to D_{1}^\prime\mu^+ \nu$, including the branching ratio, the $\mathcal{A}_{\text{FB}}$, and the $F_{\text{L}}$. The results are presented for two representative intervals of the mixing angle: $\theta_{D1} \in [-30.3^\circ, -24.9^\circ]$ (top row) and $\theta_{D1} \in [43.3^\circ, 49.9^\circ]$ (bottom row). 

\begin{itemize}
    \item For FIG.~\ref{fig:5a}, the L contribution is larger than the T contribution in the low-$q^2$ region, particularly for $q^2 \lesssim 1.5~\mathrm{GeV}^2$. However, with increasing $q^2$, the T component becomes enhanced and overtakes the L contribution over the intermediate and high-$q^2$ regions. This indicates a clear change in the polarization hierarchy from L dominance at low $q^2$ to T dominance at higher $q^2$. In contrast, FIG.~\ref{fig:5d}, shows that the L contribution remains larger than the T contribution over almost the entire $q^2$ region. The T component is strongly suppressed in this angle range and becomes comparable to the L contribution only near high $q^2$. In the highest integrated bin, $q^2\in[7,8]~\mathrm{GeV}^2$, the integrated values are $[0.013,0.021]\times10^{-4}$ and $[0.015,0.027]\times10^{-4}$ for the L and T contributions, respectively. Thus, in this endpoint bin, the T contribution becomes comparable to and slightly larger than the L contribution.

\item In FIG.~\ref{fig:5b} and FIG.~\ref{fig:5e}, the T component of $\mathcal{A}_{\text{FB}}$ dominates over the L contribution throughout the entire $q^2$ region. In FIG.~\ref{fig:5b}, the largest magnitude is observed in the range $q^2 \in [4,5]~\mathrm{GeV}^2$, with values around $[-0.364, -0.395]$, while in FIG.~\ref{fig:5e}, the maximum occurs near $q^2 \in [5,6]~\mathrm{GeV}^2$, where $\mathcal{A}_{\text{FB}}$ reaches approximately $[-0.024, -0.065]$.

\item FIGs.~\ref{fig:5c} and \ref{fig:5f} presents $F_{\text{L}}$ for both angular ranges i.e, $\theta_{D1} \in [-30.3^\circ, -24.9^\circ]$ and $\theta_{D1} \in [43.3^\circ, 49.9^\circ]$. The bin-wise values presented in the TABLE~\ref{table5} are consistent with the behavior observed in these plots across the full $q^2$ region.
\end{itemize}

\begin{figure}[H]
\caption{ The polarized and unpolarized branching ratio, lepton forward-backward asymmetry, and the  longitudinal fraction as a function of $q^2$ for $B^+ \to D_{1}^{\prime} \tau^+ \nu$ transition.}
\vspace{-0.7cm}
\begin{subfigure}{0.33\textwidth}
    \includegraphics[width=\linewidth]{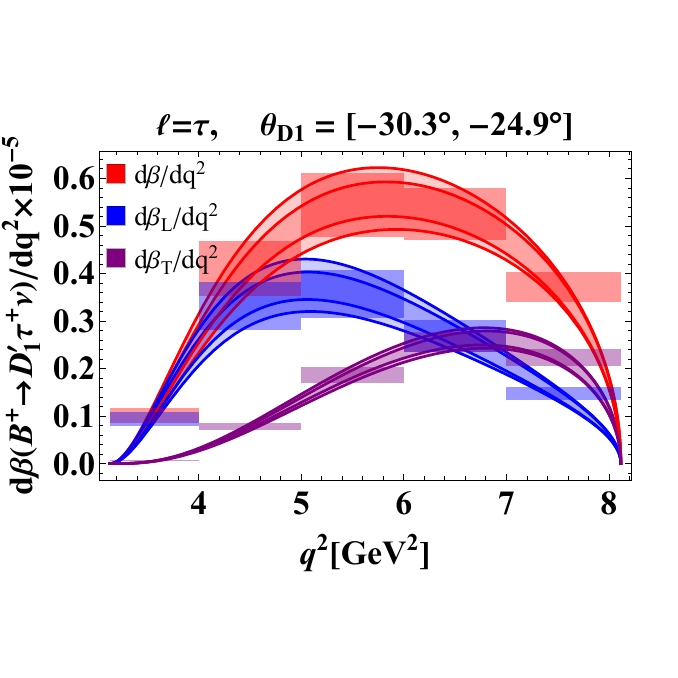}
    \vspace{-1.5cm}
    \caption{}
    \label{fig:6a}
  \end{subfigure}%
  \hfill
  \begin{subfigure}{0.33\textwidth}
    \includegraphics[width=\linewidth]{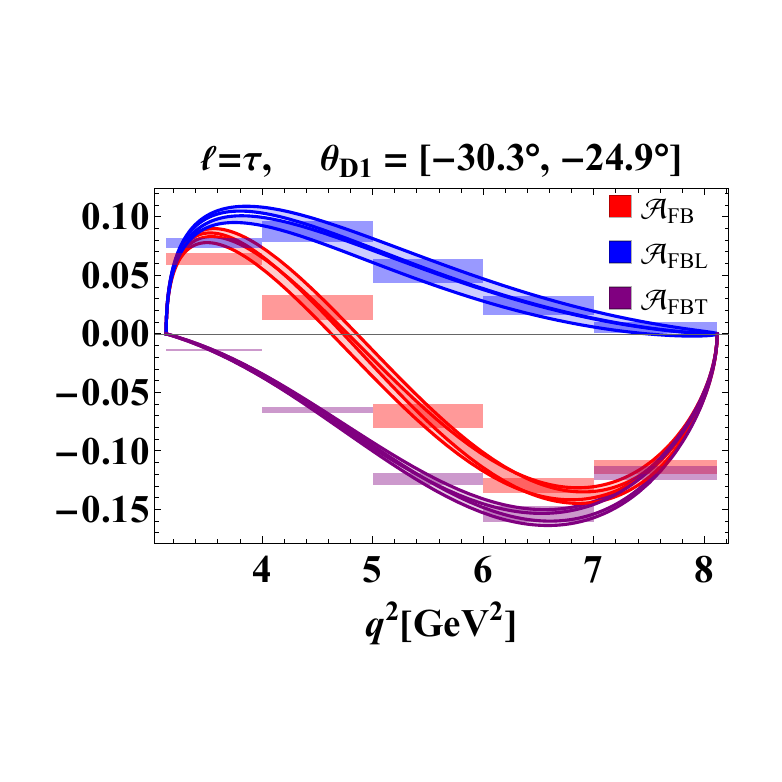}
    \vspace{-1.5cm}
    \caption{}
    \label{fig:6b}
  \end{subfigure}%
  \hfill
  \begin{subfigure}{0.33\textwidth}
    \includegraphics[width=\linewidth]{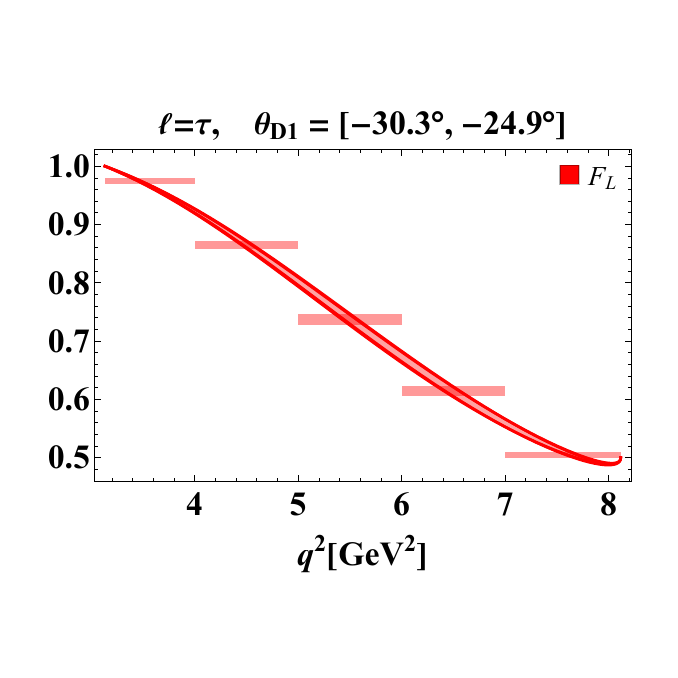}
    \vspace{-1.5cm}
    \caption{}
    \label{fig:6c}
  \end{subfigure}
 \vspace{-0.9cm} \\ 
    \begin{subfigure}{0.33\textwidth}
    \includegraphics[width=\linewidth]{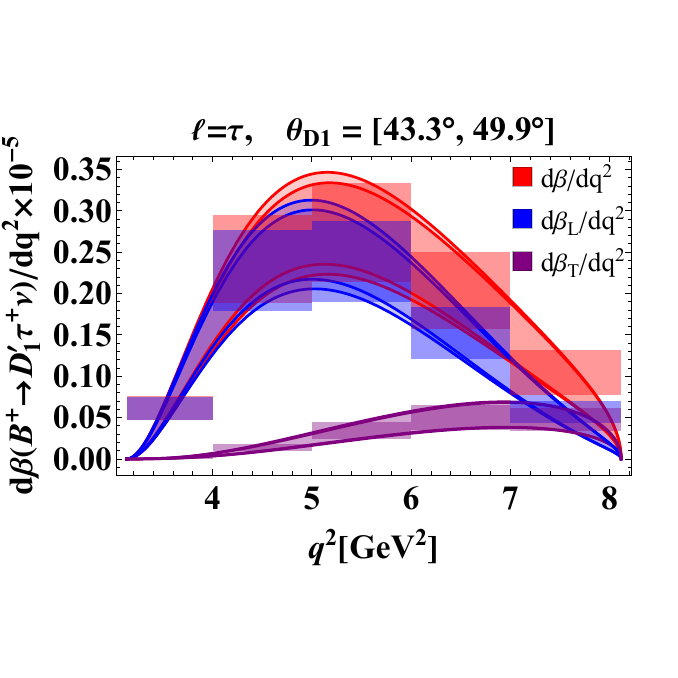}
    \vspace{-1.5cm}
    \caption{}
    \label{fig:6d}
  \end{subfigure}%
  \hfill
  \begin{subfigure}{0.33\textwidth}
    \includegraphics[width=\linewidth]{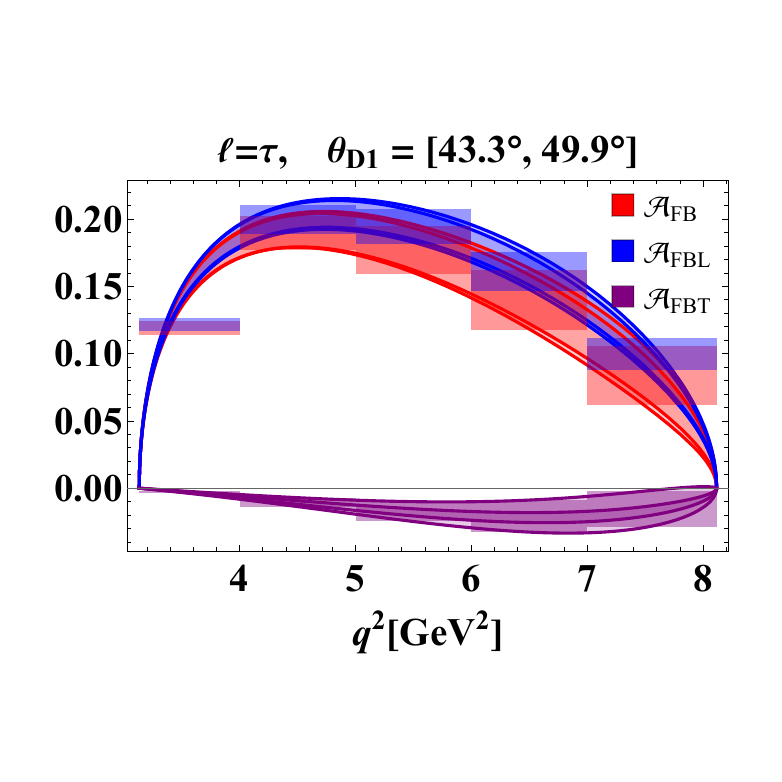}
    \vspace{-1.5cm}
    \caption{}
    \label{fig:6e}
  \end{subfigure}%
  \hfill
  \begin{subfigure}{0.33\textwidth}
    \includegraphics[width=\linewidth]{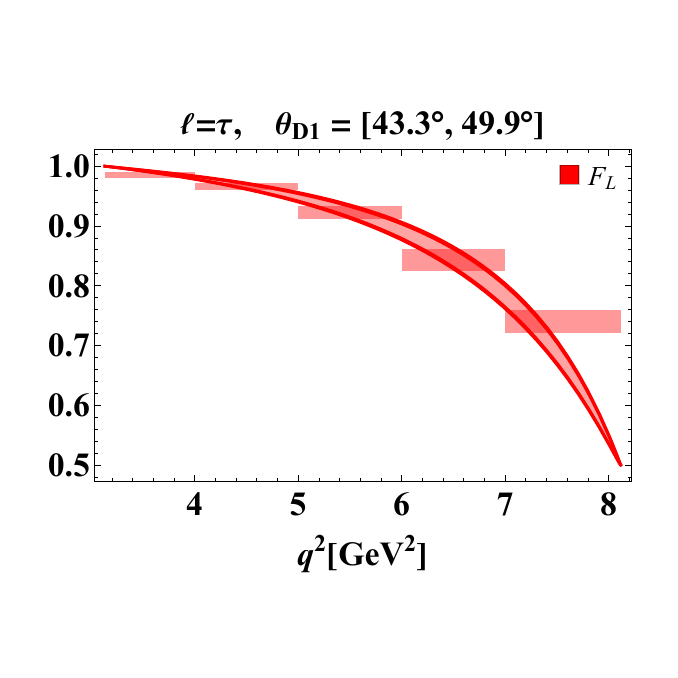}
    \vspace{-1.5cm}
    \caption{}
    \label{fig:6f}
  \end{subfigure}
\label{btod1prtuoan}
\end{figure}
FIGs.~\ref{btod1prtuoan} depict the $q^2$ dependence of the different observables for the decay $B^+ \to D_{1}^\prime\tau^+ \nu$, including the branching ratio, the $\mathcal{A}_{\text{FB}}$, and the $F_{\text{L}}$. The results are presented for two representative intervals of the mixing angle: $\theta_{D1} \in [-30.3^\circ, -24.9^\circ]$ (top row) and $\theta_{D1} \in [43.3^\circ, 49.9^\circ]$ (bottom row). 
\begin{itemize}
    \item In FIG.~\ref{fig:6a}, the L contribution dominates up to 
approximately $q^2 \simeq 6~\mathrm{GeV}^2$, beyond which the T 
component becomes comparable to, and eventually larger than, the L component. 
The corresponding numerical values in the different $q^2$ bins are reported in 
TABLE~\ref{table6}. In contrast, FIG.~\ref{fig:6d} shows that the L contribution 
remains the leading polarized component over most of the $q^2$ region. 
However, in the highest integrated bin, $q^2 \in [7-8]~\mathrm{GeV}^2$, the 
L and T contributions become comparable, as shown in the plot and listed in 
TABLE~\ref{table6}.

\item The $\mathcal{A}_{\text{FB}}$ shows a zero crossing at $q^2 \approx 4.8~\mathrm{GeV}^2$ as depicted in FIG.~\ref{fig:6b}. The T component remains small in magnitude and negative throughout the entire $q^2$ region, indicating a negligible contribution. In contrast, the L contribution is positive and clearly dominant over the full $q^2$ range. In FIG.~\ref{fig:6e}, corresponding to the positive angle range, the unpolarized $\mathcal{A}_{\text{FB}}$ does not show a zero crossing and remains positive over the whole $q^2$ region. It increases from $(0.113, 0.124)$ in the bin $q^2 = 3.13$--$4~\mathrm{GeV}^2$ to its maximum value $(0.176, 0.202)$ in the bin $q^2 = 4$--$5~\mathrm{GeV}^2$, and then decreases towards the high $q^2$, as listed in TABLE~\ref{table6}. The L contribution is positive and dominant, while the T contribution remains negative and small, $|\mathcal{A}_{\text{FBT}}| \lesssim 0.03$, over the entire $q^2$ region.

\item FIGs.~\ref{fig:6c} and \ref{fig:6f} display $F_{\text{L}}$ for both $\theta_{D1}$ ranges, and the bin-wise values presented in the TABLE~\ref{table6} are consistent with the behavior observed in these plots across the full $q^2$ region.
\end{itemize}

\subsubsection{ Observable for the decay of $B^0_s \to D^-_{s1} \ell^+ \nu_\ell$.}

\begin{figure}[H]
\caption{ The polarized and unpolarized branching ratio, lepton forward-backward asymmetry, and the  longitudinal fraction as a function of $q^2$ for $B_s^0 \to D_{s1}^{-} \mu^+ \nu$ transition.} 
\vspace{-0.7cm}
\begin{subfigure}{0.33\textwidth}
    \includegraphics[width=\linewidth]{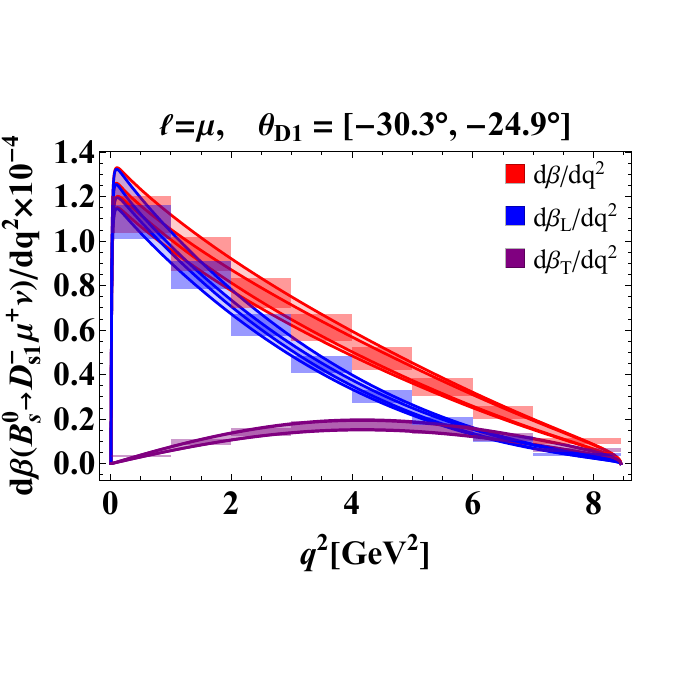}
    \vspace{-1.5cm}
    \caption{}
    \label{fig7a}
  \end{subfigure}%
  \hfill
  \begin{subfigure}{0.33\textwidth}
    \includegraphics[width=\linewidth]{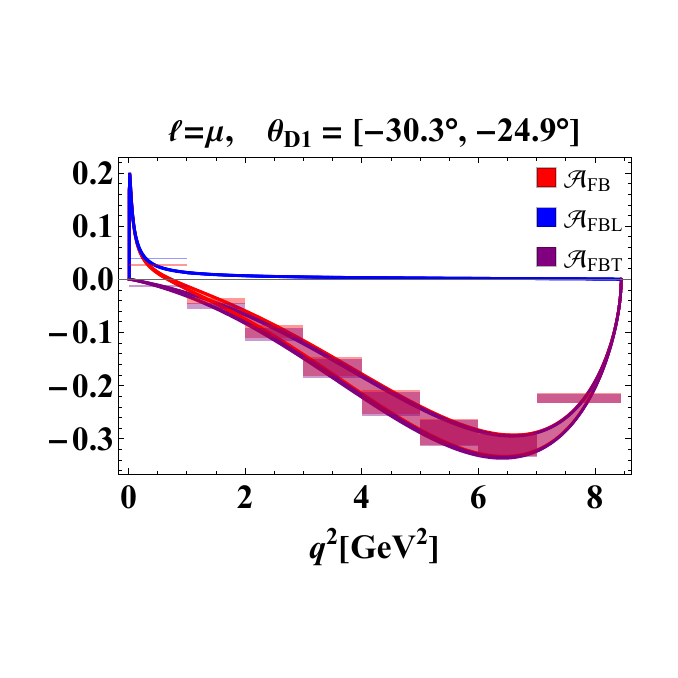}
    \vspace{-1.5cm}
    \caption{}
    \label{fig7b}
  \end{subfigure}%
  \hfill
  \begin{subfigure}{0.33\textwidth}
    \includegraphics[width=\linewidth]{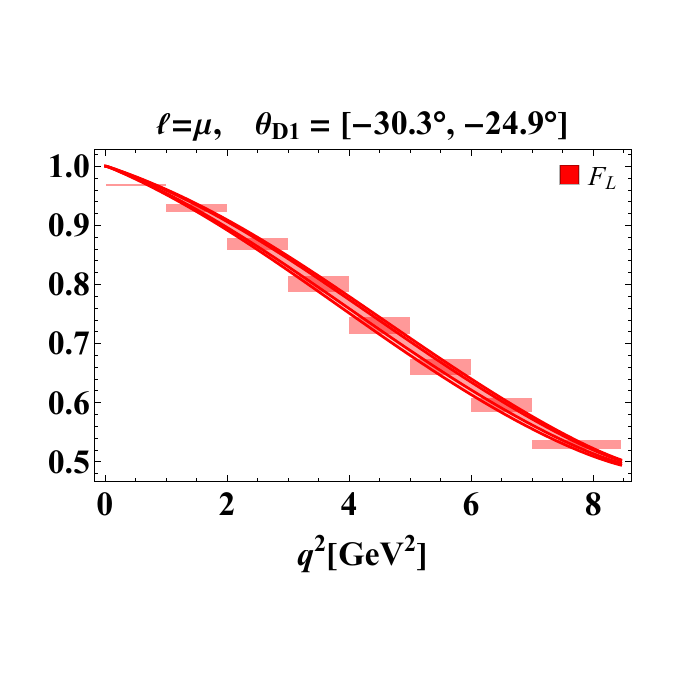}
    \vspace{-1.5cm}
    \caption{}
    \label{fig7c}
  \end{subfigure}
 \vspace{-0.9cm} \\ 
    \begin{subfigure}{0.33\textwidth}
    \includegraphics[width=\linewidth]{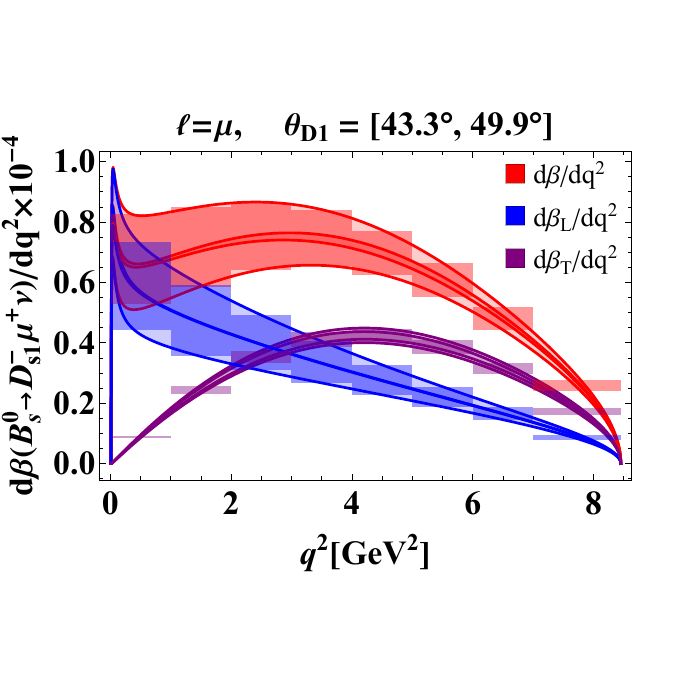}
    \vspace{-1.5cm}
    \caption{}
    \label{fig7d}
  \end{subfigure}%
  \hfill
  \begin{subfigure}{0.33\textwidth}
    \includegraphics[width=\linewidth]{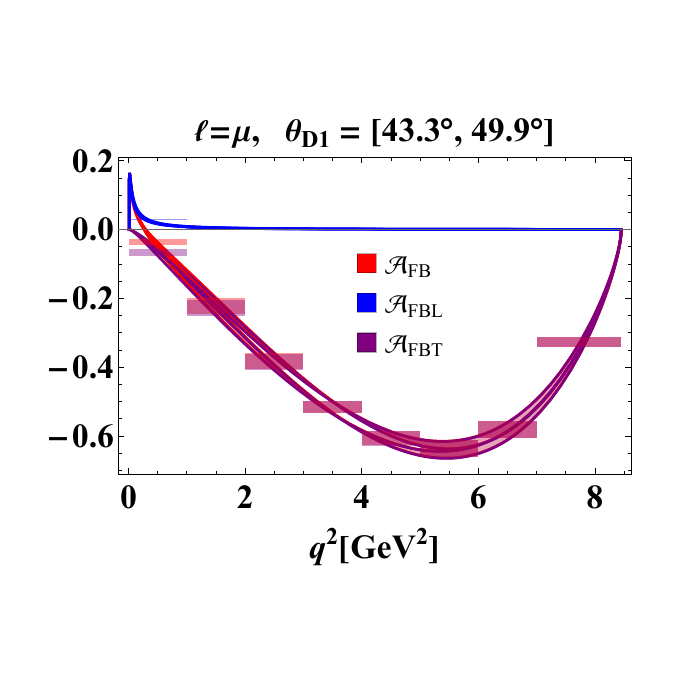}
    \vspace{-1.5cm}
    \caption{}
    \label{fig7e}
  \end{subfigure}%
  \hfill
  \begin{subfigure}{0.33\textwidth}
    \includegraphics[width=\linewidth]{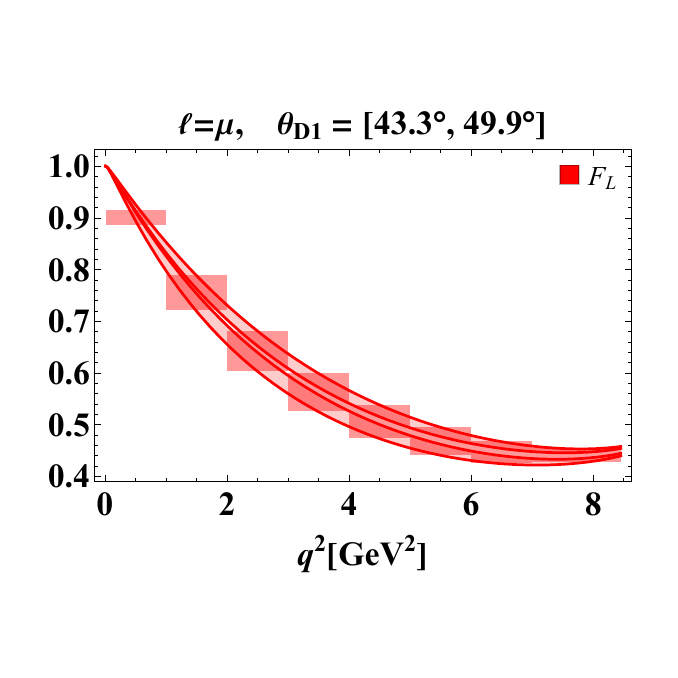}
    \vspace{-1.5cm}
    \caption{}
    \label{fig7f}
  \end{subfigure}
\label{bstods1muon}
\end{figure}

FIGs.~\ref{bstods1muon} shows the polarized and unpolarized differential branching ratio, $\mathcal{A}_{\text{FB}}$, and the $F_{\text{L}}$ as functions of $q^2$ for the decay $B_s^0 \to D_{s1}^- \mu^+ \nu$. The first row corresponds to the mixing-angle range $\theta_{D1} \in [-30.3^\circ, -24.9^\circ]$, while the second row represents $\theta_{D1} \in [43.3^\circ, 49.9^\circ]$.

\begin{itemize}
    \item In FIG.~\ref{fig7a}, the L component remains larger than the
T contribution over most of the $q^2$ region, particularly
in the low- and intermediate-$q^2$ regions. However, as $q^2$ increases, the
two polarized contributions become closer in magnitude, especially for
$q^2 \gtrsim 6~\mathrm{GeV}^2$. This behavior is also consistent with the
bin-wise numerical values reported in TABLE~\ref{table7}. For FIG.~\ref{fig7d}, the polarization pattern is different. The L component dominates only in the low-$q^2$ region, up to approximately
$q^2 \simeq 2.5~\mathrm{GeV}^2$. Beyond this point, the T contribution becomes larger than the L contribution and remains dominant over the intermediate and high-$q^2$ regions.

\item For FIG.~\ref{fig7b}, the observable shows a clear dominance of the T contribution over the L one throughout the full kinematic region. The L component remains strongly suppressed and nearly negligible in comparison. Numerically, the T contribution attains its largest magnitude in the bin $q^2=6$--$7~\mathrm{GeV}^2$, where it reaches approximately $(-0.291,-0.333)$, while the L contribution stays much smaller, around $(0.001,0.002)$. Similarly, FIG.~\ref{fig7e} exhibits the same qualitative trend, with the T contribution governing the distribution and the L part remaining suppressed across the entire $q^2$ region. In the bin $q^2 = 5\text{--}6~\mathrm{GeV}^2$, the T component takes maximum values around $(-0.612, -0.661)$, whereas the L component is significantly smaller, of the order $(\sim 10^{-4})$.

\item FIGs.~\ref{fig7c} and \ref{fig7f} display $F_{\text{L}}$ for both $\theta_{D1}$ ranges, and the bin-wise values presented in the TABLE~\ref{table7} are consistent with the behavior observed in these plots across the full $q^2$ region.
\end{itemize}

\begin{figure}[H]
\caption{ The polarized and unpolarized branching ratio, lepton forward-backward asymmetry, and the  longitudinal fraction as a function of $q^2$ for $B^0_s \to D_{s1}^{-} \tau^+ \nu$ transition.} 
\vspace{-0.7cm}
\begin{subfigure}{0.33\textwidth}
    \includegraphics[width=\linewidth]{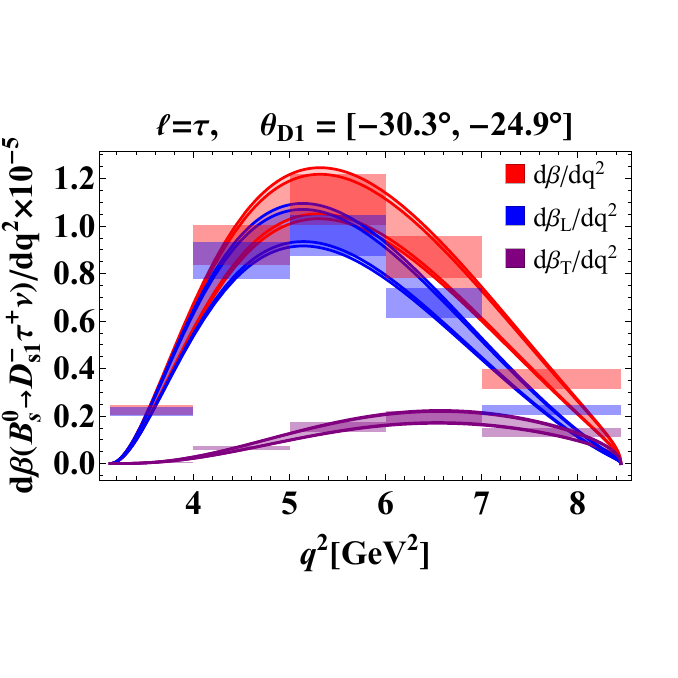}
    \vspace{-1.5cm}
    \caption{}
    \label{fig:8a}
  \end{subfigure}%
  \hfill
  \begin{subfigure}{0.33\textwidth}
    \includegraphics[width=\linewidth]{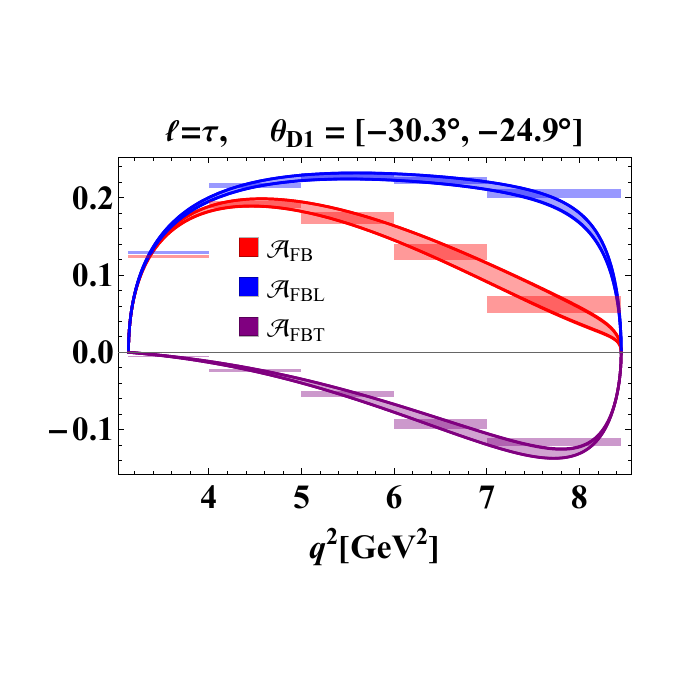}
    \vspace{-1.5cm}
    \caption{}
    \label{fig:8b}
  \end{subfigure}%
  \hfill
  \begin{subfigure}{0.33\textwidth}
    \includegraphics[width=\linewidth]{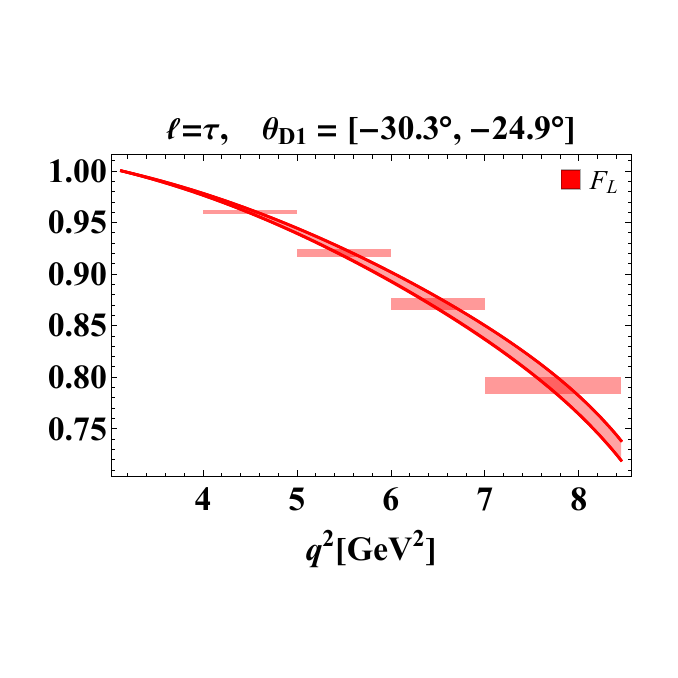}
    \vspace{-1.5cm}
    \caption{}
    \label{fig:8c}
  \end{subfigure}
 \vspace{-0.9cm} \\ 
    \begin{subfigure}{0.33\textwidth}
    \includegraphics[width=\linewidth]{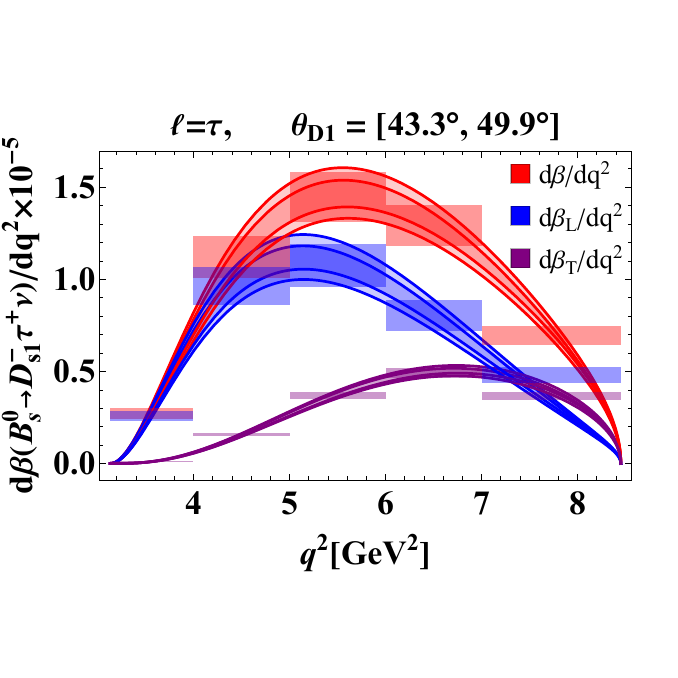}
    \vspace{-1.5cm}
    \caption{}
    \label{fig:8d}
  \end{subfigure}%
  \hfill
  \begin{subfigure}{0.33\textwidth}
    \includegraphics[width=\linewidth]{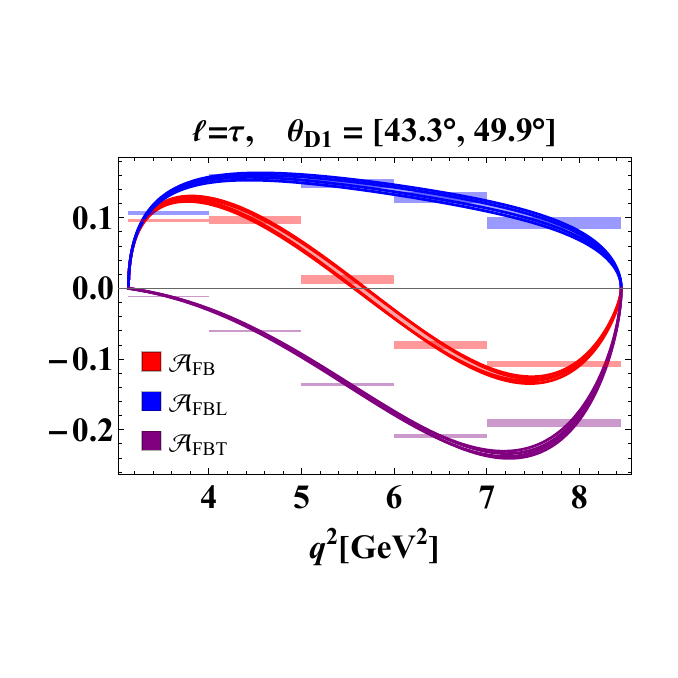}
    \vspace{-1.5cm}
    \caption{}
    \label{fig:8e}
  \end{subfigure}%
  \hfill
  \begin{subfigure}{0.33\textwidth}
    \includegraphics[width=\linewidth]{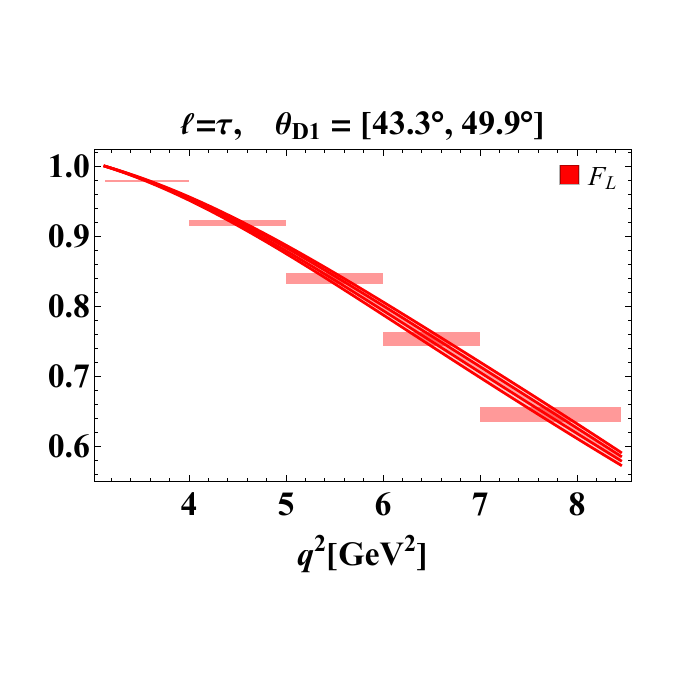}
    \vspace{-1.5cm}
    \caption{}
    \label{fig:8f}
  \end{subfigure}
  \label{btods1tauon}
\end{figure}

In FIGs.~\ref{btods1tauon}, we show the differential branching ratio (polarized and unpolarized), $\mathcal{A}_{\text{FB}}$, and the $F_{\text{L}}$ as functions of $q^2$ for the process $B_s^0 \to D_{s1}^- \tau^+ \nu$. The first (second) row corresponds to $\theta_{D1} \in [-30.3^\circ, -24.9^\circ]$ ($[43.3^\circ, 49.9^\circ]$).

\begin{itemize}
    \item For FIG.~\ref{fig:8a}, the L contribution
is the dominant polarized component over the whole $q^2$ region. It rises
rapidly from low $q^2$, reaches its maximum around
$q^2\simeq 5~\mathrm{GeV}^2$, and then decreases toward high $q^2$. The
T contribution is comparatively suppressed throughout the region, although it increases gradually and reaches a small maximum around $q^2\simeq 6.5~\mathrm{GeV}^2$. Thus, the L component remains larger than the $T$ component over the full $q^2$ range. In contrast, FIG.~\ref{fig:8d}, shows an enhanced transverse contribution compared with the Fig.~\ref{fig:8a}. The L component still dominates in the low- and intermediate-$q^2$ regions, but the T component increases significantly with $q^2$ and becomes comparable to the L component in the high-$q^2$ region. In particular, near $q^2\gtrsim 7~\mathrm{GeV}^2$, the two polarized contributions are close in magnitude, and the T component may slightly exceed the L contribution near high $q^2$. This indicates that the positive mixing angle range leads to a stronger T polarization effect at large $q^2$.

\item FIG.~\ref{fig:8b} shows that $\mathcal{A}_{\text{FB}}$ follows a behavior similar to that observed in the $B^+ \to D_1 \tau^+ \nu$ channel in FIG.~\ref{fig:4b}, with a difference in magnitude. Likewise, FIG.~\ref{fig:8e} exhibits the same pattern as seen in FIG.~\ref{fig:4e}, although the numerical values differ. The L and unpolarized components of $\mathcal{A}_{\rm FB}$ remain positive over the entire $q^2$ range, whereas the T contribution stays negative throughout. In contrast to FIG.~\ref{fig:8b}, FIG.~\ref{fig:8e} shows that the unpolarized $\mathcal{A}_{\rm FB}$ exhibits a zero crossing at approximately $q^2 \simeq 5.6~\mathrm{GeV}^2$. Nevertheless, the T component remains negative, while the L component remains positive across the full $q^2$ region.
\item FIGs.~\ref{fig:8c} and \ref{fig:8f} display $F_{\text{L}}$ for both $\theta_{D1}$ ranges, and the bin-wise values presented in the TABLE~\ref{table8} are consistent with the behavior observed in these plots across the full $q^2$ region.
\end{itemize}

\subsubsection{ Observable for the decay of $B^0_s \to D^{-\prime}_{s1} \ell^+ \nu_\ell$.}
\begin{figure}[H]
\caption{ The polarized and unpolarized branching ratio, lepton forward-backward asymmetry, and the  longitudinal fraction as a function of $q^2$ for $B_s^0 \to D_{s1}^{\prime-} \mu^+ \nu$ transition.} 
\vspace{-0.7cm}
\begin{subfigure}{0.33\textwidth}
    \includegraphics[width=\linewidth]{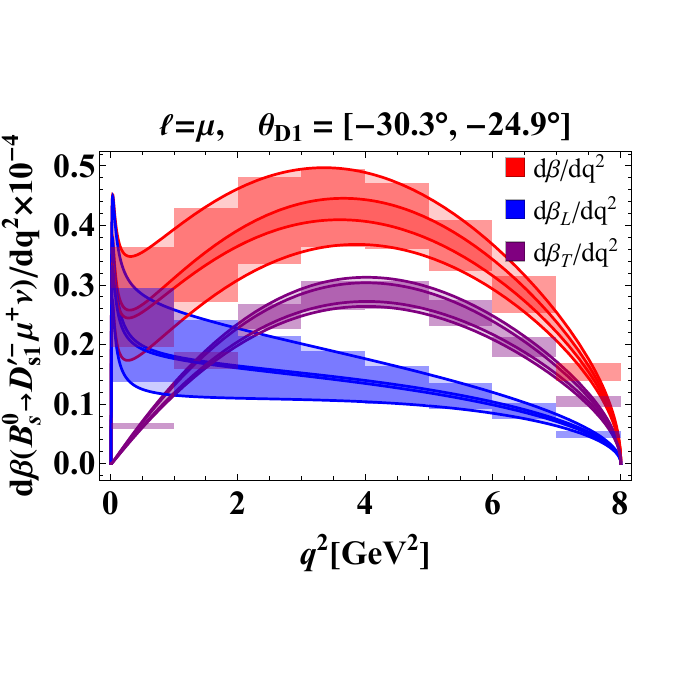}
    \vspace{-1.5cm}
    \caption{}
    \label{fig9a}
  \end{subfigure}%
  \hfill
  \begin{subfigure}{0.33\textwidth}
    \includegraphics[width=\linewidth]{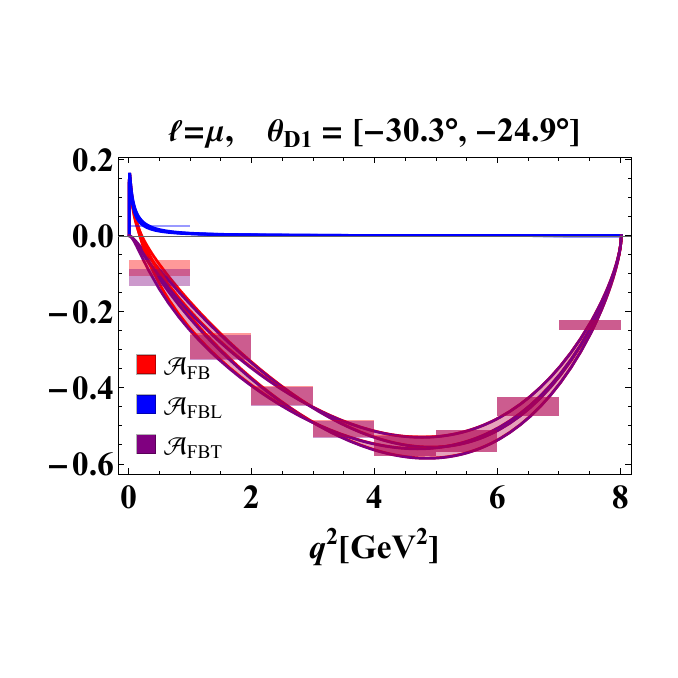}
    \vspace{-1.5cm}
    \caption{}
    \label{fig9b}
  \end{subfigure}%
  \hfill
  \begin{subfigure}{0.33\textwidth}
    \includegraphics[width=\linewidth]{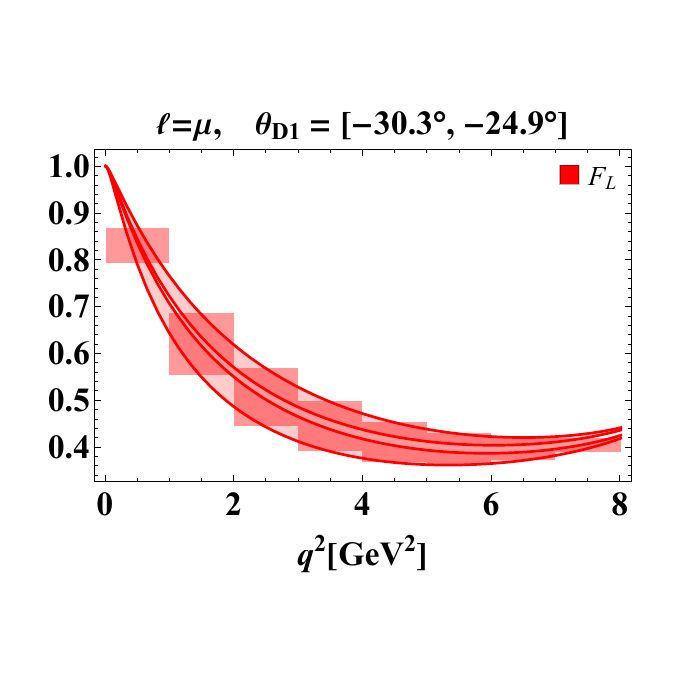}
    \vspace{-1.5cm}
    \caption{}
    \label{fig9c}
  \end{subfigure}
 \vspace{-0.9cm} \\ 
    \begin{subfigure}{0.33\textwidth}
    \includegraphics[width=\linewidth]{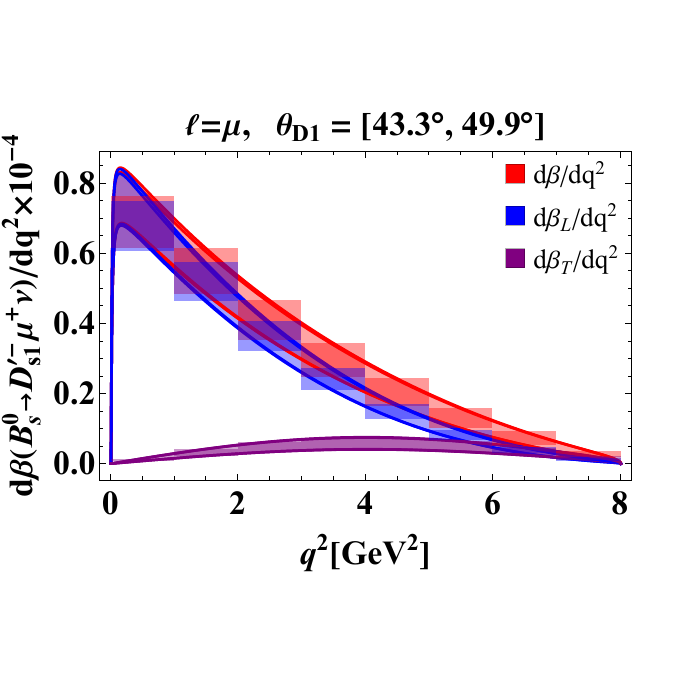}
    \vspace{-1.5cm}
    \caption{}
    \label{fig9d}
  \end{subfigure}%
  \hfill
  \begin{subfigure}{0.33\textwidth}
    \includegraphics[width=\linewidth]{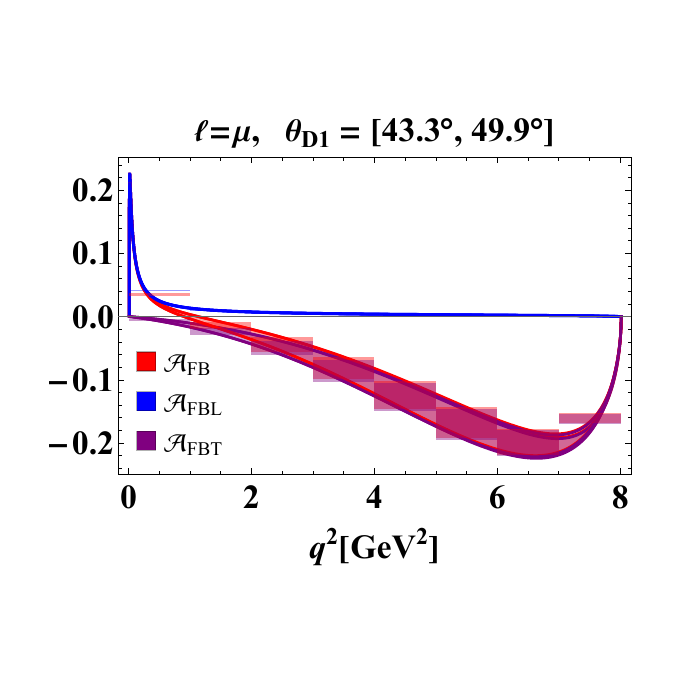}
    \vspace{-1.5cm}
    \caption{}
    \label{fig9e}
  \end{subfigure}%
  \hfill
  \begin{subfigure}{0.33\textwidth}
    \includegraphics[width=\linewidth]{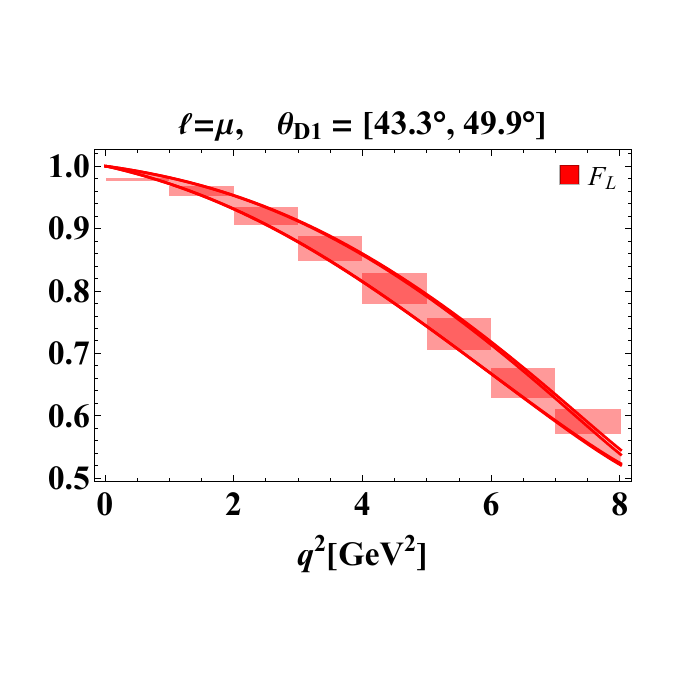}
    \vspace{-1.5cm}
    \caption{}
    \label{fig9f}
  \end{subfigure}
\label{bstods1pmuon}
\end{figure}

FIGs.~\ref{bstods1pmuon} illustrates the variation with $q^2$ of the polarized and unpolarized differential branching ratio, along with the  $\mathcal{A}_{\text{FB}}$ and the  $F_{\text{L}}$, for the decay $B_s^0 \to D_{s1}^{\prime-} \mu^+ \nu$. Results are shown for two mixing-angle intervals: $\theta_{D1} \in [-30.3^\circ, -24.9^\circ]$ (top row) and $\theta_{D1} \in [43.3^\circ, 49.9^\circ]$ (bottom row).

\begin{itemize}
    \item In FIG.~\ref{fig9a}, the L contribution dominates in the low-$q^2$ region, particularly for $q^2 \lesssim 2~\mathrm{GeV}^2$. As $q^2$ increases, the T component grows and becomes dominant in the intermediate and high-$q^2$ regions, indicating a transition from L to T dominance. In contrast, FIG.~\ref{fig9d} shows that the L contribution remains dominant over most of the $q^2$ range, with the T component significantly suppressed and becoming comparable only near the high-$q^2$ bin. In the highest bin, $q^2 \in [6-8]~\mathrm{GeV}^2$, the integrated values are $[0.013,0.060]\times10^{-4}$ and $[0.015,0.027]\times10^{-4}$ for the L and T contributions, respectively, indicating that the T contribution slightly exceeds the L one in this region.

    \item In FIG.~\ref{fig9b} and FIG.~\ref{fig9e}, corresponding to the $\mathcal{A}_{\text{FB}}$ the T component dominates over the L contribution across the entire $q^2$ region. In FIG.~\ref{fig9b}, the largest magnitude occurs in the range $q^2 \in [4-5]~\mathrm{GeV}^2$, with values around $[-0.52, -0.57]$, while in FIG.~\ref{fig9e}, the maximum is observed near $q^2 \in [6-7]~\mathrm{GeV}^2$, where $\mathcal{A}_{\text{FB}}$ reaches approximately $[-0.17, -0.22]$.

    \item The longitudinal polarization fraction $F_{\text{L}}$, presented in FIGs.~\ref{fig9c} and \ref{fig9f}, follows a behavior that is consistent with the bin-integrated results given in TABLE~\ref{table9} for both $\theta_{D1}$ ranges across the full $q^2$ range.
\end{itemize}

\begin{figure}[H]
\caption{ The polarized and unpolarized branching ratio, lepton forward-backward asymmetry, and the  longitudinal fraction as a function of $q^2$ for $B_s^0 \to D_{s1}^{\prime-} \tau^+ \nu$ transition.} 
\vspace{-0.7cm}
\begin{subfigure}{0.33\textwidth}
    \includegraphics[width=\linewidth]{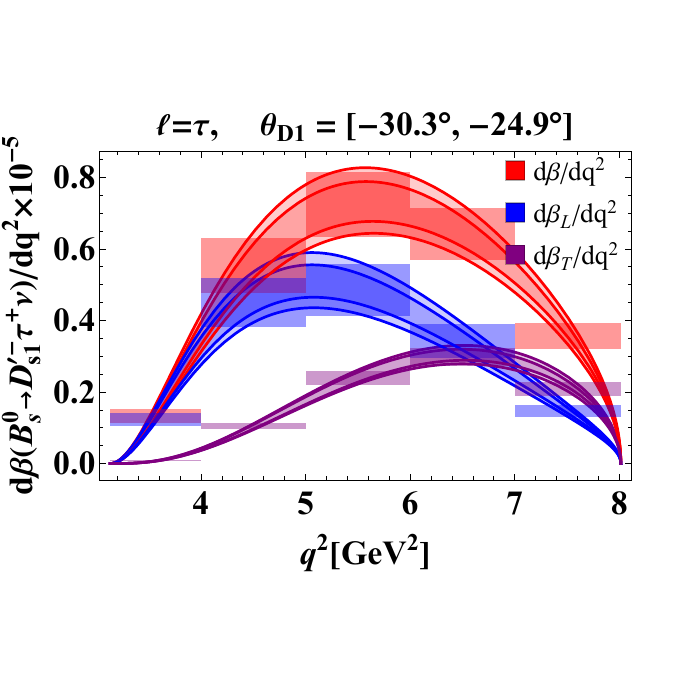}
    \vspace{-1.5cm}
    \caption{}
    \label{fig10a}
  \end{subfigure}%
  \hfill
  \begin{subfigure}{0.33\textwidth}
    \includegraphics[width=\linewidth]{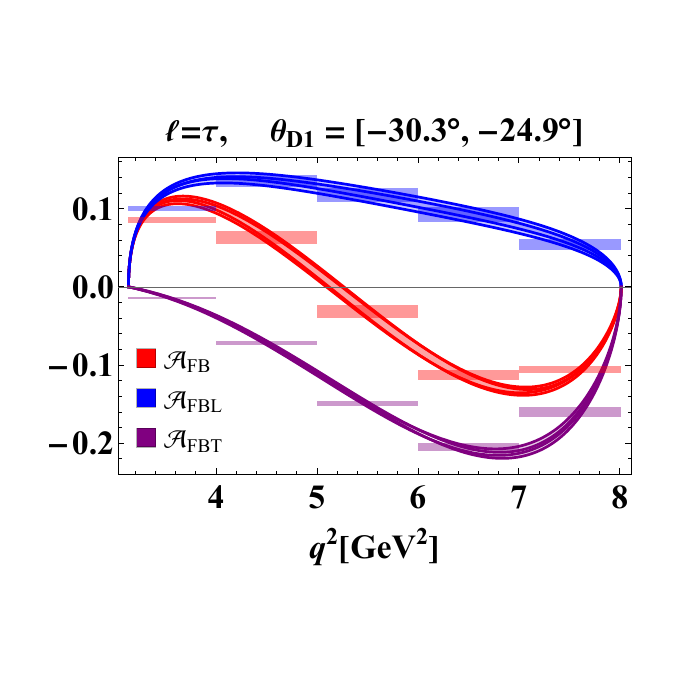}
    \vspace{-1.5cm}
    \caption{}
    \label{fig10b}
  \end{subfigure}%
  \hfill
  \begin{subfigure}{0.33\textwidth}
    \includegraphics[width=\linewidth]{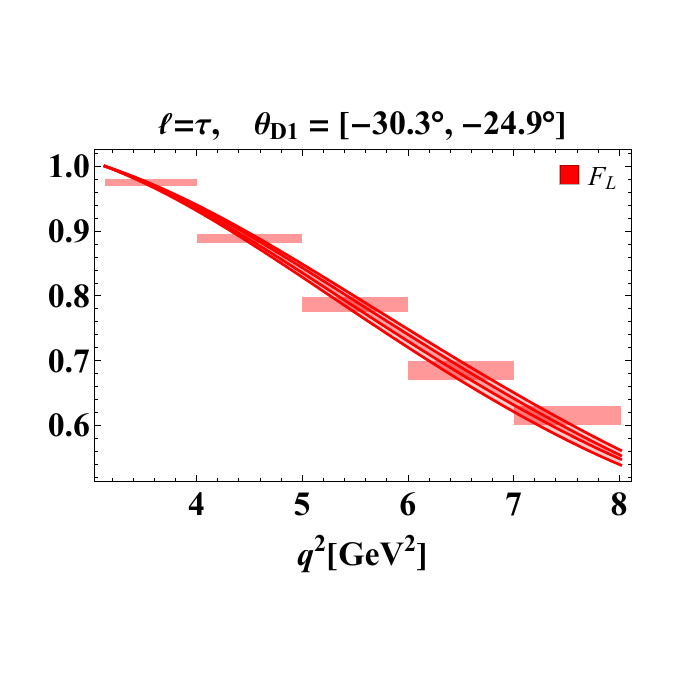}
    \vspace{-1.5cm}
    \caption{}
    \label{fig10c}
  \end{subfigure}
 \vspace{-0.9cm} \\ 
    \begin{subfigure}{0.33\textwidth}
    \includegraphics[width=\linewidth]{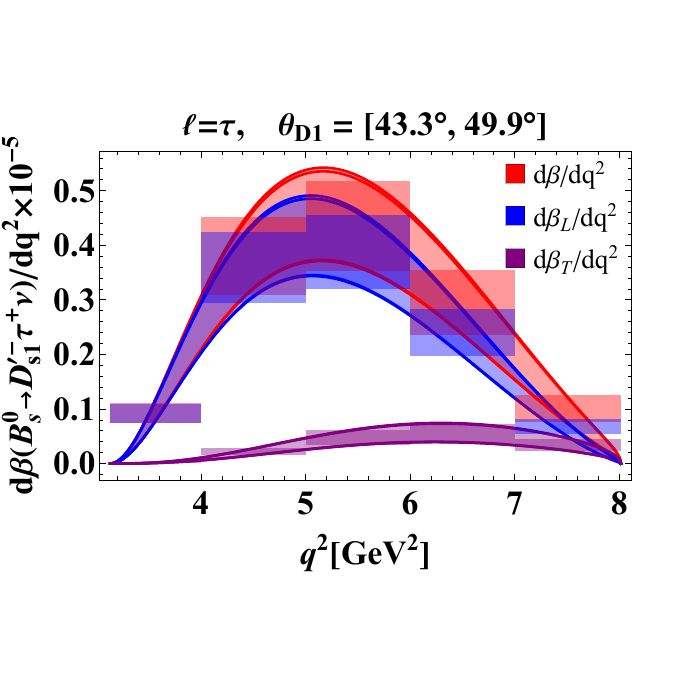}
    \vspace{-1.5cm}
    \caption{}
    \label{fig10d}
  \end{subfigure}%
  \hfill
  \begin{subfigure}{0.33\textwidth}
    \includegraphics[width=\linewidth]{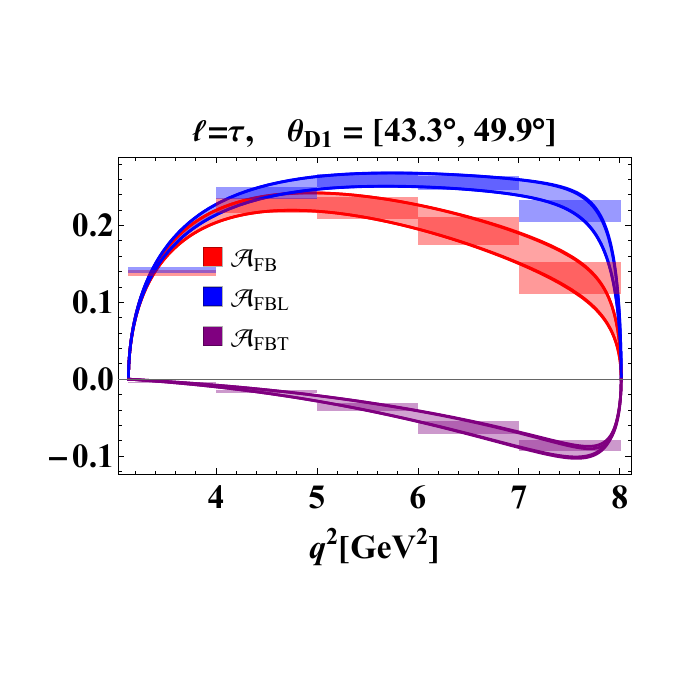}
    \vspace{-1.5cm}
    \caption{}
    \label{fig10e}
  \end{subfigure}%
  \hfill
  \begin{subfigure}{0.33\textwidth}
    \includegraphics[width=\linewidth]{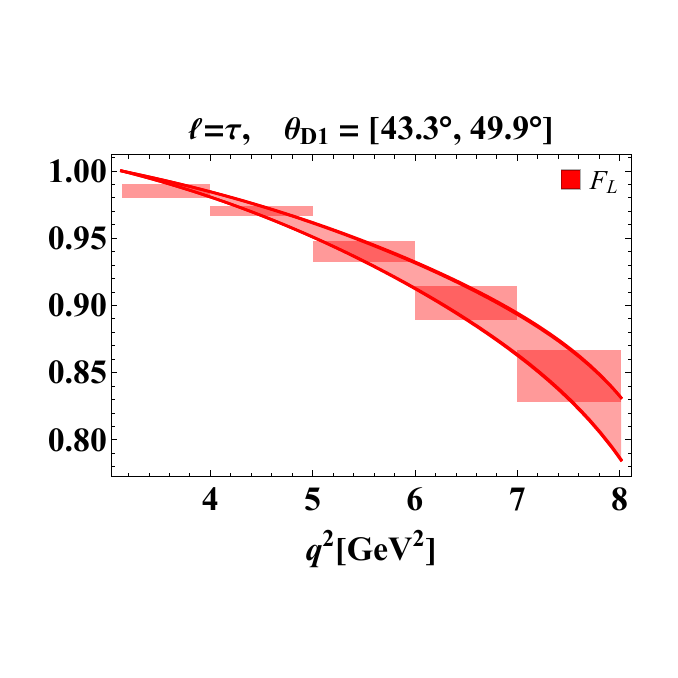}
    \vspace{-1.5cm}
    \caption{}
    \label{fig10f}
  \end{subfigure}
\label{btods1ptau}
\end{figure}

The $q^2$ distributions of the polarized and unpolarized differential branching ratio, $\mathcal{A}_{\text{FB}}$ and the  $F_{\text{L}}$ for $B_s^0 \to D_{s1}^{\prime-} \tau^+ \nu$ are displayed in FIGs.~\ref{btods1ptau}. The results are presented for two mixing-angle regions: $\theta_{D1} \in [-30.3^\circ, -24.9^\circ]$ (upper row) and $\theta_{D1} \in [43.3^\circ, 49.9^\circ]$ (lower row).
\begin{itemize}
    \item In FIG.~\ref{fig10a}, corresponding to the branching ratio the L contribution dominates up to approximately $q^2 \leq 6.5 ~\mathrm{GeV}^2$. Beyond this region, the T component increases and becomes comparable to, and eventually slightly larger than, the L contribution at higher $q^2$. This behavior closely follows that observed in the $D_{1}^\prime$ channel, with only differences in the numerical values. The corresponding bin-wise values are listed in TABLE~\ref{table10}. In contrast, FIG.~\ref{fig10d} show the L contribution remains dominant over most of the $q^2$ region. However, in the slight highest bin, $q^2 \geq 7.5~\mathrm{GeV}^2$, the L and T contributions become comparable, as also reflected in TABLE~\ref{table10}. This trend is again consistent with the behavior seen in the $D_{1}^\prime$ case.
    \item The unpolarized $\mathcal{A}_{\text{FB}}$, shown in FIG.~\ref{fig10b}, exhibits a zero crossing at around $q^2 \approx 5.1~\mathrm{GeV}^2$. The L component remains positive while the T component remains negative over the entire $q^2$ region. In FIG.~\ref{fig10e}, corresponding to the positive angle range, the unpolarized $\mathcal{A}_{\text{FB}}$ does not show a zero crossing and remains positive over the whole $q^2$ region. It increases from $(0.133, 0.142)$ in the bin $q^2 = 3.13$--$4~\mathrm{GeV}^2$ to its maximum value $(0.215, 0.236)$ in the bin $q^2 = 4$--$5~\mathrm{GeV}^2$, and then decreases towards the high $q^2$, as listed in TABLE~\ref{table10}. The L contribution is positive and dominant, while the T contribution remains negative and small, $|\mathcal{A}_{\text{FBT}}| \lesssim 0.09$. This behavior is similar to that of the $D_1^{\prime}$ channel shown in FIG.~\ref{fig:6e}, with only numerical differences.
    \item FIGs.~\ref{fig10c} and \ref{fig10f} show the variation of $F_{\text{L}}$ for both $\theta_{D1}$ ranges. The bin-wise values listed in TABLE~\ref{table10} are in good agreement with the trends observed in the plots over the entire $q^2$ region.
\end{itemize}

\subsubsection{The Ratio of Branching Fraction}
Ratios of branching fractions such as $\mathcal{R}_{(D)}$ \cite{Bigi:2016mdz, PhysRevD.95.115008, Jaiswal:2017rve} and $\mathcal{R}_{(D^*)}$ \cite{PhysRevD.95.115008, Jaiswal:2020wer, GAMBINO2019386} have been studied in both experimental and theoretical literature as sensitive probes of LFU~\cite{Lees:2012xj,Huschle:2015rga,Aaij:2015yra,Fajfer:2012vx}. Therefore, in this work, we extend the study of $\mathcal{R}_{D_{1}^{(\prime)}}$ and $\mathcal{R}_{D_{s1}^{(\prime)}}$ observables defined in Eq.~(\ref{rdratio}) as a complementary check on LFU. 

\begin{figure}[H]
\caption{The ratio of the branching fraction $\mathcal{R}_{D_1}$ and $\mathcal{R}_{D_1^\prime}$ as a function of $q^2$.}
\vspace{-0.7cm}
\centering
\begin{subfigure}{0.48\textwidth}
\includegraphics[width=\linewidth]{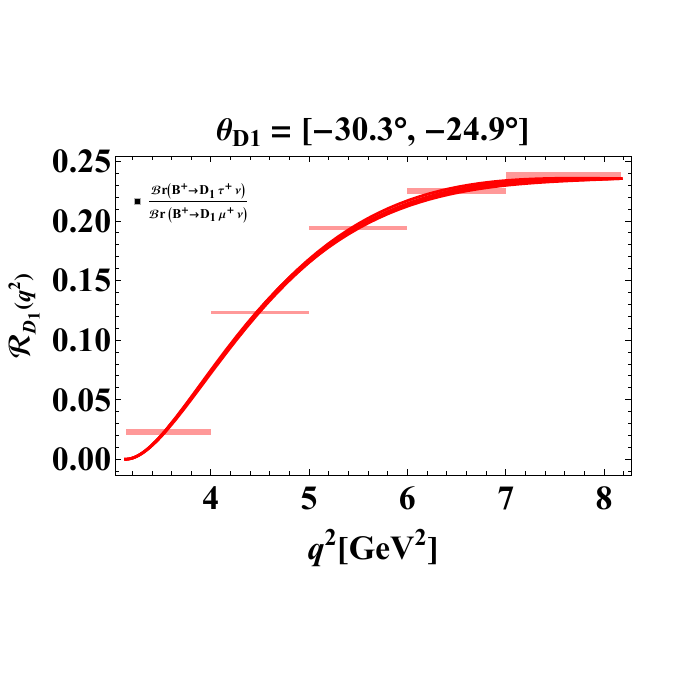}
\vspace{-1.7cm}
\caption{}
\label{rd1a}
\end{subfigure}
\begin{subfigure}{0.48\textwidth}
\includegraphics[width=\linewidth]{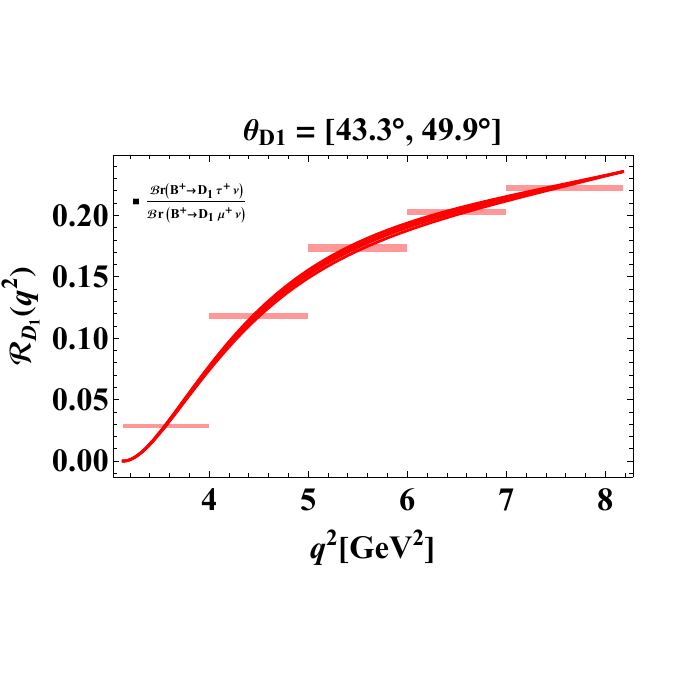}
\vspace{-1.7cm}
\caption{}
\label{rd1b}
\end{subfigure}
\vspace{-1.5cm} \\ 
\begin{subfigure}{0.48\textwidth}
\includegraphics[width=\linewidth]{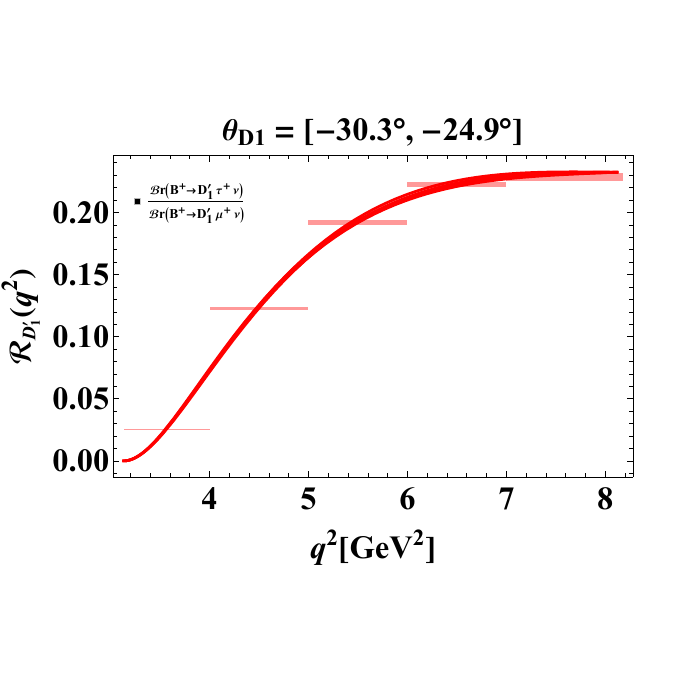}
\vspace{-1.7cm}
\caption{}
\label{rd1c}
\end{subfigure}
\begin{subfigure}{0.48\textwidth}
\includegraphics[width=\linewidth]{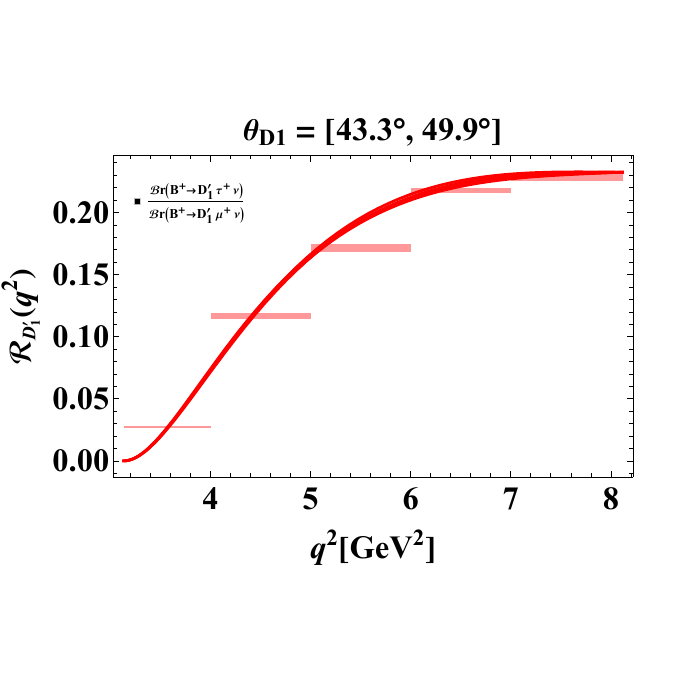}
\vspace{-1.7cm}
\caption{}
\label{rd1d}
\end{subfigure}
\label{RD1}
\end{figure}
FIGs.~\ref{rd1a} and \ref{rd1b} show the $q^2$ dependence of the ratio $\mathcal{R}(q^2)$ for the $B^+ \to D_1$ channel, while FIGs.~\ref{rd1c} and \ref{rd1d} correspond to the $B^+ \to D_1^\prime$ channel. The results are presented for both negative $\theta_{D1} \in [-30.3^\circ, -24.9^\circ]$ and positive $\theta_{D1} \in [43.3^\circ, 49.9^\circ]$ mixing angle ranges. The bin-wise values of the ratio are listed in TABLE~\ref{tableRD}.
\begin{figure}[H]
\caption{The ratio of the branching fraction $\mathcal{R}_{D_{s1}}$ and $\mathcal{R}_{D_{s1}^\prime}$ as a function of $q^2$.}
\vspace{-0.7cm}
\centering
\begin{subfigure}{0.48\textwidth}
\includegraphics[width=\linewidth]{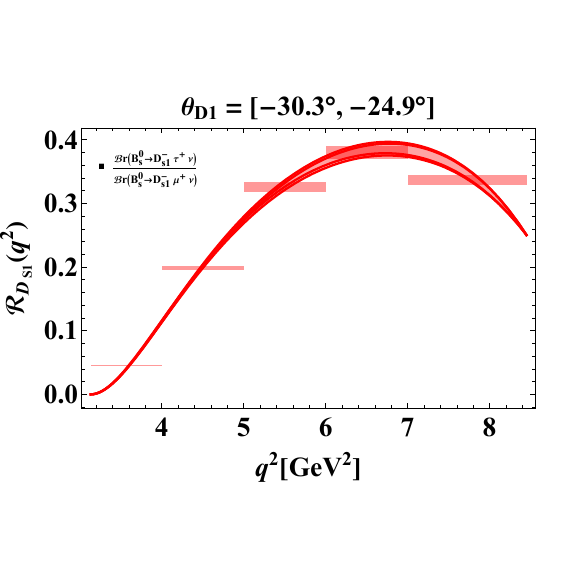}
\vspace{-1.7cm}
\caption{}
\label{rds1a}
\end{subfigure}
\begin{subfigure}{0.48\textwidth}
\includegraphics[width=\linewidth]{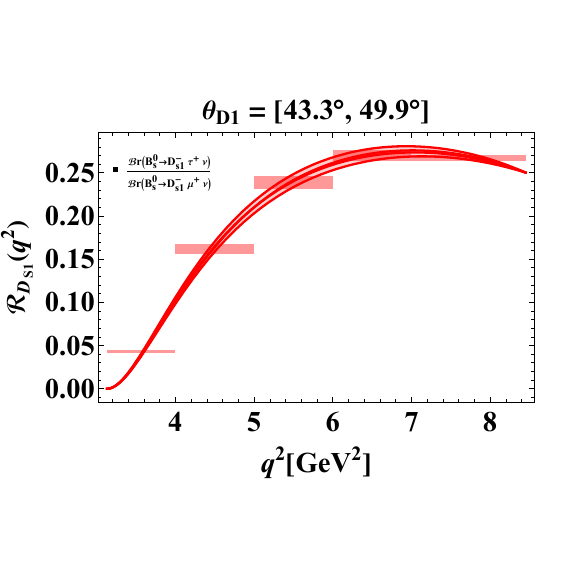}
\vspace{-1.7cm}
\caption{}
\label{rds1b}
\end{subfigure}
\vspace{-1.5cm} \\ 
\begin{subfigure}{0.48\textwidth}
\includegraphics[width=\linewidth]{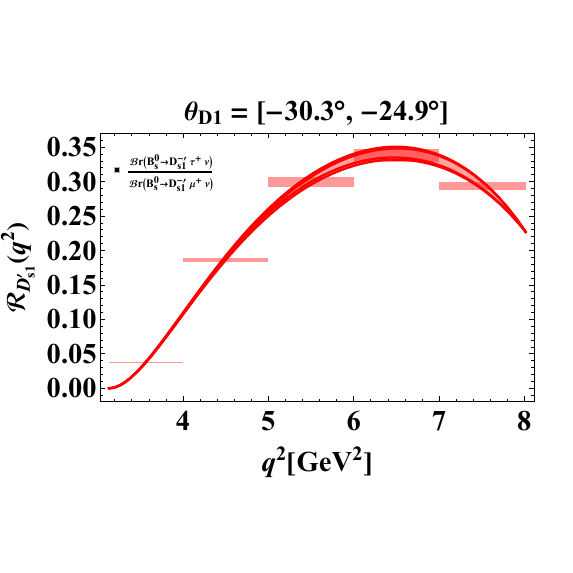}
\vspace{-1.7cm}
\caption{}
\label{rds1c}
\end{subfigure}
\begin{subfigure}{0.48\textwidth}
\includegraphics[width=\linewidth]{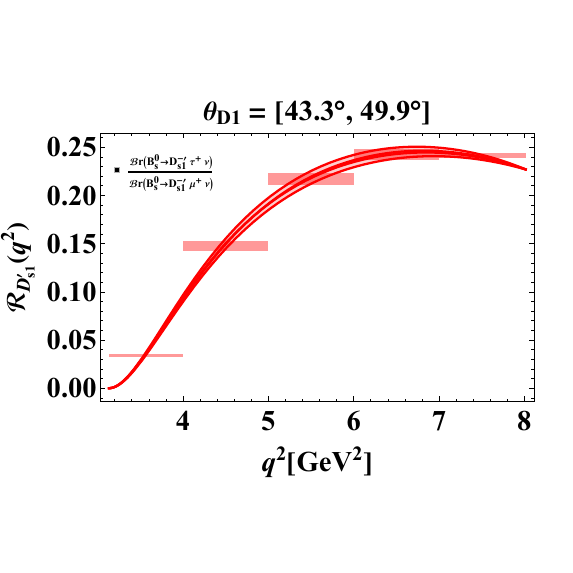}
\vspace{-1.7cm}
\caption{}
\label{rds1d}
\end{subfigure}
\label{RD2}
\end{figure}
FIGs.~\ref{rds1a} and \ref{rds1b} show the $q^2$ dependence of the ratio $\mathcal{R}(q^2)$ for the $B_s^0 \to D_{s1}^-$ channel, while FIGs.~\ref{rds1c} and \ref{rds1d} correspond to the $B_s^0 \to D^{\prime-}_{s1}$ channel. The results are presented for both negative $\theta_{D1} \in [-30.3^\circ, -24.9^\circ]$ and positive $\theta_{D1} \in [43.3^\circ, 49.9^\circ]$ mixing angle ranges. The bin-wise values of the ratio are listed in TABLE~\ref{tableRD}.

In Figs.~\ref{RD1} and~\ref{RD2}, for both cases, the ratios are suppressed in the low-$q^2$ region because of the reduced phase space available for the $\tau$ mode, and then increase toward higher $q^2$. The bin-averaged points, together with their uncertainties, follow the same overall behavior as the continuous curves, showing that the overall trend is stable within the theoretical uncertainties. Since these ratios are constructed from the same hadronic
transition in the $\tau$ and $\mu$ modes, part of the hadronic form-factor
uncertainties cancels between numerator and denominator, making
$\mathcal{R}_{D_1}$ and $\mathcal{R}_{D_{s1}}$ theoretically cleaner than the
individual branching ratios.

\subsection{Correlation Analysis for  $B^+ \to  D_1^{(\prime)}\ell^+\nu_\ell$ and
$B_s^0 \to D_{s1}^{-(\prime)}\ell^+\nu_\ell$ Decays}
\label{corelationD1}

\begin{figure}[H]
\centering
\caption{Correlations between observables for the decay channels $B^+ \to D_1 \mu^+ \nu$, $B^+ \to D_1^{\prime} \mu^+ \nu$, $B_s^0 \to D_{s1}^- \mu^+ \nu$ and $B_s^0 \to D_{s1}^{-\prime} \mu^+ \nu$.}
\vspace{1mm}

\begin{subfigure}{0.24\textwidth}
    \centering
    \includegraphics[width=\linewidth]{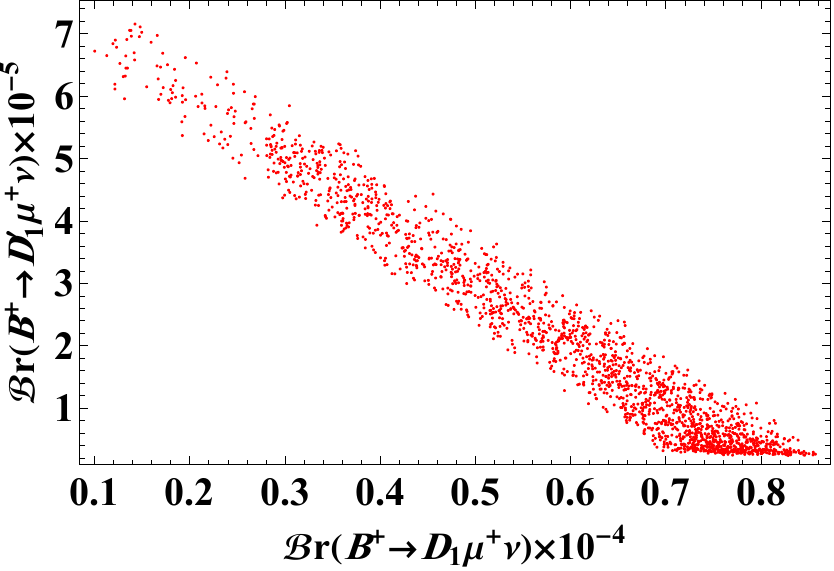}
\end{subfigure}
\begin{subfigure}{0.26\textwidth}
    \centering
    \includegraphics[width=\linewidth]{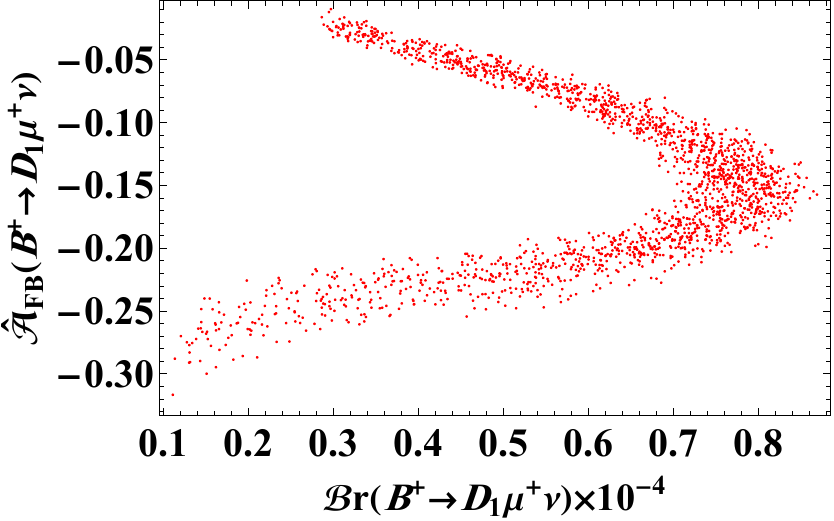}
\end{subfigure}
\begin{subfigure}{0.24\textwidth}
    \centering
    \includegraphics[width=\linewidth]{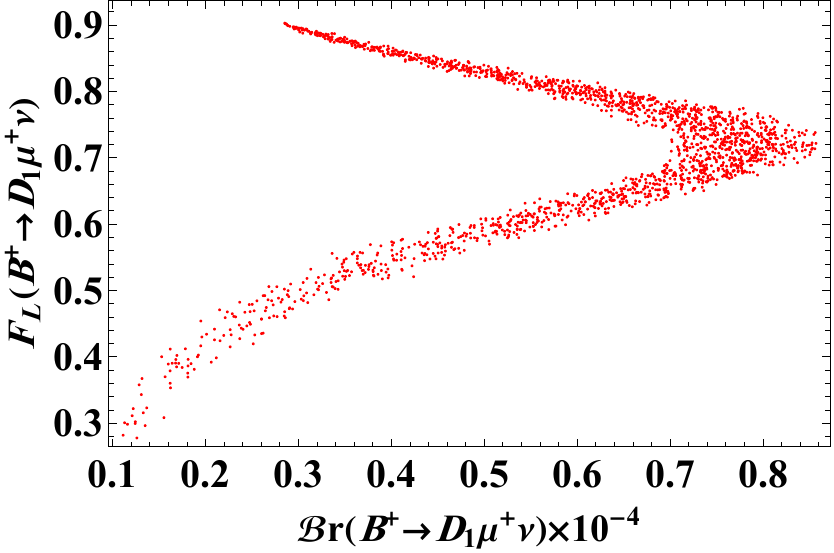}
\end{subfigure}
\begin{subfigure}{0.24\textwidth}
    \centering
    \includegraphics[width=\linewidth]{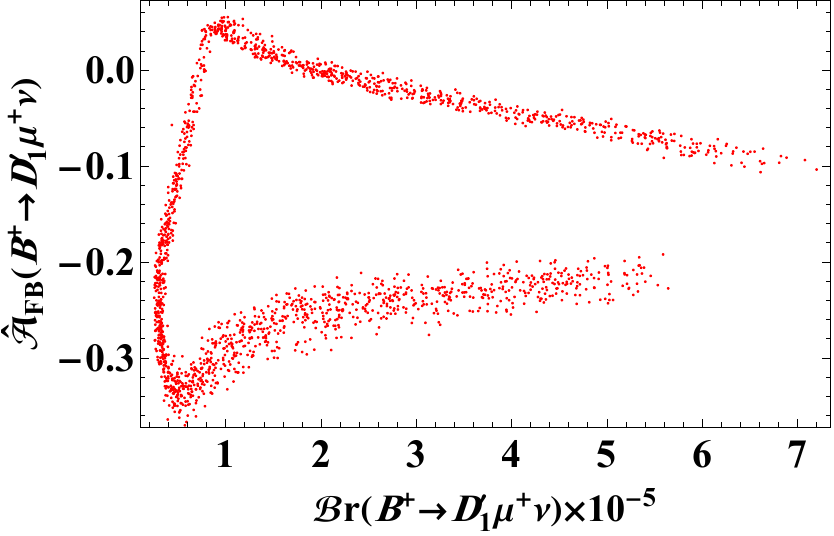}
\end{subfigure}

\vspace{0.2cm}

\begin{subfigure}{0.24\textwidth}
    \centering
    \includegraphics[width=\linewidth]{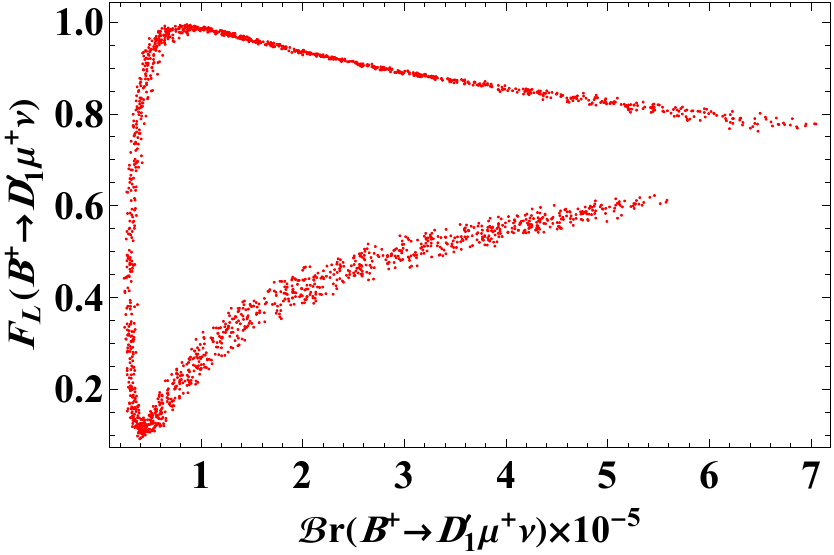}
\end{subfigure}
\begin{subfigure}{0.24\textwidth}
    \centering
    \includegraphics[width=\linewidth]{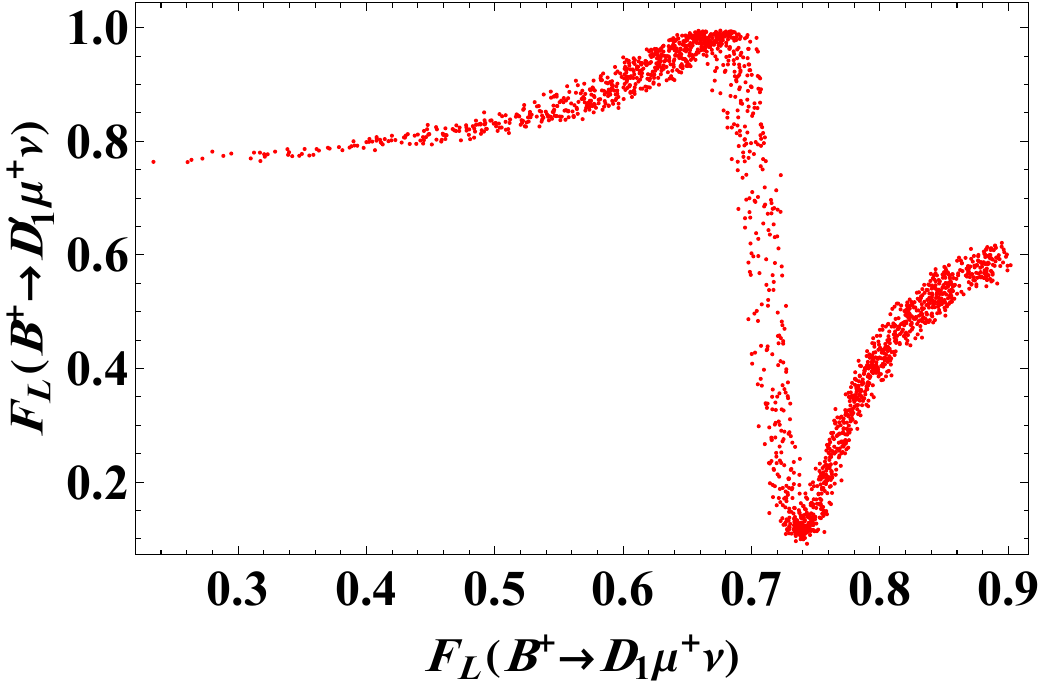}
\end{subfigure}
\begin{subfigure}{0.25\textwidth}
    \centering
    \includegraphics[width=\linewidth]{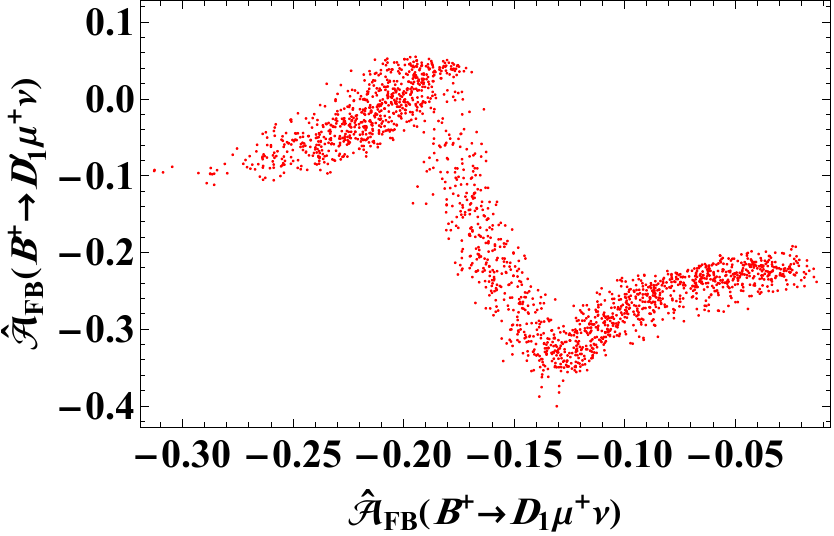}
\end{subfigure}
\begin{subfigure}{0.24\textwidth}
    \centering
    \includegraphics[width=\linewidth]{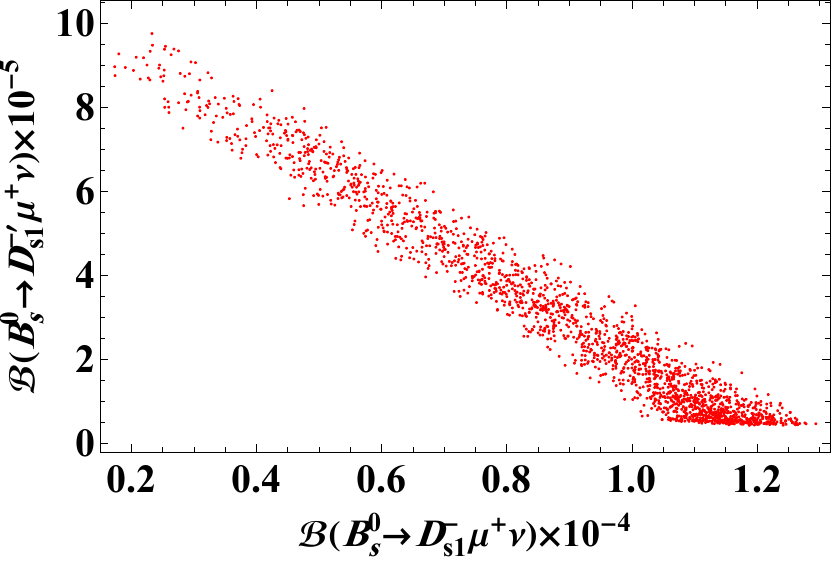}
\end{subfigure}

\vspace{0.2cm}

\begin{subfigure}{0.25\textwidth}
    \centering
    \includegraphics[width=\linewidth]{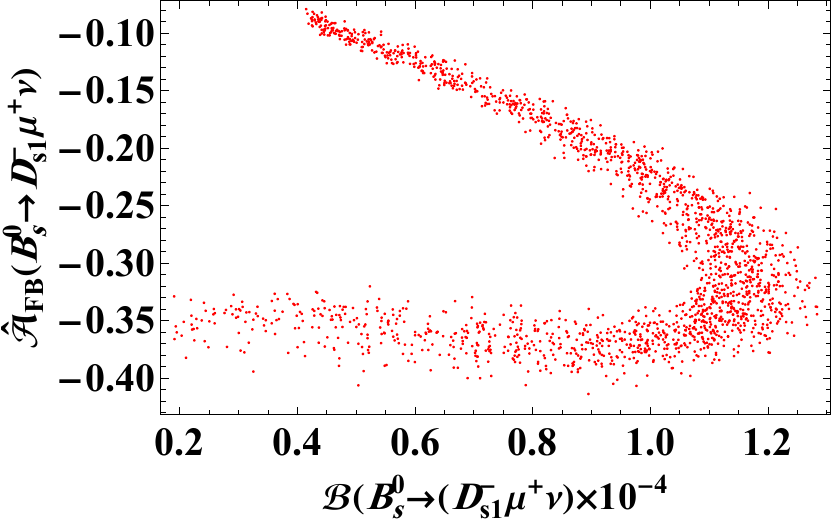}
\end{subfigure}
\begin{subfigure}{0.24\textwidth}
    \centering
    \includegraphics[width=\linewidth]{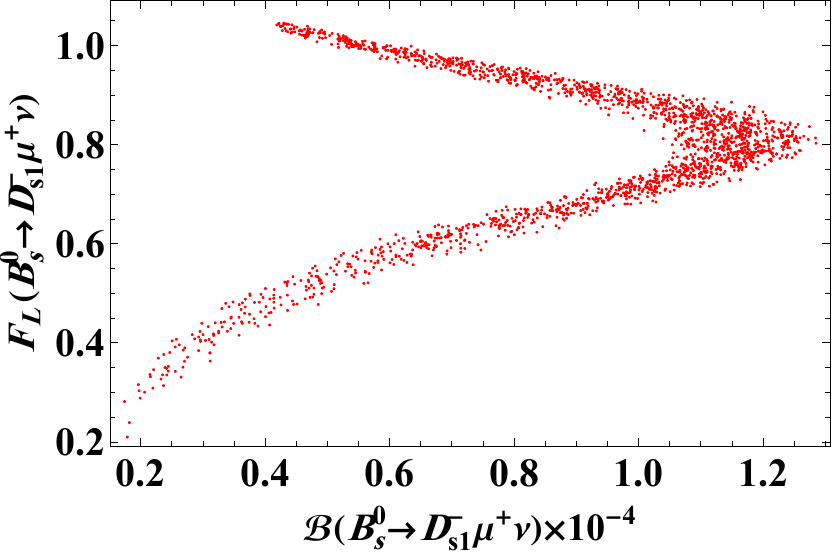}
\end{subfigure}
\begin{subfigure}{0.25\textwidth}
    \centering
    \includegraphics[width=\linewidth]{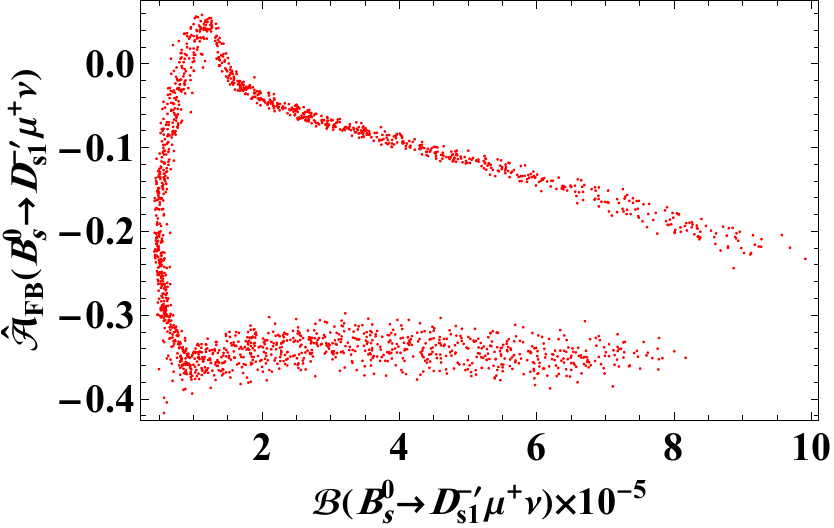}
\end{subfigure}

\vspace{0.2cm}

\begin{subfigure}{0.24\textwidth}
    \centering
    \includegraphics[width=\linewidth]{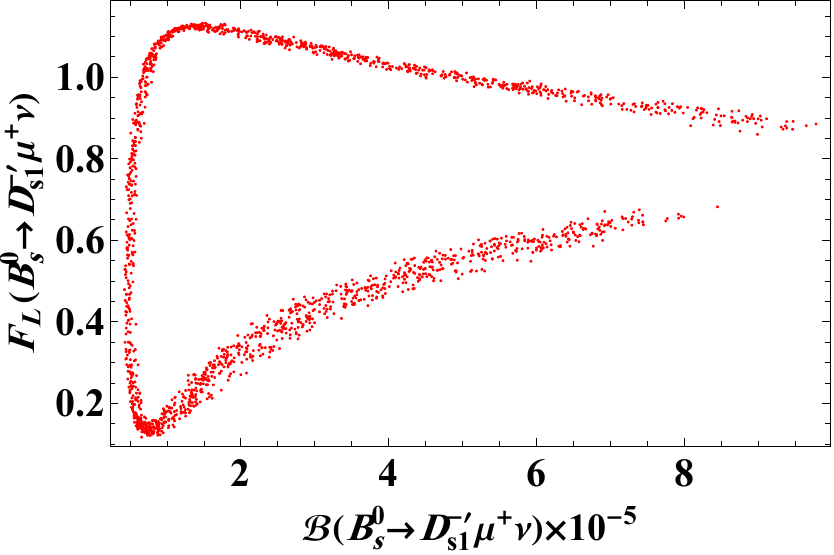}
\end{subfigure}
\begin{subfigure}{0.25\textwidth}
    \centering
    \includegraphics[width=\linewidth]{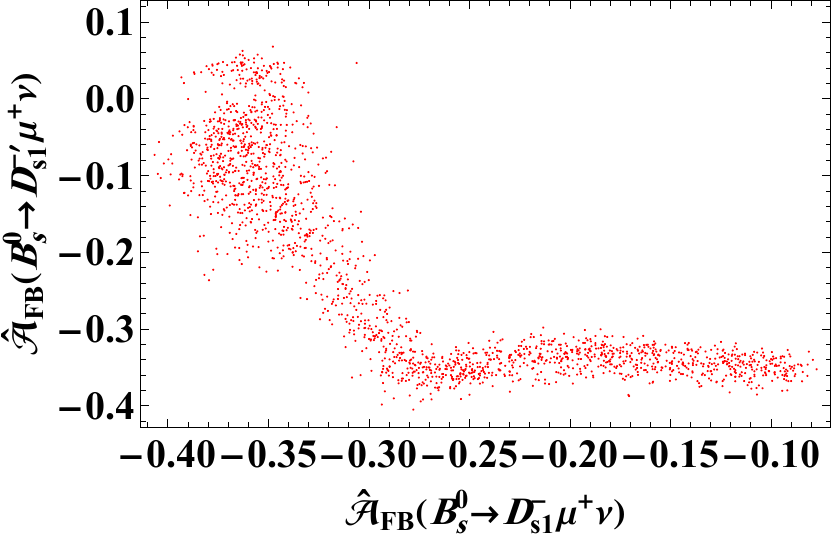}
\end{subfigure}
\begin{subfigure}{0.24\textwidth}
    \centering
    \includegraphics[width=\linewidth]{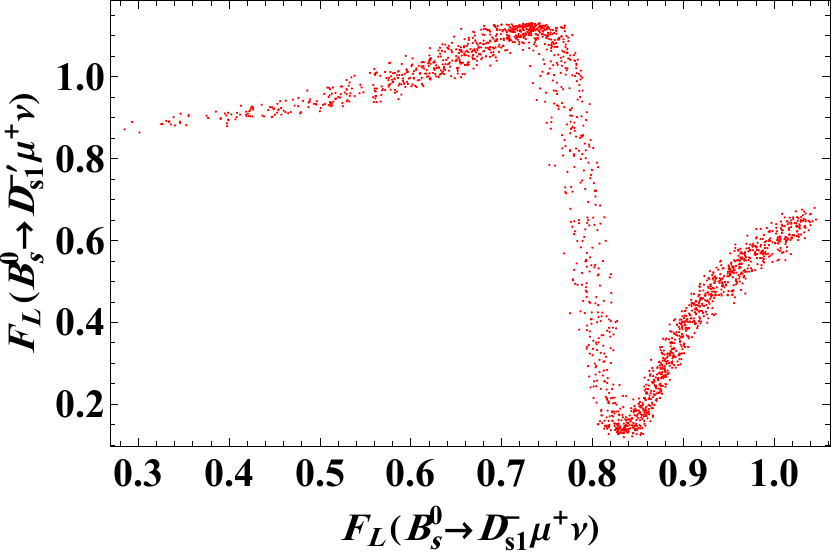}
\end{subfigure}
\label{corelation_combined}
\end{figure}

In this section, we analyze the correlations among the key observables, namely the branching ratio $d\mathcal{B}/dq^2$, the lepton forward-backward asymmetry $\mathcal{A}_{\rm FB}$, and the longitudinal polarization fraction $F_L$, for the decay channels $B^+ \to D_1 \mu^+ \nu_\mu$ and $B^+ \to D_1' \mu^+ \nu_\mu$. These correlations are obtained by scanning over the full allowed range of the mixing angle $\theta_{D_1} \sim [-45.0^\circ, 80.0^\circ]$ and are shown in FIG.~\ref{corelation_combined}.
The branching ratios of the two mixed axial vector states show an anti-correlated behavior, reflecting the complementary contributions of the $j_\ell=1/2$ and $j_\ell=3/2$ components.
The correlations involving $\mathcal{A}_{\rm FB}$ and $F_L$ show a curved structures rather than simple linear trends, indicating that these observables are strongly affected by the variation of the mixing angle. Overall, the correlation patterns show that the branching ratios, forward-backward asymmetries, and longitudinal polarization fractions provide complementary sensitivity to the axial-vector mixing structure.

\clearpage
\section{Conclusion}
\label{concl}
In this work, we have investigated the $b \to c \ell \nu$ mediated $B$ decays
 $B^+ \to  D_1^{(\prime)}\ell^+\nu_\ell$ and
$B_s^0 \to D_{s1}^{-(\prime)}\ell^+\nu_\ell$, where $\ell=\mu $ and $ \tau $ within the SM framework, using the effective weak Hamiltonian. The physical axial-vector states $D_{1}^{(\prime)}$ and $D_{s1}^{\prime}$ are treated as mixtures of the $D_{1}^{3/2}$ and $D_{1}^{1/2}$ states, respectively, and the dependence on the mixing angle $\theta_{D1}$ is explicitly taken into account. The elements of the hadronic matrix associated with the $B^+ \to  D_1^{\prime}\ell^+\nu_\ell$ and
$B_s^0 \to D_{s1}^{-(\prime)}\ell^+\nu_\ell$ transition are parametrized in terms
of the helicity form factors that are dependent on $q^2$. By using CLFQM form factors we have numerically calculated the differential branching ratio $d\mathcal{B}/dq^2$, the longitudinal polarization fraction $F_L$, the lepton forward-backward asymmetry $\mathcal{A}_{\text{FB}}$, and the lepton flavor universality ratios $R_{D_{1}^{\prime}}$ and $R_{D_{s1}^{\prime}}$. We have also presented the polarized branching ratios and forward-backward asymmetries corresponding to longitudinal and transverse final state mesons. The numerical analysis is carried out by considering the mixing angle in two regions, $\theta_{D1} \in [-30.3^\circ, -24.9^\circ]$ and $\theta_{D1} \in [43.3^\circ, 49.9^\circ]$. We sequentially discuss the implications of the $d\mathcal{B}/dq^2$ (polarized and unpolarized), $\mathcal{A}_{\text{FB}}$ (polarized and unpolarized) and $F_L$ over a wide range of angles. At present, experimental data are available only for the branching ratios of the $B^+ \to  D_1^{\prime}\ell^+\nu_\ell$ channels. In addition to these, we provide detailed predictions for the corresponding observables in the $B_s^0 \to D_{s1}^{-(\prime)}\ell^+\nu_\ell$ channels, for which no experimental measurements are currently available. These results offer a consistent SM baseline and can serve as important inputs for future experimental studies, helping to test the sensitivity of different observables to the mixing angle and to further clarify the existing tension between theoretical predictions and experimental observations. Furthermore, we have presented bin-averaged predictions of the observables using a bin width of $\Delta q^2 = 1~\text{GeV}^2$, which are particularly useful for future experimental analyses.  Finally, our study extends the analysis to the $R_{D_{1}^{(\prime)}}$ and $R_{D_{s1}^{(\prime)}}$ observables as complementary probes of LFU, and also discusses the correlations among different observables. Precise measurements of these observables, together with polarization-dependent quantities, will not only help constrain the mixing angle but also provide deeper insight into the existing tensions between theoretical predictions and experimental observations, thereby offering an important avenue for testing the SM in semileptonic $b \to c \ell \nu$ transitions.

The main points of the present analysis can be summarized as follows. (i) All
observables are calculated within the two mixing-angle windows allowed by the
experimental data and also over a wider range of $\theta_{D1}$, and the
form-factor uncertainties are shown in all figures as error bands. (ii) Besides
the unpolarized observables, the forward-backward asymmetry is given separately
for the longitudinally and transversely polarized final state meson, which was not
given in Ref.~\cite{Yang:2025abu}. (iii) All observables are also calculated
bin-wise with a bin width of $\Delta q^2 = 1~\text{GeV}^2$, whereas
Ref.~\cite{Yang:2025abu} gives only integrated values. (iv) The branching ratios
in the two allowed windows are close to each other, while $\mathcal{A}_{\text{FB}}$
and $F_L$ are quite different. For the decay $B^+ \to D_1 \mu^+ \nu$, for example,
$\mathcal{A}_{\text{FB}}$ varies within $[-0.257, -0.263]$ in the negative window
and within $[0.002, 0.019]$ in the positive one, while $F_L$ varies within
$[0.828, 0.810]$ and $[0.581, 0.543]$, respectively. A measurement of these
angular observables can therefore distinguish between the two windows, which is
not possible with the branching ratios alone. (v) Our integrated branching ratios
are compared with other theoretical approaches and with the available
experimental data in TABLE~\ref{tabcomparebr}. These results can serve as a SM
reference for future measurements of semileptonic $B_{(s)}$ decays into orbitally
excited charmed mesons.

It is also important to mention that our prediction for
$\mathcal{B}(B^+ \to \bar{D}_1^{0}\mu^+\nu_\mu)$ is larger than the measured
value. This difference mainly comes from the hadronic input. The branching ratio is
proportional to the square of the transition form factors, and the form factors of
the broad $j_\ell = 1/2$ state are not well determined, as can be seen in
TABLE~\ref{tabcompareff}. In addition, the branching ratio depends on the mixing
angle, which is not precisely known, and the experimental value is not measured
directly, but it is extracted by using the branching fraction of the
$D_1 \to D^{*} \pi$ decay. It is worth noting that all the theoretical predictions
given in TABLE~\ref{tabcomparebr} are larger than the measured value for this
channel. Therefore, this difference is not specific to our calculation.

Finally, we comment on whether the deviations observed in the axial vector
channels can be interpreted as signals of new physics. In our opinion, they are
more likely of hadronic origin. These deviations appear in the decays with light
leptons in the final state, where new physics effects in the $b \to c \ell \nu$
transition are strongly constrained by the well measured $B \to D^{(*)} \ell \nu$
decays. Moreover, the difference between theory and experiment is of the same size
as the hadronic uncertainties. As can be seen in TABLE~\ref{tabcompareff},
different quark model calculations of the $j_\ell = 1/2$ form factors differ by
more than a factor of two, and the $D_1^{\prime}$ is a broad state, which makes
the extraction of its branching ratio more difficult. The angular observables
$\mathcal{A}_{\text{FB}}$ and $F_L$, together with the ratios
$\mathcal{R}_{D_1^{(\prime)}}$ and $\mathcal{R}_{D_{s1}^{(\prime)}}$, are less
sensitive to the form factors than the branching ratios, and their measurement can
therefore help to separate the hadronic effects from possible new physics.
\clearpage
\appendix
\section*{Appendix}
\begin{table}[H]
\caption{Predictions of binned observables for the decay channel $B^+ \to D_1 \mu^+ \nu $, including the unpolarized differential branching ratio $\left\langle d\mathcal{B}/dq^2\right\rangle$, the lepton forward-backward asymmetry $\left\langle \mathcal{A}_{\text{FB}}\right\rangle$ and the longitudinal polarization fraction $\left\langle F_{\mathrm{L}}\right\rangle$, given within the SM framework in bins of width $\Delta q^2 = 1~\text{GeV}^2$. The differential branching ratio and the forward-backward asymmetry are also given separately for a longitudinally (L) and a transversely (T) polarized final-state meson. Each entry is a pair $(x_{\min},x_{\max})$ giving the smallest and the largest value the observable takes when the mixing angle is varied over the quoted window and the form-factor parameters are varied within their errors. The upper block of the table corresponds to $\theta_{D1} \in [-30.3^\circ, -24.9^\circ]$ and the lower block to $\theta_{D1} \in [43.3^\circ, 49.9^\circ]$. The differential branching ratios are given in units of $10^{-4}~\text{GeV}^{-2}$.}
\centering
\resizebox{\textwidth}{!}{%
\begin{tabular}{|c|ccc|ccc|c|}
\hline
\hline
\multicolumn{8}{c}{\textbf{$[-30.3^\circ, -24.9^\circ]$ }} \\
\hline
$q^2=$ GeV  & $\left\langle d\mathcal{B}/dq^2\right\rangle$ & $\left\langle d\mathcal{B}_{\text{L}}/dq^2\right\rangle$ &  $\left\langle d\mathcal{B}_{\text{T}}/dq^2\right\rangle$ &  $\left\langle \hat{\mathcal{A}}_{\text{FB}}\right\rangle$ & $\left\langle \hat{\mathcal{A}}_{\text{FBL}}\right\rangle$ & $\left\langle \hat{\mathcal{A}}_{\text{FBT}}\right\rangle$ &$\left\langle \hat{F}_{\text{L}}\right\rangle$ \\
\hline

$[0.01-1]$   & (0.993, 1.096) & (0.963, 1.057) & (0.029, 0.039) & (0.021, 0.025) & (0.036, 0.037) &(-0.012, -0.015)  & (0.961, 0.966) \\
$[1-2]$     & (0.824, 0.922) & (0.750, 0.825) & (0.073, 0.097) & (-0.031, -0.043) & (0.007, 0.008) &(-0.039, -0.050)  & (0.918, 0.931) \\
$[2-3]$     & (0.652, 0.744) & (0.555, 0.616) & (0.096, 0.128) & (-0.060, -0.080) & (0.003, 0.004) &  (-0.064, -0.083)&(0.866, 0.887)  \\
$[3-4]$     &(0.503, 0.585)  & (0.396, 0.445) &(0.106, 0.140)  & (-0.080, -0.107) & (0.001, 0.002)& (-0.083, -0.109) & (0.815, 0.840) \\
$[4-5]$    & (0.376, 0.448) & (0.269, 0.307) & (0.107, 0.140)  &  (-0.091, -0.122)& (0.001, 0.0012) & (-0.092, -0.123) &(0.760, 0.788)  \\
$[5-6]$   & (0.272, 0.330) & (0.117, 0.199) & (0.100, 0.131)  & (-0.091, -0.122) & $(\sim 10^{-4})$ &(-0.093, -0.123)  &  (0.698, 0.726)\\
$[6-7]$   & (0.183, 0.228) &(0.097, 0.116)  & (0.086, 0.111) & (-0.078, -0.104) & $(\sim 10^{-4})$ & (-0.078, -0.105) &  (0.626, 0.651)\\
$[7-8.17]$   & (0.105, 0.133) & (0.043, 0.054) & (0.061, 0.078) &  (-0.051, -0.069)& $(\sim 10^{-4})$ & (-0.052, -0.069) & (0.530, 0.540) \\
\hline
\multicolumn{8}{c}{\textbf{$[43.3^\circ, 49.9^\circ]$ }} \\ \hline

$[0.01-1]$   & (0.563, 0.727) & (0.489, 0.641) &(0.074, 0.085)  & (-0.025, -0.034) & (-0.031, 0.028) &  (-0.055, -0.065)& (0.881, 0.899) \\
$[1-2]$     & (0.586, 0.730) &(0.394, 0.511)  & (0.191, 0.219) & (-0.173, -0.197) &(0.004, 0.004)  &(-0.178, -0.202)  &  (0.725, 0.763)\\
$[2-3]$     & (0.595, 0.721) &  (0.336, 0.426)& (0.259, 0.294) &(-0.274, -0.304)  &(0.001, 0.002)  & (-0.276, -0.306) & (0.626, 0.667) \\
$[3-4]$     &  (0.575, 0.681)&  (0.284, 0.353)&(0.290, 0.327)  & (-0.335, -0.367) & $(\sim 10^{-4})$ &  (-0.355, -0.368)&  (0.563, 0.602)\\
$[4-5]$    & (0.529, 0.615) & (0.236, 0.286) & (0.293, 0.328) &(-0.354, -0.386)  & $(\sim 10^{-4})$ & (-0.330, -0.386) & (0.522, 0.556) \\
$[5-6]$   & (0.461, 0.527) & (0.189, 0.223) &  (0.272, 0.303)&(-0.330, -0.360)  & $(\sim 10^{-4})$ & (-0.264, -0.360) & (0.494, 0.521) \\
$[6-7]$   & (0.368, 0.414) & (0.139, 0.160) &  (0.228, 0.253)& (-0.263, -0.288) & $(\sim 10^{-4})$ & (-0.155, -0.289) &(0.477, 0.496)  \\
$[7-8.17]$   &(0.243, 0.269)  & (0.085, 0.095) & (0.157, 0.173) &  (-0.155, -0.171)&  $(\sim 10^{-4})$&(-0.155, -0.171)  & (0.480, 0.490) \\
\hline
\hline
\end{tabular}}
\label{table3}
\end{table}

\begin{table}[H]
\caption{Same as in Table~\ref{table3} but for  $B^+ \to D_1 \tau^+ \nu $. The differential branching ratios are given in units of $10^{-5}$ $\text{GeV}^{-2}$.}
\centering
\resizebox{\textwidth}{!}{%
\begin{tabular}{|c|ccc|ccc|c|}
\hline
\hline
\multicolumn{8}{c}{\textbf{ $[-30.3^\circ, -24.9^\circ]$}} \\
\hline
$q^2=$ GeV  & $\left\langle d\mathcal{B}/dq^2\right\rangle$ & $\left\langle d\mathcal{B}_{\text{L}}/dq^2\right\rangle$ &  $\left\langle d\mathcal{B}_{\text{T}}/dq^2\right\rangle$ &  $\left\langle \hat{\mathcal{A}}_{\text{FB}}\right\rangle$ & $\left\langle \hat{\mathcal{A}}_{\text{FBL}}\right\rangle$ & $\left\langle \hat{\mathcal{A}}_{\text{FBT}}\right\rangle$ &$\left\langle \hat{F}_{\text{L}}\right\rangle$ \\
\hline
$[3.13-4]$     &(0.011, 0.014) &(0.011, 0.013) & $(\sim 10^{-4})$ & (0.101, 0.105) & (0.105, 0.109) & (-0.004, -0.005) &(0.970, 0.980)    \\
$[4-5]$    & (0.045, 0.054) &(0.041, 0.051)&(0.003, 0.004)  & (0.142, 0.152) & (0.163, 0.171) & (-0.018, -0.022) &    (0.945, 0.951)\\
$[5-6]$   & (0.052, 0.063) &(0.043, 0.052)& (0.008, 0.011) &  (0.108, 0.126)& (0.150, 0.160) &  (-0.033, -0.042)&   (0.880, 0.892) \\
$[6-7]$   & (0.041, 0.050)&(0.029, 0.035) & (0.012, 0.015) &(0.058, 0.080)  & (0.115, 0.126) & (-0.041, -0.057) &  (0.780, 0.797)  \\
$[7-8.17]$   & (0.020, 0.026) &(0.010, 0.013)& (0.010, 0.013) & (0.013, 0.032) &(0.067, 0.075)  & (-0.041, -0.054)&  (0.680, 0.700)\\
\hline
\multicolumn{8}{c}{\textbf{ $[43.3^\circ, 49.9^\circ]$}} \\
\hline

$[3.13-4]$     & (0.015, 0.018) &(0.014, 0.017) &(0.001, 0.002)  & (0.069, 0.076) &(0.081, 0.008)  &(-0.010, -0.011)  &  (0.930, 0.940)\\
$[4-5]$    & (0.060, 0.073) &(0.050, 0.062)&  (0.009, 0.011)   &(0.043, 0.057)  & (0.099, 0.112) & (0.053, -0.057) &(0.887, 0.895)  \\
$[5-6]$   & (0.078, 0.092) & (0.054, 0.066)&(0.023, 0.026)    & (-0.027, -0.041) & (0.069, 0.084) &(-0.107, -0.115)  &(0.772, 0.786)  \\
$[6-7]$   & (0.073, 0.084) & (0.040, 0.047)& (0.033, 0.036)  & (-0.093, -0.106) &(0.037, 0.050)  &(-0.038, -0.050)  &(0.650, 0.664) \\
$[7-8.17]$   &(0.045, 0.051)&  (0.018, 0.021) & (0.026, 0.029)   & (-0.096, -0.108) &  (0.013, 0.021)&  (-0.114, -0.125)&(0.501, 0.511)  \\
\hline \hline
\end{tabular}}
\label{table4}
\end{table}

\begin{table}[H]
\caption{Same as in Table~\ref{table3} but for  $B^+ \to D_1^\prime \mu^+ \nu $. The differential branching ratios are given in units of $10^{-4}$ $\text{GeV}^{-2}$.}
\centering
\resizebox{\textwidth}{!}{
\begin{tabular}{|c|ccc|ccc|c|}
\hline
\hline
\multicolumn{8}{c}{\textbf{$[-30.3^\circ, -24.9^\circ]$ }} \\
\hline
$q^2=$ GeV  & $\left\langle d\mathcal{B}/dq^2\right\rangle$ & $\left\langle d\mathcal{B}_{\text{L}}/dq^2\right\rangle$ &  $\left\langle d\mathcal{B}_{\text{T}}/dq^2\right\rangle$ &  $\left\langle \hat{\mathcal{A}}_{\text{FB}}\right\rangle$ & $\left\langle \hat{\mathcal{A}}_{\text{FBL}}\right\rangle$ & $\left\langle \hat{\mathcal{A}}_{\text{FBT}}\right\rangle$ &$\left\langle \hat{F}_{\text{L}}\right\rangle$ \\
\hline
$[0.01-1]$   & (0.245, 0.355) & (0.193, 0.290) &  (0.052, 0.064)& (-0.052, -0.069) & (0.025, 0.029) &  (-0.079, -0.097)&  (0.813, 0.850)\\
$[1-2]$     &(0.293, 0.399)  & (0.158, 0.233) & (0.135, 0.165)&(-0.217, -0.250)  & (0.002, 0.003) & (-0.220, -0.254) & (0.599, 0.665) \\
$[2-3]$     & (0.330, 0.431) &  (0.146, 0.207)& (0.184, 0.224) &  (-0.309, -0.344)&(0.001, 0.002)  &(-0.311, -0.345)  & (0.503, 0.566) \\
$[3-4]$     &(0.341, 0.433)  &(0.134, 0.184)  &(0.206, 0.249)  & (-0.357, -0.389) & $(\sim 10^{-4})$  & (-0.357, -0.390) & (0.457, 0.511) \\
$[4-5]$    & (0.330, 0.409) & (0.122, 0.160) &(0.208, 0.249)  &  (-0.363, -0.395)& $(\sim 10^{-4})$  & (-0.330, -0.395) & (0.437, 0.481) \\
$[5-6]$   & (0.299, 0.363) &(0.105, 0.134)  &(0.193, 0.228) &(-0.330, -0.361)  & $(\sim 10^{-4})$ & (-0.330, -0.361) &  (0.430, 0.464)\\
$[6-7]$   & (0.244, 0.291)& (0.084, 0.103) &   (0.160, 0.188)& (-0.147, -0.163) &$(\sim 10^{-4})$   & (-0.259, -0.287 )&  (0.435, 0.458)\\
$[7-8.12]$   & (0.156, 0.183) & (0.047, 0.056) & (0.092, 0.108) & (-0.147, -0.163) & $(\sim 10^{-5})$  & (-0.147, -0.163) & (0.450, 0.470) \\
\hline
\multicolumn{8}{c}{\textbf{$[43.3^\circ, 49.9^\circ]$ }} \\ \hline
$[0.01-1]$   & $(0.619, 0.772)$ & (0.755, 0.610) &  (0.008, 0.016)  & (0.030, 0.034) & (0.038, 0.040) &(-0.005, -0.008)& (0.972, 0.978) \\
$[1-2]$     &(0.493, 0.625)  & (0.472, 0.583) &  (0.020, 0.041)& (-0.007, -0.019) & (0.007, 0.008) & (-0.015, -0.027) &(0.949, 0.968)  \\
$[2-3]$     & (0.362, 0.476) & (0.335, 0.422) &  (0.027, 0.053)&(-0.018, -0.039)  & (0.003, 0.004) & (-0.023, -0.044) & (0.914, 0.946) \\
$[3-4]$     & (0.256, 0.350) &  (0.226, 0.292)&   (0.030, 0.058)&  (-0.025, -0.054)& (0.001, 0.002) &  (-0.028, -0.057)&(0.876, 0.919)  \\
$[4-5]$    & (0.174, 0.249) & (0.143, 0.190) &   (0.030, 0.058)& (-0.026, -0.063) & (0.001, 0.002) &(-0.028, -0.065)&(0.830, 0.883) \\
$[5-6]$   &(0.112, 0.169)  &(0.082, 0.115)  &(0.029, 0.054)  &  (-0.023, -0.064)&  (0.001, 0.002)& (-0.024, -0.065) & (0.796, 0.830) \\
$[6-7]$   &(0.066, 0.107)  & (0.040, 0.060) &(0.025, 0.046) &  (-0.014, -0.055)& $(\sim 10^{-4}) $& (-0.015, -0.056) &  (0.687, 0.747)\\
$[7-8.12]$   &  (0.031, 0.055)&  (0.013, 0.021)&(0.015, 0.027)  & (-0.002, -0.035) &  $(\sim 10^{-4}) $& (-0.002, -0.035) &  $(0.610, 0.640)$\\
\hline
\hline
\end{tabular}}
\label{table5}
\end{table}

\begin{table}[H]
\caption{Same as in Table~\ref{table4} but for $B^+ \to D_1^\prime \tau^+ \nu $. The differential branching ratios are given in units of  $10^{-5}$ $\text{GeV}^{-2}$. }
\centering
\resizebox{\textwidth}{!}{%
\begin{tabular}{|c|ccc|ccc|c|}
\hline
\hline
\multicolumn{8}{c}{\textbf{ $[-30.3^\circ, -24.9^\circ]$}} \\
\hline
$q^2=$ GeV  & $\left\langle d\mathcal{B}/dq^2\right\rangle$ & $\left\langle d\mathcal{B}_{\text{L}}/dq^2\right\rangle$ &  $\left\langle d\mathcal{B}_{\text{T}}/dq^2\right\rangle$ &  $\left\langle \hat{\mathcal{A}}_{\text{FB}}\right\rangle$ & $\left\langle \hat{\mathcal{A}}_{\text{FBL}}\right\rangle$ & $\left\langle \hat{\mathcal{A}}_{\text{FBT}}\right\rangle$ &$\left\langle \hat{F}_{\text{L}}\right\rangle$ \\
\hline

$[3.13-4]$     & (0.086, 0.117) & (0.079, 0.109) &  (0.006, 0.008)&  (0.058, 0.068)& (0.072, 0.082) &(-0.013, -0.014)  & (0.970, 0.980)  \\
$[4-5]$    &(0.352, 0.478)  & (0.281, 0.382) & (0.071, 0.085) & (0.011, 0.032) & (0.078, 0.096)& (-0.062, -0.067) &(0.857, 0.871)  \\
$[5-6]$   &(0.477, 0.610)  &(0.306, 0.407)  &(0.170, 0.203)  &  (-0.060, -0.080)&  (0.043, 0.063)& (-0.119, -0.128) &(0.727, 0.746) \\
$[6-7]$   & (0.469, 0.578) & (0.235, 0.302) &  (0.234, 0.276)&(-0.123, -0.135)  & (0.015, 0.032) & (-0.147, -0.160) & (0.605, 0.622)  \\
$[7-8.12]$   &(0.339, 0.402)  & (0.132, 0.161) &  (0.206, 0.241)&(-0.107, -0.119)  & (0.001, 0.009) & (-0.113, -0.125) & (0.501, 0.511)\\
\hline

\multicolumn{8}{c}{\textbf{ $[43.3^\circ, 49.9^\circ]$}} \\
\hline

$[3.13-4]$     &(0.047, 0.076)  &(0.074, 0.046)  & (0.001, 0.002) & (0.113, 0.124) & (0.116, 0.126) &  (-0.002, -0.003)& (0.981, 0.990) \\
$[4-5]$    & (0.188, 0.294) & (0.179, 0.276) &  (0.009, 0.018)& (0.176, 0.202) &(0.189, 0.210)  &(-0.006, -0.013)  &(0.960, 0.971)  \\
$[5-6]$   &  (0.213, 0.332)& (0.190, 0.287) &(0.023, 0.045)  &  (0.159, 0.195)&(0.181, 0.207)  & (-0.009, -0.024) & (0.911, 0.933) \\
$[6-7]$   & (0.156, 0.249) & (0.121, 0.184) &  (0.035, 0.065)& (0.117, 0.162) & (0.146, 0.175) & (-0.008, -0.032) & (0.824, 0.861) \\
$[7-8.12]$   & (0.077, 0.131) & (0.043, 0.070)& (0.033, 0.061) & (0.061, 0.105) & (0.087, 0.111) &(-0.002, -0.028)  &  (0.721, 0.759)\\
\hline \hline
\end{tabular}}
\label{table6}
\end{table}

\begin{table}[H]
\caption{Same as in Table~\ref{table3} but for  $B_s^0 \to D_{s1}^{-} \mu^+ \nu $. The differential branching ratios are given in units of  $10^{-4}$ $\text{GeV}^{-2}$.}
\centering
\resizebox{\textwidth}{!}{%
\begin{tabular}{|c|ccc|ccc|c|}
\hline
\hline
\multicolumn{8}{c}{\textbf{$[-30.3^\circ, -24.9^\circ]$ }} \\
\hline
$q^2=$ GeV  & $\left\langle d\mathcal{B}/dq^2\right\rangle$ & $\left\langle d\mathcal{B}_{\text{L}}/dq^2\right\rangle$ &  $\left\langle d\mathcal{B}_{\text{T}}/dq^2\right\rangle$ &  $\left\langle \hat{\mathcal{A}}_{\text{FB}}\right\rangle$ & $\left\langle \hat{\mathcal{A}}_{\text{FBL}}\right\rangle$ & $\left\langle \hat{\mathcal{A}}_{\text{FBT}}\right\rangle$ &$\left\langle \hat{F}_{\text{L}}\right\rangle$ \\
\hline

$[0.01-1]$   & (1.036, 1.201) & (1.007, 1.163) &  (0.028, 0.037)& (0.024, 0.028) & (0.038, 0.039) & (-0.011, -0.014) & (0.966, 0.970) \\
$[1-2]$     &(0.866, 1.017)  & (0.784, 0.909) & (0.082, 0.108) &(-0.034, -0.047)  &(0.008, 0.009)  &(-0.044, -0.056)  & (0.922, 0.936) \\
$[2-3]$     &(0.697, 0.833)  & (0.575, 0.673) & (0.122, 0.160) & (-0.085, -0.109) & (0.005, 0.006) & (-0.091, -0.115) &(0.858, 0.879)  \\
$[3-4]$     & (0.551, 0.671) & (0.406, 0.481) &  (0.145, 0.190)& (-0.145, -0.181) &  (0.003, 0.004)& (-0.149, -0.185) & (0.787, 0.814) \\
$[4-5]$    & (0.523, 0.551) & (0.272, 0.328) &(0.148, 0.194)  & (-0.208, -0.253) & (0.002, 0.003) &(-0.211, -0.256)  & (0.715, 0.744) \\
$[5-6]$   &(0.304, 0.385)  & (0.171, 0.210) &(0.133, 0.175)  & (0.001, 0.002) & (-0.264, -0.313) & (-0.264, -0.313) & (0.646, 0.674) \\
$[6-7]$   &  (0.200, 0.259)& (0.096, 0.121) & (0.103, 0.137) &(-0.262, -0.311)  &  (0.001, 0.002)&(-0.291, -0.333)  & (0.583, 0.607) \\
$[7-8.45]$   & (0.086, 0.115) & (0.034, 0.045) & (0.051, 0.069) &  (-0.214, -0.232)&  $(\sim 10^{-4})$&  (-0.214, -0.233)& 0.521, 0.537 \\
\hline
\multicolumn{8}{c}{\textbf{$[43.3^\circ, 49.9^\circ]$ }} \\ \hline

$[0.01-1]$   &(0.526, 0.827)  & (0.443, 0.734) & (0.083, 0.091) &(-0.028, -0.046)  & (0.027, 0.031) &-0.056, -0.077  & (0.886, 0.915) \\
$[1-2]$     &  (0.585, 0.849)&(0.355, 0.592) & (0.230, 0.257) & (-0.198, -0.245) & (0.004, 0.005) & (-0.204, -0.250) & (0.721, 0.789) \\
$[2-3]$     & (0.640, 0.864) &  (0.308, 0.492)&(0.322, 0.371) &(-0.357, -0.405)  & (0.002, 0.003) & (-0.360, -0.407) & (0.604, 0.682) \\
$[3-4]$     &(0.654, 0.839)  & (0.266, 0.405) & (0.387, 0.434) & (-0.497, -0.532) & (0.001, 0.002) & (-0.498, -0.534) &  (0.600, 0.604)\\
$[4-5]$    &(0.624, 0.771)  &(0.227, 0.326)  &(0.396, 0.444)  &(-0.584, -0.626)  & $(\sim 10^{-4})$ & (-0.585, -0.627) & (0.473, 0.539) \\
$[5-6]$   & (0.551, 0.663) & (0.187, 0.254) &(0.364, 0.408)  &(-0.611, -0.660)  &  $(\sim 10^{-4})$& (-0.612, -0.661) &(0.441, 0.495)  \\
$[6-7]$   &(0.441, 0.518)  &(0.144, 0.185)  &(0.296, 0.332)  & (-0.556, -0.605) & $(\sim 10^{-4})$ & (-0.556, -0.606) & (0.425, 0.467) \\
$[7-8.45]$   & (0.227, 0.239) &(0.078, 0.094)  &  (0.161, 0.182)&  (-0.311, -0.342)& $(\sim 10^{-4})$ & (-0.311, -0.342) &(0.427, 0.454)  \\
\hline
\hline
\end{tabular}}
\label{table7}
\end{table}

\begin{table}[H]
\caption{Same as in Table~\ref{table4} but for $B_s^0 \to D_{s1}^{-} \tau^+ \nu $. The differential branching ratios are given in units of  $10^{-5}$ $\text{GeV}^{-2}$. }
\centering
\resizebox{\textwidth}{!}{%
\begin{tabular}{|c|ccc|ccc|c|}
\hline
\hline
\multicolumn{8}{c}{\textbf{ $[-30.3^\circ, -24.9^\circ]$}} \\
\hline
$q^2=$ GeV  & $\left\langle d\mathcal{B}/dq^2\right\rangle$ & $\left\langle d\mathcal{B}_{\text{L}}/dq^2\right\rangle$ &  $\left\langle d\mathcal{B}_{\text{T}}/dq^2\right\rangle$ &  $\left\langle \hat{\mathcal{A}}_{\text{FB}}\right\rangle$ & $\left\langle \hat{\mathcal{A}}_{\text{FBL}}\right\rangle$ & $\left\langle \hat{\mathcal{A}}_{\text{FBT}}\right\rangle$ &$\left\langle \hat{F}_{\text{L}}\right\rangle$ \\
\hline

$[3.13-4]$     &(0.203, 0.146)  &(0.198, 0.329)  & (0.005, 0.007) &(0.122, 0.125)  &(0.127, 0.130)  & (-0.004, -0.005) &  (0.987, 0.988)\\
$[4-5]$    &(0.834, 1.005)  & (0.776, 0.930) &(0.057, 0.074) & (0.186, 0.196) & (0.212, 0.218) &  (-0.022, -0.025)&  (0.958, 0.962)\\
$[5-6]$   & (1.005, 1.217) & (0.873, 1.044)& (0.132, 0.172) &(0.165, 0.181)  & (0.223, 0.231) &  (-0.050, -0.057)& (0.916, 0.924) \\
$[6-7]$   & (0.781, 0.956) &  (0.614, 0.737)& (0.166, 0.219) & (0.118, 0.140) &  (0.217, 0.226)& (-0.085, -0.089) & (0.865, 0.877) \\
$[7-8.45]$   & (0.315, 0.398) & (0.202, 0.247) &(0.112, 0.166)  & (0.050, 0.073) &(0.200, 0.211)  &  (-0.111, -0.121)&  (0.783, 0.800)\\
\hline

\multicolumn{8}{c}{\textbf{ $[43.3^\circ, 49.9^\circ]$}} \\
\hline

$[3.13-4]$     & (0.244, 0.301) & (0.230, 0.285) & (0.014, 0.015) &(0.093, 0.098)  &  (0.104, 0.108)&(-0.010, -0.011)  & (0.977, 0.979) \\
$[4-5]$    & (1.007, 1.233) & (0.858, 1.067) &(0.149, 0.165)  & (0.091, 0.102) &(0.151, 0.162)  &(-0.058, -0.061) &(0.914, 0.923)  \\
$[5-6]$   & (1.310, 1.583) & (0.960, 1.192) &(0.350, 0.390)  &(0.006, 0.019)  & (0.141, 0.154) &(-0.133, -0.138)  & (0.831, 0.847) \\
$[6-7]$   & (1.183, 1.404) &(0.719, 0.885)  & (0.464, 0.518) & (-0.070, -0.080) &(0.121, 0.134)  & (-0.206, -0.211) &(0.742, 0.763)  \\
$[7-8.45]$   & (0.644, 0.748) & (0.434, 0.522) &(0.344, 0.388) & (-0.102, -0.110)&(0.084, 0.101)  &  (-0.184, -0.192)& (0.635, 0.656) \\
\hline \hline
\end{tabular}}
\label{table8}
\end{table}

\begin{table}[H]
\caption{Same as in Table~\ref{table3} but for  $B_s^0 \to D_{s1}^{\prime -} \mu^+ \nu $. The differential branching ratios are given in units of $10^{-4}$ $\text{GeV}^{-2}$.}
\centering
\resizebox{\textwidth}{!}{%
\begin{tabular}{|c|ccc|ccc|c|}
\hline
\hline
\multicolumn{8}{c}{\textbf{$[-30.3^\circ, -24.9^\circ]$ }} \\
\hline
$q^2=$ GeV  & $\left\langle d\mathcal{B}/dq^2\right\rangle$ & $\left\langle d\mathcal{B}_{\text{L}}/dq^2\right\rangle$ &  $\left\langle d\mathcal{B}_{\text{T}}/dq^2\right\rangle$ &  $\left\langle \hat{\mathcal{A}}_{\text{FB}}\right\rangle$ & $\left\langle \hat{\mathcal{A}}_{\text{FBL}}\right\rangle$ & $\left\langle \hat{\mathcal{A}}_{\text{FBT}}\right\rangle$ &$\left\langle \hat{F}_{\text{L}}\right\rangle$ \\
\hline

$[0.01-1]$   &(0.195, 0.363)  &(0.136, 0.294)  &(0.058, 0.068)  & (-0.062, -0.104) & (0.027, 0.028) & (-0.087, -0.131) & (0.794, 0.867) \\
$[1-2]$     &(0.270, 0.429)  &(0.111, 0.241)  &(0.159, 0.188)  & (-0.256, -0.323) & (0.003, 0.004) & (-0.259, -0.327) & (0.554, 0.685)  \\
$[2-3]$     &(0.334, 0.480)  &(0.108, 0.213)  &(0.159, 0.267)  & (-0.394, -0.445) & (0.001, 0.002) & (-0.396, -0.446) & (0.444, 0.596)  \\
$[3-4]$     &(0.464, 0.494)  &(0.105, 0.188)  &(0.258, 0.306)  & (-0.485, -0.530) & $(\sim 10^{-4})$ & (-0.486, -0.531) & (0.391, 0.497)  \\
$[4-5]$    &(0.359, 0.470)  &(0.100, 0.163)  &(0.259, 0.307)  & (-0.525, -0.578) & $(\sim 10^{-4})$ & (-0.525, -0.579) & (0.367, 0.453)  \\
$[5-6]$   &(0.322, 0.409)  &(0.091, 0.135)  &(0.231, 0.274)  & (-0.510, -0.567) & $(\sim 10^{-4})$ & (-0.510, -0.568) & (0.362, 0.428)  \\
$[6-7]$   &(0.253, 0.314)  &(0.091, 0.102)  &(0.178, 0.212)  & (-0.422, -0.474) & $(\sim 10^{-4})$ & (-0.423, -0.475) & (0.371, 0.420)  \\
$[7-8.01]$   &(0.137, 0.168)  &(0.043, 0.055)  &(0.094, 0.113)  & (-0.220, -0.248) & $(\sim 10^{-4})$ & (-0.220, -0.249) & (0.390, 0.420)  \\
\hline
\multicolumn{8}{c}{\textbf{$[43.3^\circ, 49.9^\circ]$ }} \\ \hline

$[0.01-1]$   &(0.613, 0.762)  &(0.606, 0.747)  &(0.007, 0.014)  & (0.032, 0.037) & (0.039, 0.041) & (-0.004, -0.007) & (0.975, 0.980)  \\
$[1-2]$     &(0.484, 0.615)  &(0.642, 0.574)  &(0.021, 0.041)  & (-0.008, -0.018) & (0.009, 0.010) & (-0.018, -0.029) & (0.952, 0.968)  \\
$[2-3]$     &(0.353, 0.467)  &(0.320, 0.405)  &(0.033, 0.061)  & (-0.033, -0.055) & (0.005, 0.006) & (-0.039, -0.061) & (0.905, 0.934)  \\
$[3-4]$     &(0.248, 0.344)  &(0.208, 0.271)  &(0.039, 0.072)  & (-0.063, -0.098) & (0.004, 0.005) & (-0.068, -0.102) & (0.847, 0.887)  \\
$[4-5]$    &(0.165, 0.243)  &(0.125, 0.170)  &(0.039, 0.073)  & (-0.101, -0.146) & (0.003, 0.004) & (-0.105, -0.149) & (0.779, 0.828)  \\
$[5-6]$   &(0.101, 0.159)  &(0.067, 0.096)  &(0.033, 0.063)  & (-0.143, -0.192) & (0.002, 0.003) & (-0.146, -0.195) & (0.705, 0.757)  \\
$[6-7]$   &(0.053, 0.092)  &(0.029, 0.046)  &(0.023, 0.045)  & (-0.177, -0.219) & (0.001, 0.002) & (-0.179, -0.221) & (0.628, 0.676)  \\
$[7-8.01]$   &(0.019, 0.036)  &(0.008, 0.014)  &(0.010, 0.021)  & (-0.152, -0.168) & $(\sim 10^{-4})$ & (-0.153, -0.169) & (0.570, 0.616)  \\
\hline
\hline
\end{tabular}}
\label{table9}
\end{table}

\begin{table}[H]
\caption{Same as in Table~\ref{table4} but for $B_s^0 \to D_{s1}^{\prime -} \tau^+ \nu $. The differential branching ratios are given in units of $10^{-5}$ $\text{GeV}^{-2}$. }
\centering
\resizebox{\textwidth}{!}{%
\begin{tabular}{|c|ccc|ccc|c|}
\hline
\hline
\multicolumn{8}{c}{\textbf{ $[-30.3^\circ, -24.9^\circ]$}} \\
\hline
$q^2=$ GeV  & $\left\langle d\mathcal{B}/dq^2\right\rangle$ & $\left\langle d\mathcal{B}_{\text{L}}/dq^2\right\rangle$ &  $\left\langle d\mathcal{B}_{\text{T}}/dq^2\right\rangle$ &  $\left\langle \hat{\mathcal{A}}_{\text{FB}}\right\rangle$ & $\left\langle \hat{\mathcal{A}}_{\text{FBL}}\right\rangle$ & $\left\langle \hat{\mathcal{A}}_{\text{FBT}}\right\rangle$ &$\left\langle \hat{F}_{\text{L}}\right\rangle$ \\
\hline

$[3.13-4]$     &(0.114, 0.153)  &(0.105, 0.142)  &(0.009, 0.010)  & (0.081, 0.089) & (0.096, 0.103) & (-0.013, -0.015) & (0.970, 0.980)  \\
$[4-5]$    &(0.477, 0.630)  &(0.381, 0.517)  &(0.096, 0.112)  & (0.055, 0.071) & (0.128, 0.143) & (-0.069, -0.074) & (0.881, 0.896)  \\
$[5-6]$   &(0.632, 0.814)  &(0.413, 0.556)  &(0.218, 0.257)  & (-0.020, -0.040) & (0.107, 0.126) & (-0.146, -0.152) & (0.774, 0.798)  \\
$[6-7]$   &(0.569, 0.715)  &(0.295, 0.413)  &(0.273, 0.323)  & (-0.106, -0.118) & (0.101, 0.107) & (-0.200, -0.209) & (0.669, 0.699)  \\
$[7-8.01]$   &(0.319, 0.392)  &(0.129, 0.164)  &(0.189, 0.227)  & (-0.101, -0.109) & (0.047, 0.061) & (-0.153, -0.166) & (0.606, 0.630)  \\
\hline

\multicolumn{8}{c}{\textbf{ $[43.3^\circ, 49.9^\circ]$}} \\
\hline

$[3.13-4]$     &(0.075, 0.111)  &(0.073, 0.108)  &(0.001, 0.002)  & (0.133, 0.142) & (0.137, 0.145) & (-0.002, -0.003) & (0.980, 0.990)  \\
$[4-5]$    &(0.309, 0.452)  &(0.294, 0.424)  &(0.015, 0.028)  & (0.215, 0.236) & (0.234, 0.250) & (-0.014, -0.018) & (0.966, 0.973)  \\
$[5-6]$   &(0.352, 0.517)  &(0.319, 0.455)  &(0.033, 0.062)  & (0.208, 0.236) & (0.249, 0.267) & (-0.030, -0.041) & (0.932, 0.947)  \\
$[6-7]$   &(0.234, 0.354)  &(0.196, 0.282)  &(0.037, 0.071)  & (0.174, 0.210) & (0.245, 0.264) & (-0.054, -0.071) & (0.888, 0.914)  \\
$[7-8.01]$   &(0.076, 0.125)  &(0.053, 0.080)  &(0.022, 0.044)  & (0.111, 0.152) & (0.204, 0.232) & (-0.078, -0.094) & (0.827, 0.866)  \\
\hline \hline
\end{tabular}}
\label{table10}
\end{table}

\begin{table}[H]
\caption{Bin-wise values of the ratio $\mathcal{R}_{D_{(s)1}^{(\prime)}}(q^2)$ for the decay $B_{(s)}^{0,+} \to D_{(s)1}^{(\prime)} \tau^+ \nu/B_{(s)}^{0,+} \to D_{(s)1}^{(\prime)} \mu^+ \nu$ in different $q^2$ intervals.}
\centering
\resizebox{0.7\textwidth}{!}{%
\begin{tabular}{|c|cc|cc|}
\hline
\hline
\multicolumn{5}{c}{\textbf{$[-30.3^\circ, -24.9^\circ]$}} \\
\hline
$q^2$ (GeV$^2$) & $\langle \mathcal{R}_{D_1} \rangle$ & $\langle \mathcal{R}_{D_1^\prime} \rangle$ & $\langle \mathcal{R}_{D_{s1}} \rangle$& $\langle \mathcal{R}_{D^\prime_{s1}} \rangle$ \\
\hline

$[3.13-4]$   &(0.021, 0.025)  & (0.024, 0.025) &(0.044, 0.046) &(0.036, 0.037) \\
$[4-5]$      & (0.122, 0.124) & (0.121, 0.123 )& (0.195, 0.201)&(0.183, 0.189) \\
$[5-6]$      & (0.192, 0.195) & (0.190, 0.193)& (0.318, 0.332)&(0.292, 0.306) \\
$[6-7]$      & (0.222, 0.227) &  (0.220, 0.224)& (0.370, 0.390) &(0.328, 0.346)\\
$[7-(m_{B_{(s)}^{0,+}}-m_{D_{(s)1}^{(\prime)}})^2]$   & (0.234, 0.241) &(0.225, 0.232)  & (0.329, 0.343) &(0.287, 0.298)\\
\hline

\multicolumn{5}{c}{\textbf{$[43.3^\circ, 49.9^\circ]$}} \\
\hline

$[3.13-4]$   &(0.027, 0.031)  &(0.026, 0.027)  & (0.040, 0.044)&  (0.32, 0.035)\\
$[4-5]$      &(0.115, 0.120)  & (0.114, 0.119) & (0.155, 0.167) &(0.142, 0.152)\\
$[5-6]$      & (0.170, 0.176) &(0.168, 0.174)  &(0.231, 0.246)  &(0.210, 0.223)\\
$[6-7]$      &(0.200, 0.205)  & (0.216, 0.220) &(0.263, 0.276 ) &(0.237, 0.248)\\
$[7-(m_{B_{(s)}^{0,+}}-m_{D_{(s)1}^{(\prime)}})^2]$   & (0.220, 0.225) & (0.225, 0.232) &  (0.263, 0.270)&(0.239, 0.244)\\
\hline \hline
\end{tabular}}
\label{tableRD}
\end{table}

\section{Decay Kinematics and Scalar Products for Unpolarized Final State meson}
\label{APPA}
The relevant non zero scalar products among the four momenta of initial meson, final state meson, charge lepton and neutrino i.e., $p^\mu$, $k^\mu$, $p_1^\mu$ and $p_2^\mu$  respectively are defined as
\begin{align}
p_{1}^{\mu} p_{1\,\mu} = m_\ell^2, \quad 
p_{2}^{\mu} p_{2\mu} = 0, \quad
p^{\mu} p_{\mu} = m_{B_{(s)}^{0,+}}^2, \quad
k^{\mu} k_{\mu} = m_{D^{(\prime)}_{(s)1}}^2, \quad
q^{\mu} q_{\mu} = q^2, \quad p_{1}^{\mu} p_{2\mu} = \frac{q^2 - m_\ell^2}{2}.
\end{align}
The scalar product between the charged lepton with the final state meson is given by,
\begin{align} 
p^\mu k_\mu =
\frac{m_{B_{(s)}^{0,+}}^2 + m_{D^{(\prime)}_{(s)1}}^2 - q^2}{2}, \quad k^{\mu} p_{1\mu} =
\frac{(q^2+m_\ell^2)\,(m_{B_{(s)}^{0,+}}^2-m_{D^{(\prime)}_{(s)1}}^2-q^2) - (q^2-m_\ell^2)\sqrt{\lambda(m_{B_{(s)}^{0,+}}^2, m_{D^{(\prime)}_{(s)1}}^2, q^2)}\,\cos{\theta}}{4q^2}.
\end{align}
with the momentum transfer defined as $q^\mu = p^\mu - k^\mu$ along with $P^\mu=p^\mu+k^\mu$, the
following scalar products then hold,
\begin{align}
 k^\mu q_\mu&= k^\mu p_\mu - k^\mu k_\mu, \quad
p^\mu q_\mu = p^\mu p_\mu - p^\mu k_\mu, \quad
p_{1}^{\mu} q_{\mu} = p^{\mu} p_{1\mu} - k^{\mu} p_{1\mu}, \quad p_{2}^{\mu} q_{\mu} =  p^{\mu} p_{2\mu} - k^{\mu} p_{2\mu}, \\  p^{\mu} p_{1\mu} &=
p_{1}^{\mu} p_{1\mu} +
p_{1}^{\mu} p_{2\mu} +
p_{1}^{\mu} k_{\mu}, \quad  p^{\mu} p_{2\mu} =
p^{\mu} p_{\mu} -
p^{\mu} p_{1\mu} -
p^{\mu} k_{\mu}, \quad k^{\mu} p_{2\,\mu} =
k^\mu p_\mu - k^\mu k_\mu - k^{\mu} p_{1\mu}.
\end{align}
\section{Decay Kinematics and Scalar Products for Longitudinal polarized Final State meson}
\label{longiA}
For the longitudinal polarized final state meson the non zero scalar products are given, 
\begin{align}
k \cdot \varepsilon^*(k)& =k \cdot \varepsilon(k) = 0, \quad\varepsilon^*(k) \cdot \varepsilon(k) = -1, \\
 p \cdot \varepsilon(k)& = p \cdot \varepsilon^*(k) =  q \cdot \varepsilon(k) = q \cdot \varepsilon^*(k)=\frac{\sqrt{\lambda(m_{B_{(s)}^{0,+}}^2, m_{D^{(\prime)}_{(s)1}}^2, q^2)}}{2m_{D^{(\prime)}_{(s)1}}^2},  \\
    p_1 \cdot \varepsilon(k)&= p_1 \cdot \varepsilon^{*}(k) = \frac{1}{m_{D^{(\prime)}_{(s)1}}} \left( \frac{\sqrt{\lambda(m_{B_{(s)}^{0,+}}^2, m_{D^{(\prime)}_{(s)1}}^2, q^2)}}{4} -\frac{1}{4}\sqrt{1 - \frac{m_\ell^2}{q^2}} (m_{B_{(s)}^{0,+}}^2 - m_{D^{(\prime)}_{(s)1}}^2 - q^2) \cos{\theta}  \right), \\
       p_2 \cdot \varepsilon(k)& =p_2 \cdot \varepsilon^*(k)= \frac{1}{m_{D^{(\prime)}_{(s)1}}} \left( \frac{\sqrt{\lambda(m_{B_{(s)}^{0,+}}^2, m_{D^{(\prime)}_{(s)1}}^2, q^2)}}{4} +\frac{1}{4}\sqrt{1 - \frac{m_\ell^2}{q^2}} (m_{B_{(s)}^{0,+}}^2 - m_{D^{(\prime)}_{(s)1}}^2- q^2) \cos{\theta} \right). 
\end{align}
\section{Decay Kinematics and Scalar Products for Transverse polarized Final State meson}
\label{tansA}
For the transverse polarized final state meson the non zero scalar products are given, 
\begin{align}
\varepsilon^*(k)\cdot \varepsilon(k)& = -1, \quad \varepsilon_{\mu\nu\rho\sigma} k^\mu p^\nu \varepsilon^{*\rho}(k) \varepsilon^\sigma(k) = 0, \quad
    \varepsilon_{\mu\nu\rho\sigma} p_1^\mu p_2^\nu \varepsilon^{*\rho}(k) \varepsilon^\sigma(k) = 0,\\
    p_1 \cdot \varepsilon(k)& =   p_1 \cdot \varepsilon^*(k) =-\frac{\sqrt{q^2 - m_\ell^2}}{2\sqrt{2}} \sqrt{1 - \cos^2{\theta}}, \quad
    p_2 \cdot \varepsilon(k) = p_2 \cdot \varepsilon^*(k) = \frac{\sqrt{q^2 - m_\ell^2}}{2\sqrt{2}} \sqrt{1 - \cos^2{\theta}}, \\
     p \cdot \varepsilon(k) &= p \cdot \varepsilon^*(k) = \frac{\sqrt{\lambda(m_{B_{(s)}^{0,+}}^2, m_{D^{(\prime)}_{(s)1}}^2, q^2)}}{2m_{D^{(\prime)}_{(s)1}}}, \quad
    q \cdot \varepsilon(k) =q \cdot \varepsilon^*(k) = \frac{\sqrt{\lambda(m_{B_{(s)}^{0,+}}^2, m_{D^{(\prime)}_{(s)1}}^2, q^2)}}{2m_{D^{(\prime)}_{(s)1}}},
\\
    p_1 \cdot \varepsilon(k) &= p_1 \cdot \varepsilon^*(k) = \frac{1}{m_{D^{(\prime)}_{(s)1}}} \left( \frac{\sqrt{\lambda(m_{B_{(s)}^{0,+}}^2, m_{D^{(\prime)}_{(s)1}}^2, q^2)}}{4} - \frac{\sqrt{q^2 - m_\ell^2}}{4\sqrt{s}} (m_{B_{(s)}^{0,+}}^2 - m_{D^{(\prime)}_{(s)1}}^2 - q^2) \cos{\theta} \right), \\
    p_2 \cdot \varepsilon(k)& = p_2 \cdot \varepsilon^*(k) = \frac{1}{m_{D^{(\prime)}_{(s)1}}} \left( \frac{\sqrt{\lambda(m_{B_{(s)}^{0,+}}^2, m_{D^{(\prime)}_{(s)1}}^2, q^2)}}{4} + \frac{\sqrt{q^2 - m^2}}{4\sqrt{q^2}} (m^2_{B_{(s)}^{0,+}} - m_{D^{(\prime)}_{(s)1}}^2 - q^2) \cos{\theta} \right).\\
    \varepsilon_{\mu\nu\rho\sigma} p_1^\mu k^\nu p^\rho \varepsilon^\sigma(k) 
   & =\varepsilon_{\mu\nu\rho\sigma} k^\mu p^\nu p_1^\rho \varepsilon^\sigma(k)= \varepsilon_{\mu\nu\rho\sigma} k^\mu p^\nu p_2^\rho \varepsilon^{*\sigma}(k)=-\frac{iG\sqrt{q^2 - m_\ell^2}}{4\sqrt{2}} \sqrt{\lambda(m_{B_{(s)}^{0,+}}^2, m_{D^{(\prime)}_{(s)1}}^2, q^2)} \sqrt{1 - \cos^2{\theta}}, \\
    \varepsilon_{\mu\nu\rho\sigma} k^\mu p^\nu p_1^\rho \varepsilon^{*\sigma}(k)&= \varepsilon_{\mu\nu\rho\sigma} p_2^\mu k^\nu p^\rho \varepsilon^\sigma(k) 
    =\varepsilon_{\mu\nu\rho\sigma} k^\mu p^\nu p_2^\rho \varepsilon^\sigma(k) 
    = \frac{iG\sqrt{q^2 - m_\ell^2}}{4\sqrt{2}} \sqrt{\lambda(m_{B_{(s)}^{0,+}}^2, m_{D^{(\prime)}_{(s)1}}^2, q^2)} \sqrt{1 - \cos^2{\theta}}.
\end{align}
The remaining scalar products are zero. The G correspond to the transverse polarization component $+$ and $-.$
\clearpage

\bibliographystyle{refstyle}
\bibliography{references}

\end{document}